\documentclass[acmtog,table,nonacm]{acmart}
\usepackage[ruled,linesnumbered]{algorithm2e} 

\usepackage{subfigure}
\usepackage{epstopdf}
\usepackage{wrapfig}
\usepackage{multirow}
\usepackage{mathrsfs}
\usepackage{colortbl}
\usepackage{BOONDOX-cal}
\usepackage{url}
\usepackage{tikz}
\usepackage{makecell}
\usepackage{cleveref}
\usepackage{todonotes}

\Crefname{equation}{Eq.}{Eqs.}
\Crefname{figure}{Fig.}{Figs.}

\newcommand{\TOMODIFY}[1]{\textcolor{red}{#1}}

\newcommand{\x}{\mathbf{x}}
\newcommand{\y}{\mathbf{y}}
\newcommand*\Laplace{\mathop{}\!\mathbin\bigtriangleup}

\AppendGraphicsExtensions{.tga}

\acmJournal{TOG
}
\begin{document}
\setcopyright{none} 
\author{Tianyu Wang}
\affiliation{\institution{FaceUnity} \city{Hangzhou} \state{Zhejiang} \country{China}}
\email{wtyatzoo@zju.edu.cn}
\authornote{Corresponding author}
\author{Jiong Chen}
\affiliation{\institution{LTCI, Telecom Paris, Institut Polytechnique de Paris} \city{Paris} \country{France}}
\email{chenjiong1991@126.com}
\author{Dongping Li}
\affiliation{\institution{FaceUnity} \city{Hangzhou} \state{Zhejiang} \country{China}}
\email{dongpingli@faceunity.com}
\author{Xiaowei Liu}
\affiliation{\institution{FaceUnity} \city{Hangzhou} \state{Zhejiang} \country{China}}
\email{liuxiaowei_TH@hotmail.com}
\author{Huamin Wang}
\affiliation{\institution{Style3D} \city{Hangzhou} \state{Zhejiang} \country{China}}
\email{wanghmin@gmail.com}
\author{Kun Zhou}
\affiliation{\institution{State Key Lab of CAD\&CG, Zhejiang University} \city{Hangzhou} \state{Zhejiang} \country{China}}
\email{kunzhou@acm.org}

\title{Fast GPU-Based Two-Way Continuous Collision Handling}

\begin{teaserfigure}
    \subfigure[The initial state]{\includegraphics[width=1.74in]{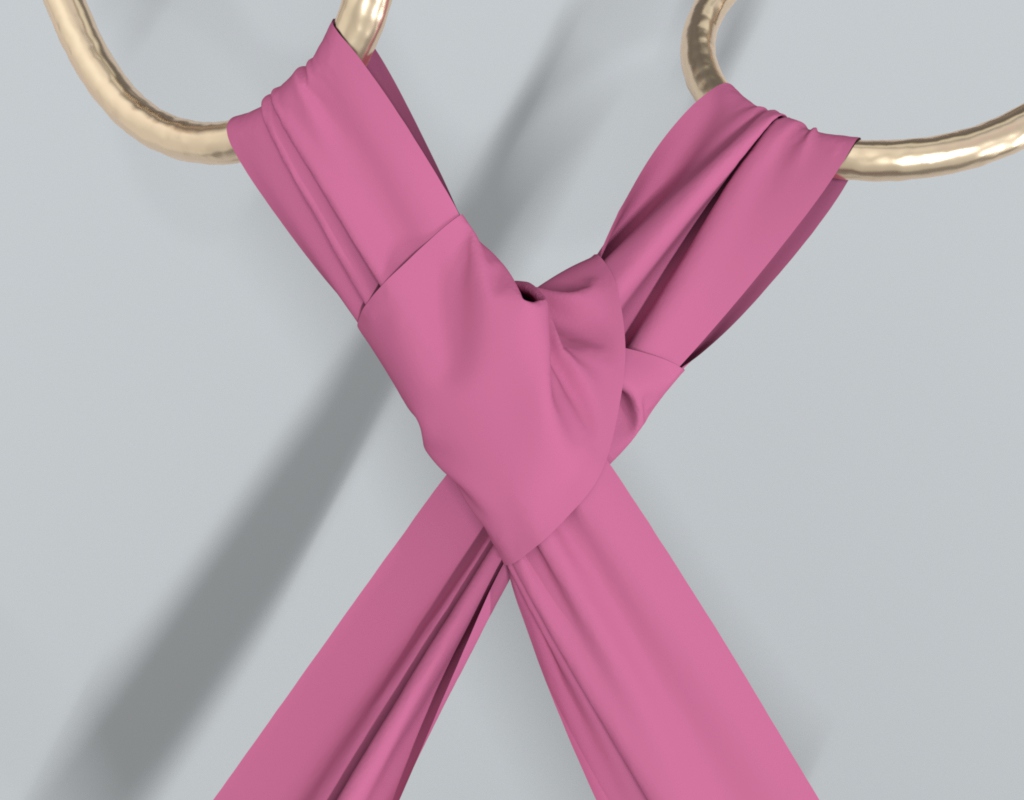}}
    \subfigure[The tightening state]{\includegraphics[width=1.74in]{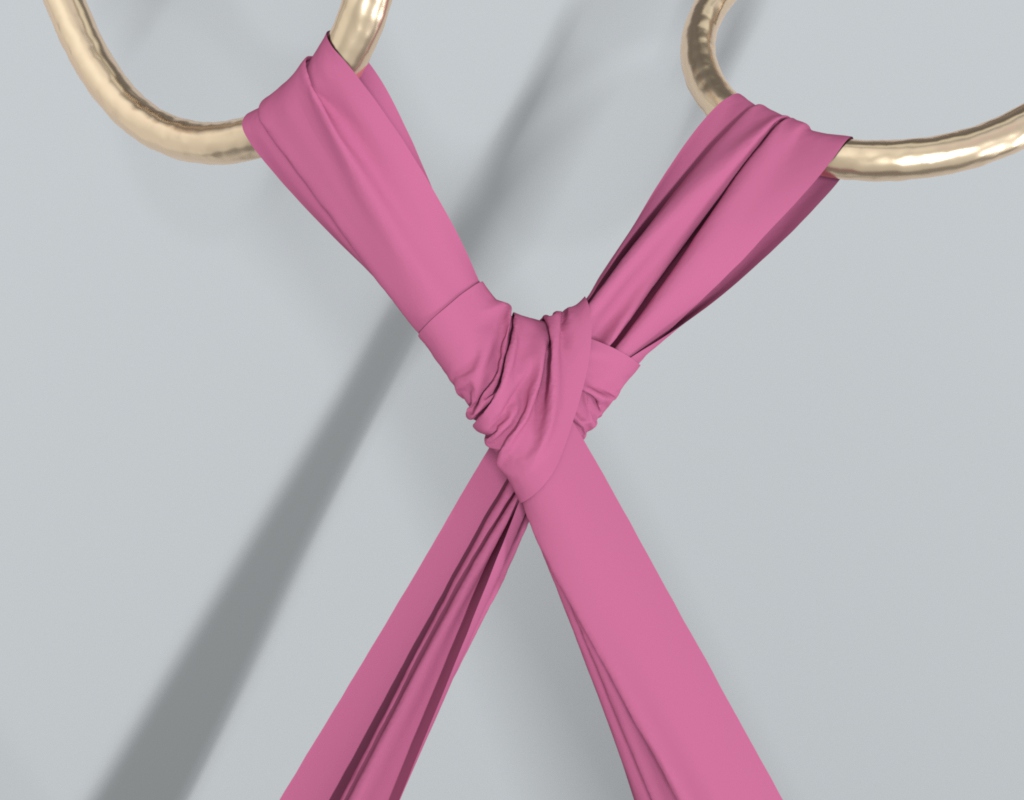}}     
    \subfigure[The final state]{\includegraphics[width=1.74in]{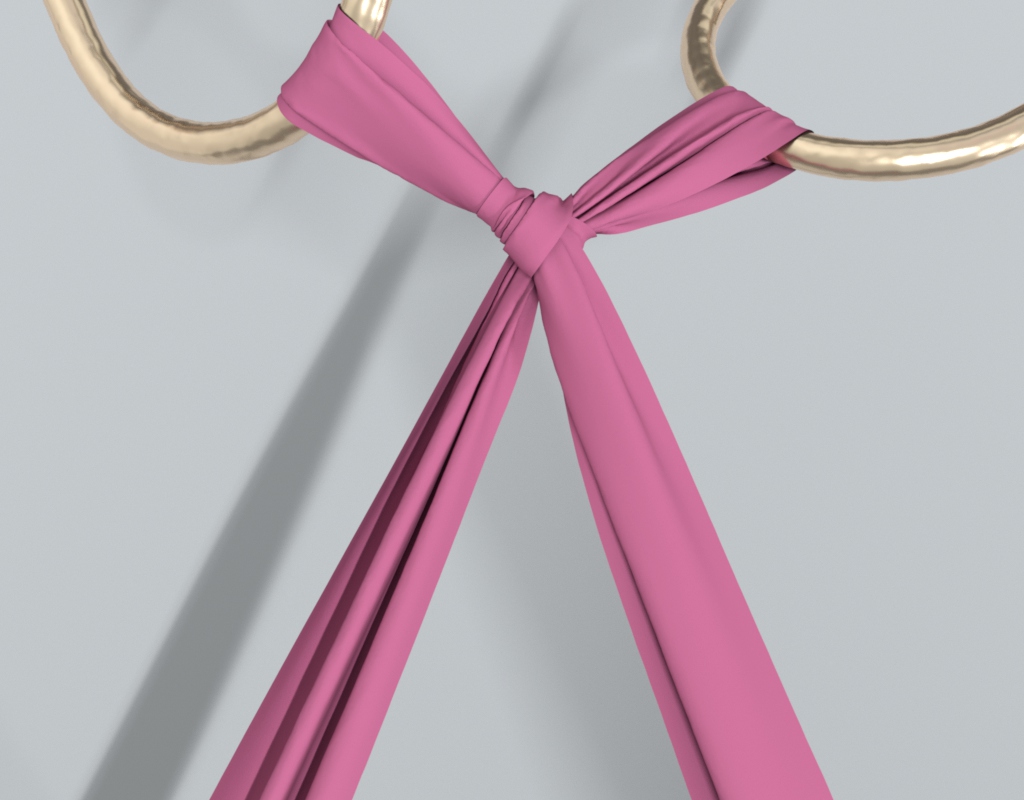}} 
    \subfigure[The final state in a  closeup]{\includegraphics[width=1.74in]{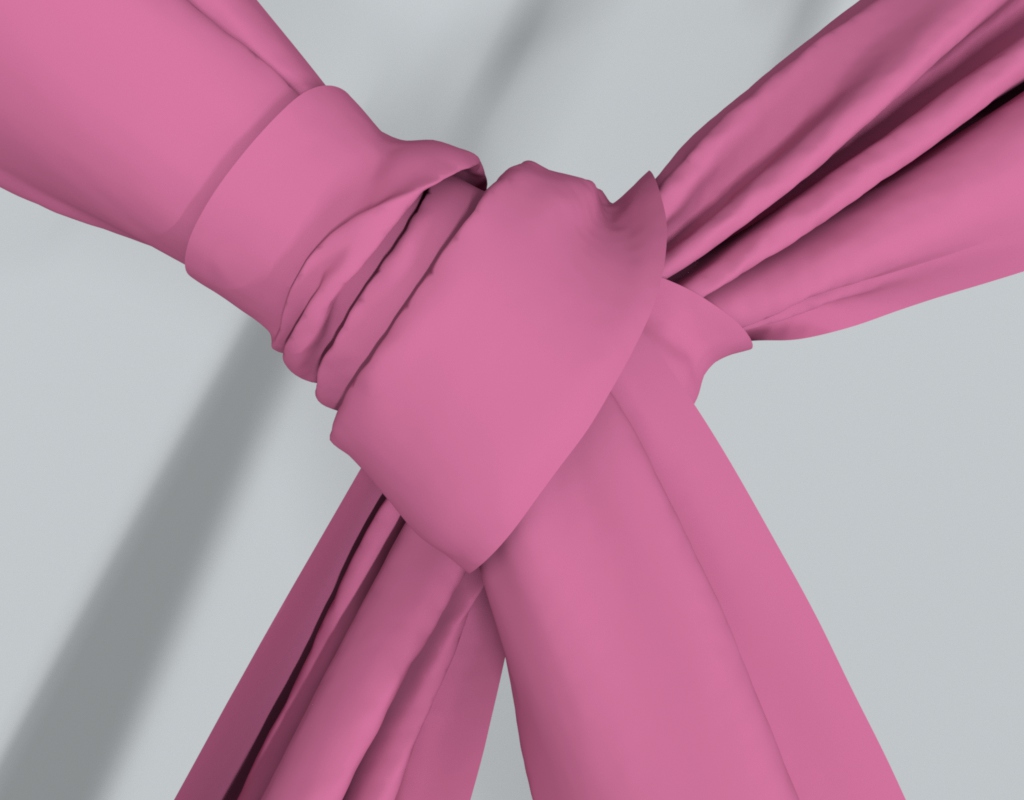}} \\ 
    \subfigure[The initial state]{\includegraphics[width=1.74in]{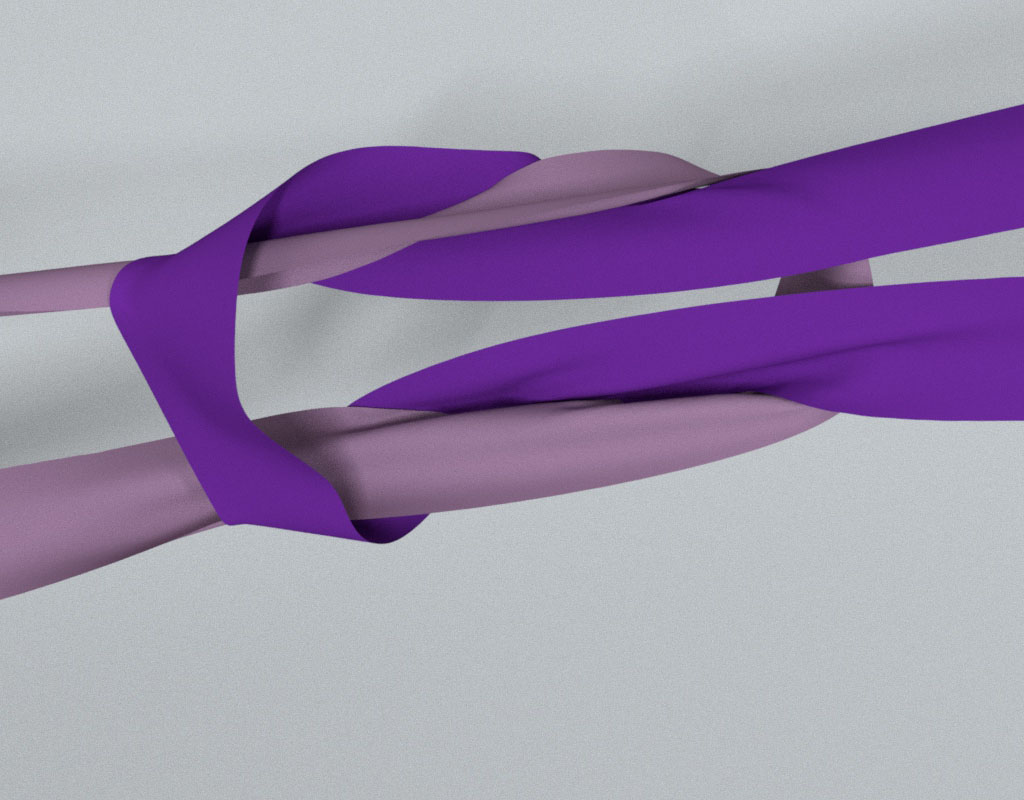}}
    \subfigure[The tightening state]{\includegraphics[width=1.74in]{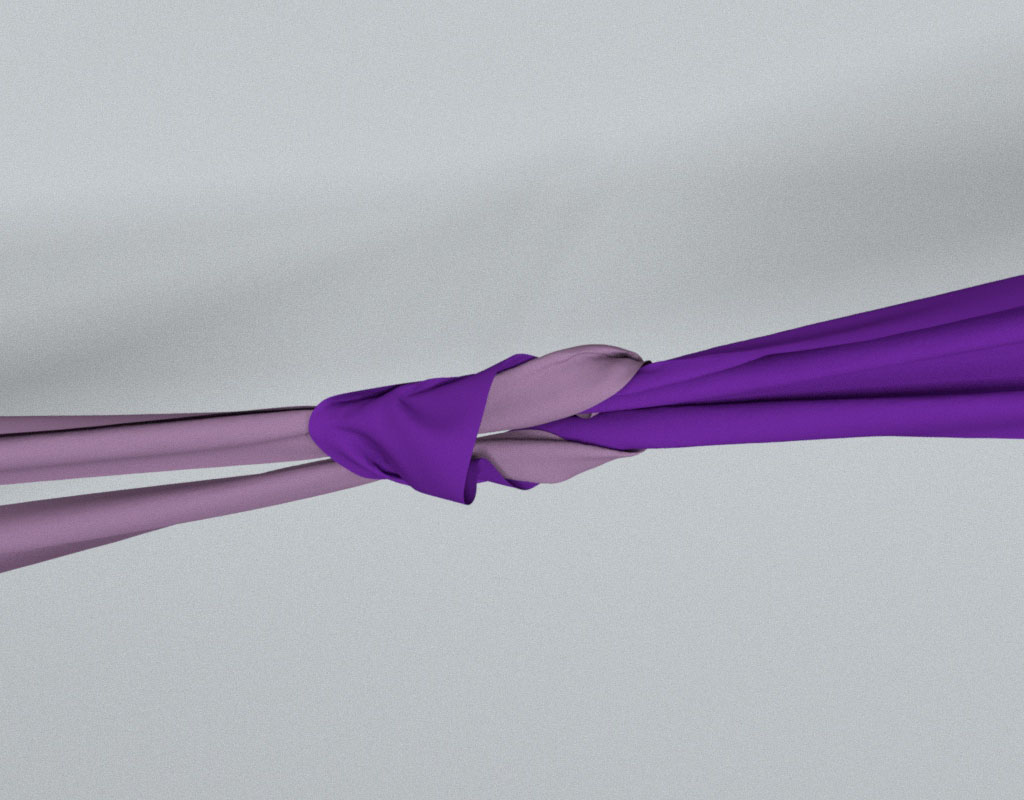}}
    \subfigure[The final state]{\includegraphics[width=1.74in]{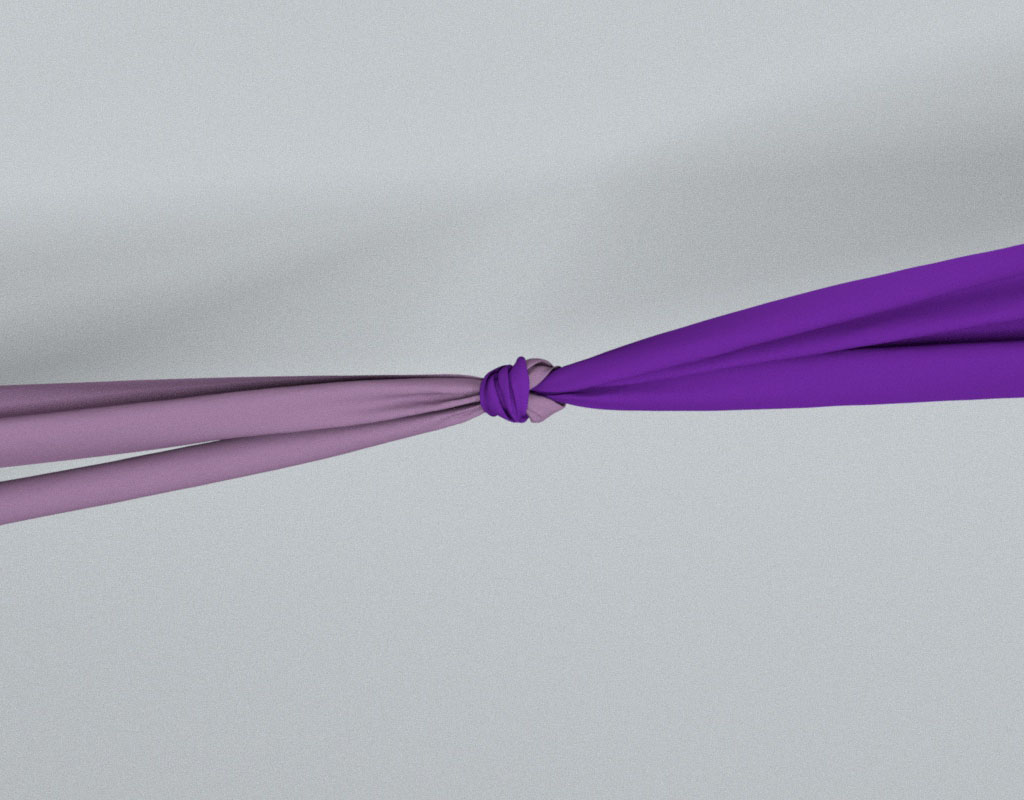}}
    \subfigure[The final state in a closeup]{\includegraphics[width=1.74in]{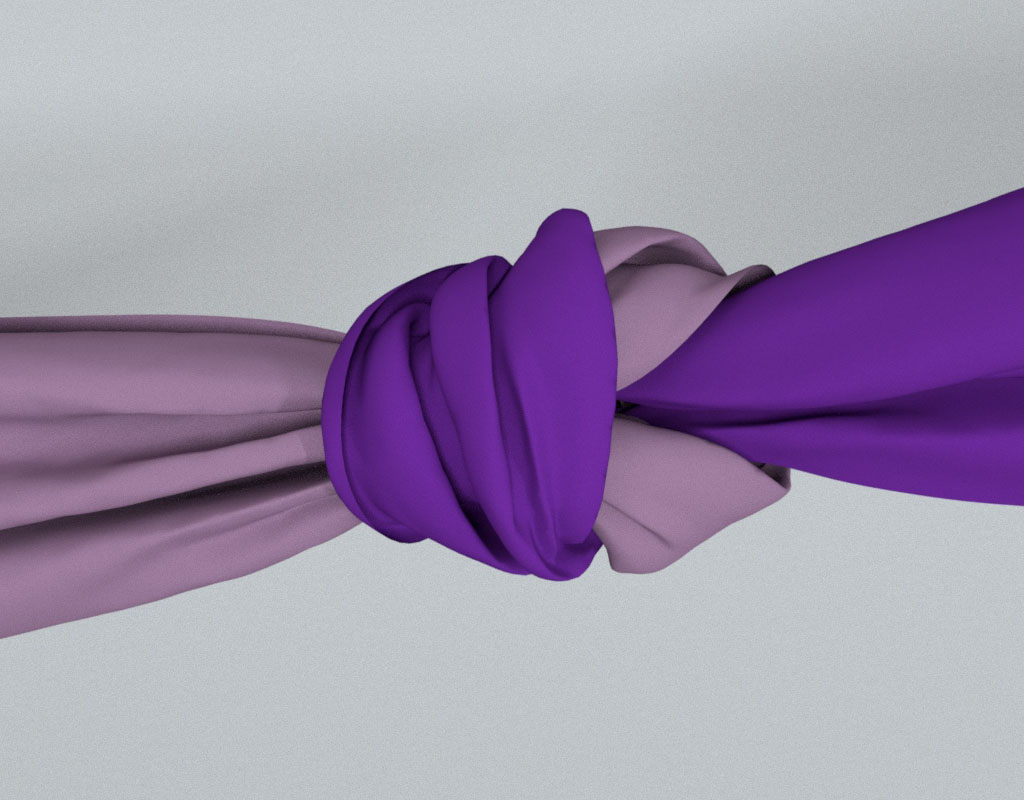}} \\     \vspace{-0.12in}
    \caption{\textbf{Knotting.} The bow knot example (on the top, with 142K triangles) and the reef knot example (on the bottom, with 71K triangles) are presented. In this work, we develop a two-way method for \emph{safe and fast} collision handling in deformable body simulation.
      Thanks to this method, our simulator can robustly handle complex collision contacts in these two examples at 4 to 17 FPS and 10 to 21 FPS.
    }
    \label{fig:teaser}
    \vspace{0.12in}
  \end{teaserfigure}

  \begin{abstract}
    Step-and-project is a popular way to simulate non-penetrated
    deformable bodies in physically-based animation. First integrating
    the system in time regardless of contacts and post resolving
    potential intersections practically strike a good balance between
    plausibility and efficiency. However, existing methods could be
    defective and unsafe when the time step is large, taking risks of
    failures or demands of repetitive collision testing and resolving
    that severely degrade performance. In this paper, we propose a
    novel two-way method for fast and reliable continuous collision
    handling. Our method launches the optimization at both ends of the
    intermediate time-integrated state and the previous
    intersection-free state, progressively generating a piecewise-linear path
    and finally reaching a feasible solution for the next time
    step. Technically, our method interleaves between a forward step
    and a backward step at a low cost, until the result is
    conditionally converged. Due to a set of unified volume-based
    contact constraints, our method can flexibly and reliably handle a
    variety of codimensional deformable bodies, including volumetric
    bodies, cloth, hair and sand. The experiments show that our method
    is safe, robust, physically faithful and numerically efficient,
    especially suitable for large deformations or large time steps.
   
\end{abstract}


\maketitle

\section{Introduction}
\label{sec:intro}
The simulation of intersection-free deformable body dynamics can be formulated as a constrained optimization problem~\cite{kane1999finite,martin2011example}:
 \begin{equation}
	\x^{t+1} = \mathop{\arg\min}\limits_{\x} {E\big(\x,\x^{t},\mathbf{v}^{t} \big) }, \quad\; 
	\mathrm{s.t.} \;\; \mathcal X\big(\x^{t}, \x\big) \subset \Omega,
	\label{eq:objective}
\end{equation}
in which $\x^{t}, \mathbf{v}^{t} \in \mathbb R^{3N}$ are the stacked position and velocity vectors of $N$ vertices at time $t$, $E(\x,\x^{t},\mathbf{v}^{t} )$ is the dynamics objective, $\mathcal X(\x^{t}, \x)$ is a sufficiently short {\color{black}linear or piecewise linear} path from $\x^{t}$ to $\x$, and $\Omega$ is the feasible intersection-free region.
Li et al.~\shortcite{li2020incremental,li2020codimensional} showed that augmenting the objective function with a smoothed Log-barrier-based contact energy term to convert the original constrained problem into an unconstrained one and then using Newton's method with a continuous-collision-detection-based pre-filtered line search strategy \cite{smith2015bijective} could get globally convergent solutions.

However, their techniques are computationally expensive, due to the
frequent launching of the high cost dynamics solver and {\color{black}truncated} step
sizes needed for keeping the path within $\Omega$.  A common
strategy~\cite{harmon2008robust,narain2012adaptive,tang2016cama,tang2018clotha,Tang:2018:PPS,li2020p}
for more efficient simulation is to divide the optimization into two
steps:
\begin{equation}
\left\{	\begin{array}{l}
         \y^{[k+1]} = \mathop{\arg\min}\limits_{\y} Q_k\big(\y,\x^{[k]},\x^{t},\mathbf{v}^{t} \big),\\ 
	\x^{[k+1]} =  \mathop{\arg\min}\limits_{\x} {D\big(\x, \y^{[k+1]} \big) }, \quad\; 
	\mathrm{s.t.} \;\; \mathcal X\big(\x^{[k]}, \x\big) \subset \Omega.
	\end{array}
	\label{eq:step_and_project}
\right.
\end{equation}
In each iteration, \textcolor{black}{firstly the dynamics solver forms
  a quadratic model $Q_k$ (or a linear model
  $L_k$~\cite{wang2016descent,Wu:2020:ASF}) of
  $E(\x,\x^{t},\mathbf{v}^{t} )$ at $\mathbf x^{[k]} \in \Omega$} to
compute a target state $\mathbf y^{[k+1]}$, and then
\textcolor{black}{a collision handling module addresses all potential
  intersections to obtain a feasible state $\mathbf
  x^{[k+1]}$}. \textcolor{black}{Here $D(\mathbf x, \mathbf y^{[k+1]}
  )$ is a metric measuring the distance between $\mathbf x$ and
  $\mathbf y^{[k+1]}$}, which is supposed to be significantly simpler
than $E(\mathbf x,\mathbf x^{t},\mathbf v^{t} )$. After several
iterations of Eq.~\eqref{eq:step_and_project}, the algorithm reports
$\mathbf x^{[k+1]}$ as $\mathbf x^{t+1}$ for Eq.~\eqref{eq:objective}.
This strategy avoids frequent launching of high-cost solvers such
as~\cite{li2020incremental,li2020codimensional}, but it sacrifices
part of accuracy in exchange for performance since this alternating
approach might not converge to a local minimum.
{\color{black}Li et al.~\shortcite{li2020codimensional} showed that this inaccuracy could be 
observed from the wrinkling or jittering artifacts and parameter tuning could alleviate this problem.}

\setlength{\columnsep}{0.16in}
\begin{wrapfigure}[11]{r}{1.55in}\vspace*{-1mm}
	\centering
	\vspace{-0in}  
	\includegraphics[width=1.55in]{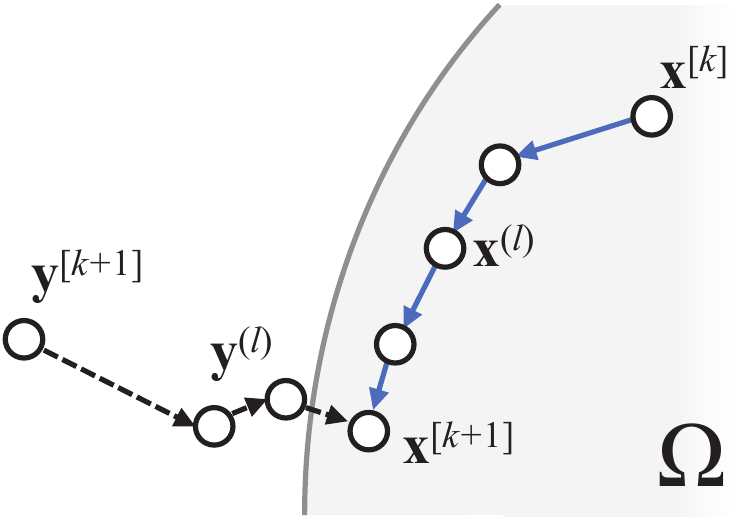}
	\vspace{-6mm}
	\caption{\textbf{The two-way collision handling approach.}}
	\label{fig:our_opt}
\end{wrapfigure}
Intuitively, the collision handling module projects $\mathbf
y^{[k+1]}$ back to $\Omega$, but it also needs to ensure that a
sufficiently short path exists to avoid tunneling artifacts as shown
in Fig.~\ref{fig:tunneling}.  In the past,
researchers~\cite{harmon2008robust,narain2012adaptive,tang2016cama,tang2018clotha,Tang:2018:PPS,li2020p}
parameterized this path as a line segment and evaluated $\mathcal
X(\mathbf x^{[k]}, \mathbf x) \subset \Omega$ by continuous collision
detection (CCD) tests. This solution is generally plausible when
$\mathbf y^{[k+1]}$ is close to $\mathbf x^{[k]}$, but it is no longer
the case as two states go far away from each other, where finding
$\mathbf x^{[k+1]}$ on the line segment that passes all of the CCD
tests can become extremely
difficult~\cite{tang2018clotha}. \textcolor{black}{Besides, how to
  measure the distance for projection is also crucial for realistic
  simulation.} Using mass-weighted $L_2$ norm is straightforward
\textcolor{black}{for a minimal change in post-response kinetic
  energy~\cite{harmon2008robust,narain2012adaptive,tang2016cama,tang2018clotha,Tang:2018:PPS,li2020p}},
but it can severely distort local elements to resolve all contacts
(see~\Cref{fig:cmp_arcsim,fig:cmp_with_impact_zone}), resulting in
\textcolor{black}{spuriously over-stretched artifacts} or oscillations
over time.

In this paper, we develop a collision handling algorithm for the
step-and-project method, capable of finding a quality solution
$\mathbf x^{[k+1]}$ at a low cost. Our key idea is a two-way approach
shown in Fig.~\ref{fig:our_opt}. In this approach, we iteratively
solve two steps: a \emph{backward step} finding a sequence of targets
$\{\mathbf y^{(l)}\}$ by impact zone optimization as a guidance for the evolution of $\x$, and a
\emph{forward step} generating the actual path $\mathcal X(\mathbf
x^{[k]}, \mathbf x) \subset \Omega$ by \textcolor{black}{conservative
  vertex advancement} towards the guidance. As the forward step finds
a sequence of states $\{ \mathbf x^{(l)} \}$, satisfying $(1-t)\mathbf
x^{(l)}+t\mathbf x^{(l+1)}\in \Omega$ for any $t\!\in\![0, 1]$ and
$l$, we essentially form $\mathcal X(\mathbf x^{[k]}, \mathbf x)$ as a
\emph{piecewise linear} path rather than linear
ones~\cite{harmon2008robust,narain2012adaptive,tang2016cama,tang2018clotha,Tang:2018:PPS,li2020p}
by relaxing the \textcolor{black}{restriction} on the searching
space. Aiming at both efficiency and quality for resolving contacts,
we make the following technical contributions.
\begin{itemize}	
        {\item {\it An inexact backward step.} \hspace{0.12in} We formulate impact zone optimization as a linear complementary problem and solve it inexactly by a small number of iterations in each backward step. In addition, we introduce \textcolor{black}{soft unilateral constraints on edge length} to effectively eliminate oscillations caused by large local element deformations.} 
        {\item {\it A lightweight forward step.} \hspace{0.12in} Instead of using CCD tests,  we use fast discrete distance evaluation to calculate safe asynchronous step sizes, which enables to keep the path generated in each forward step safely staying inside $\Omega$.}
\end{itemize}
We show that our two-way collision handling algorithm can be
conveniently implemented on a GPU and integrated into our in-house
GPU-based deformable body simulator. Coupled with volume-based contact
constraints, our simulator is capable of simulating a variety of
codimensional examples (see Figs.~\ref{fig:teaser} and \ref{fig:cod}),
including volumetric bodies, cloth, hair and sand.  The experiments
show our system is safe, fast, GPU-friendly, and robust against large
time steps and deformations.

\section{Related Work}
\vspace{0.06in}
\paragraph{Discrete collision handling.}   \hspace{0.12in} 
Researchers~\cite{Baraff:2003:UC,Volino:2006:RSC,Wicke:2006:UCB} developed discrete collision handling methods to remove intersections at the end of every time step. If a discrete collision handling method fails to eliminate all of the intersections, it can still repeat the process in the next time step, hopefully achieving the intersection-free state later.  Therefore a discrete collision handling method is robust regardless of the time step.  But as the time step increases, it becomes less likely to remove all of the intersections, which leads to long-lasting penetration artifacts in simulation.

Many physics-based simulators apply repulsion forces among proximity pairs to lessen the likelihood of collisions.  Broadly speaking, this repulsion approach is a discrete collision handling method, as it calculates repulsive forces based on proximity distances discretely evaluated in time. Thanks to its simplicity, the repulsion approach is widely used in GPU-based simulation~{\color{black}\cite{Stam:2009:TUD,fratarcangeli2016vivace,wang2016descent}} alone, without support from any other method. To help the repulsion approach achieve intersection-free guarantee, Wu et al.~\shortcite{Wu:2020:ASF} presented a fail-safe Log-barrier repulsion phase, whose usage should be minimized due to a large computational overhead.

\paragraph{Continuous collision handling.}   \hspace{0.12in} The main difference between discrete collision handling and continuous collision handling is that discrete collision handling tries to eliminate intersections at the end of the time step, while continuous collision handling must resolve all of the intersections at any time. A typical continuous collision handling method contains two components: continuous collision detection (CCD) and applying collision responses. {\color{black} Continuous detection of a vertex-triangle or edge-edge collision involves solving a cubic equation, which could be prone to errors~\cite{Provot:1997:CSC,brochu2012efficient,wang2014defending,tang2014fast}, especially in the single-precision floating-point computation environment.} 
{\color{black}Recently, Yuksel~\shortcite{Cem_2022} presented an efficient and robust method for finding real roots of cubic and higher-order polynomials and Lan et al.~\shortcite{lan_2022} provided a re-designment of the CCD root finding procedure on GPU.}
For the robustness of CCD queries, we recommend~\cite{wang2021large} for a more detailed discussion.
{\color{black} In addition to the possible robustness issue, the typical spatial acceleration structures used for CCD, such as bounding volume hierarchies (BVHs) and bounding volume traversal trees (BVTTs), could be hard to parallelize~\cite{tang2011collision,tang2016cama}.}

But compared with CCD, obtaining the right collision responses is an
even greater challenge. Bridson et
al.~\shortcite{bridson2002robust,Bridson:2003:SCF} initially used
geometric impulses as responses and the rigid impact zone technique as
a failsafe.  {\color{black}To avoid the locking artifacts caused by the rigid impact
  zone technique, Harmon et al.~\shortcite{harmon2008robust} calculated
  collision responses by non-rigid impact zone optimization. ARCSim~\cite{narain2012adaptive} used this method as its inner collision handling component and provided an open-sourced implementation\footnote{http://graphics.berkeley.edu/resources/ARCSim/} based a combination of BVH-based CCD and an augmented Lagrangian solver for collision response.
  CAMA~\cite{tang2016cama}, PSCC~\cite{Tang:2018:PPS}, I-Cloth~\cite{tang2018clotha}, and P-Cloth~\cite{li2020p} further improved its performance on GPU(s) in two aspects.

\begin{itemize}
	{\item {\it Faster CCD on GPU(s)}. \hspace{0.12in} In detail, this benefits from localized BVTT front propagation exploiting spatio-temporal coherence~\cite{tang2016cama}, parallel normal cone culling with spatial hashing~\cite{Tang:2018:PPS}, incremental collision detection with spatial hashing exploiting spatio-temporal coherence~\cite{tang2018clotha}, and distributing the incremental collision detection on multiple GPUs~\cite{li2020p}.}
	{\item {\it More GPU-friendly non-rigid impact zone solver}. \hspace{0.12in}  In detail, this benefits from assembling all of the impacts into one linear system to perform inelastic projection~\cite{tang2016cama}, paralleling the augmented Lagrangian method of ARCSim in a gradient-descent manner~\cite{tang2018clotha}, and further paralleling the augmented Lagrangian method on multiple GPUs~\cite{li2020p}.
        }
\end{itemize}
We note that although the performance is continuously improved in these step-and-project methods, both the
assumption of linear paths and simply measuring the
projection distance via mass-weighted $L_2$ norm
  are inherited, and even the specific choice of the augmented Lagrangian method for collision response is inherited in~\cite{narain2012adaptive,tang2018clotha,li2020p}.
}

Recently, Log-barrier-based methods~\cite{Wu:2020:ASF,li2020incremental,li2020codimensional} have emerged to be popular choices for collision handling in physical animation. They usually employ a smoothed~\cite{li2020incremental,li2020codimensional} or unsmoothed~\cite{Wu:2020:ASF} Log-barrier potential term to augment the objective, so that the potential term could be extremely large to ``push'' against the boundary of the feasible region. However, it is not enough for guaranteeing the state stays in the feasible region and they usually employ CCD to compute an upper bound of the safe distance to make sure no intersection happens.
Clearly, the strength of these methods is its safety, but notoriously slow for two reasons.
\begin{itemize}
	{\item {\it Frequent tests}. \hspace{0.12in} The method needs to repetitively test the vertices in every optimization step, to make sure that no intersection occurs.}
	{\item {\it Barrier functions}. \hspace{0.12in}  Barrier functions push $\mathbf x$ sufficiently against the feasible region boundary $\partial \Omega$, only when $\mathbf x$ gets close to $\partial \Omega$.
        }
\end{itemize}

We note that many continuous collision handling methods may require considerable implementation efforts to get accelerated on GPUs~\cite{tang2016cama,Tang:2018:PPS,Lauterbach:2010:HGO,li2020p}, due to their high dependency on sequential tasks~\cite{li2020codimensional} and high complexity~\cite{tang2018clotha}. 

\paragraph{Asynchronous steppings.}   \hspace{0.12in} 
The conservative advancement approach~\cite{mirtich1994impulse,von1990geometric} for rigid body collision handling suffers from the small stepping issue, as it requires all of the bodies to take the same step size.  Mirtich~\shortcite{mirtich2000timewarp} addressed this issue by allowing rigid bodies to take different step sizes, while still respecting causality. Researchers~\cite{Thomaszewski:2008:ACS,Harmon:2009:ACM} later investigated this idea for asynchronous collision handling of cloth, and explored several speedup options~\cite{Harmon:2011:AIP,brochu2012efficient}.
Our method is also asynchronous: vertices away from collisions can take large step sizes to reach their targets fast.  More importantly, it avoids CCD tests and it is naturally free of performance or robustness issues associated with them.

\paragraph{Broad-phase collision culling.}   \hspace{0.12in}  Broad-phase collision culling is important to collision handling methods, as it avoids unnecessary collision tests for collision-free primitive pairs. In general, collision culling techniques fall into two categories: those based on BVHs~\cite{Lauterbach:2010:HGO,Wang:2017:ERS,Tang:2011:VFC,Tang:2010:CCE} and those based on spatial hashing~\cite{teschner2003optimized,Pabst:2010:FSC,Zheng:2012:ESC,Barbic:2010:SSC,Tang:2018:PPS}. While GPU implementations of both categories have been investigated before, GPU-based spatial hashing is arguably more popular, thanks to its simplicity and parallelizability.  Our method is orthogonal to collision culling techniques and it can adopt more advanced ones later.

\paragraph{Frictional contacts} \hspace{0.12in} How to simulate frictional contacts, especially frictional self contacts, is another challenging problem in deformable body simulation. The popular velocity filtering approach~\cite{bridson2002robust,Muller:2008:HPB} is simple, fast, but not so physically plausible, as it handles collisions and frictions in separate processes. Recently, researchers~\cite{daviet2020simple,verschoor2019efficient,bertails2011nonsmooth,macklin2019non,ly2020projective,li2020incremental} are interested in handling collisions and frictions together through joint optimization. While our work does not consider friction, we plan to borrow their ideas for simulating plausible frictional contacts in the future.

\begin{algorithm}[t]
	\SetAlgoNoLine
	\KwIn{the current state $\mathbf x^{[k]}$, the target state ${\mathbf y}^{[k+1]}$, \hspace{0.3in} the proximity search bound $\left[D^{\min},\, D^{\max} \right]$, the step number limit $L$ and the termination condition $\epsilon$.}
	$\mathbf x^{(0)} \gets \mathbf x^{[k]}$\;
	$\mathbf y^{(0)} \gets {\mathbf y}^{[k+1]}$\;
	$D^{(0)} \gets 0$\;
    $\mathbf r^{(0)} \gets \mathbf 1$\;	
	$\mathcal P \gets \emptyset$\;
	\For{$l=0...L$ \label{lst:line:max_iteration}} 
	{
	    \If{$D^{(l)}<D^{\min}$}
	    {
		   $\mathcal P \gets \mathtt {Proximity\_Search}\big(\mathbf x^{(l)},\, D^{\max} \big)$\;
		   $D^{(l)} \gets D^{\max}$\;
	    }
    	$\mathbf y^{(l+1)} \gets \mathtt{Backward}\big(\mathbf y^{(l)}, \mathbf x^{(l)}, \mathbf y^{[k+1]}, \mathcal P   \big)$~~// Sec.~\ref{sec:backward}\;
	\mbox{$\big\{\mathbf x^{(l+1)}, \mathbf r^{(l+1)} \big\} \gets \mathtt{Forward} \big( \mathbf x^{(l)}, \mathbf r^{(l)}, \mathbf y^{(l+1)} - \mathbf x^{(l)}, \mathcal P\big)$~~// Sec.~\ref{sec:forward}\;}
        
 	$D^{(l+1)} \gets D^{(l)} - 2 \max\limits_i \big\| \mathbf x_i^{(l+1)} - \mathbf x_i^{(l)}\big\|$\;
        \If{$\big\| \mathbf r^{(l+1)} \big\|_{\infty}< \epsilon$ \label{lst:line:converge} } 
	 	 {
	 	    {\bf break}\; 
	 	 }
    }

	$\mathbf x^{[k+1]} \gets \mathbf x^{(l+1)}$\;
	\caption{A two-way method}
	\label{alg:one}
\end{algorithm}

\section{A Two-Way Framework}
As we discussed in Section~\ref{sec:intro}, restricting to a linear
path and simply taking mass-weighted $L_2$ norm as the distance
function can be inappropriate for post-projection,
especially when $\| \mathbf y^{[k+1]}-\mathbf x^{[k]} \|$
\textcolor{black}{is} large. Thus, we formulate \textcolor{black}{the}
collision handling as the following optimization
problem:\vspace*{-1mm}
\begin{subequations}
  \begin{align}
    &\x^{[k+1]} = \mathop{\arg\min}\limits_\mathbf{x} \frac{1}{2}\big\|\mathbf{x}- {\mathbf y}^{[k+1]}\big\|^2_{\mathbf M}, \quad\\
    &\text{s.t.}  \left\{
  \begin{array}{l}
    \mathcal X\big(\x^{([k])}, \x\big) \subset \Omega\\
    \mathbf c(\x, \y^{[k+1]}) \geq \mathbf 0,
    \label{eq:cons_our_opt}
  \end{array}
      \right.   
  \end{align}
  \label{eq:our_opt}
\end{subequations}
in which $\mathbf M \in \mathbb R^{3N \times 3N}$ is the scaled lumped
mass matrix~\cite{tang2018clotha}. The major differences between our
formulation and the ones
in~\cite{harmon2008robust,narain2012adaptive,tang2016cama,tang2018clotha,Tang:2018:PPS,li2020p}
\textcolor{black}{are} twofold rooted in Eq.~\eqref{eq:cons_our_opt}:
(i) the path $\mathcal X(\x^{([k])}, \x)$ here should be piecewise
linear and sufficiently short,
while~\cite{harmon2008robust,narain2012adaptive,tang2016cama,tang2018clotha,Tang:2018:PPS,li2020p}
restricts it as a linear segment; and (ii) additional edge length
constraints $\mathbf c(\x, \y^{[k+1]})$ are introduced (see
Section~\ref{edge_length_constraint} for details) to avoid spuriously
large deformations of local elements, which compensates for possible
loss of shape preservation by only considering the mass-weighted Euclidean
distance. In fact, \textcolor{black}{the projection distance jointly
  defined by the mass-weighted $L_2$ norm plus edge length constraints
  is more consistently measured as the actual energy, in a sense that
  we are approximately minimizing the change of both kinetic energy
  and potential before and after projection.} Numerically, such
treatment does not complicate the objective function, allowing us to use
existing fast iterative techniques as described below.

Our two-way method aims to find a good approximate solution subject to
Eq.~\eqref{eq:our_opt} at a low cost. In this method, two sets of
intermediate variables $\{\mathbf y^{(l)}\}$ and $\{\mathbf x^{(l)}\}$
are introduced and updated alternately in two steps: a backward step
starting at $\mathbf y^{(0)}=\mathbf y^{[k+1]}$, aiming to inexactly
and progressively project $\mathbf y^{(l)}$ back to $\Omega$ as a
target, and a forward step starting at $\mathbf x^{(0)}=\mathbf
x^{[k]}$, aiming to move \textcolor{black}{from} the current state $\mathbf x^{(l)}\in \Omega$
towards $\y^{(l+1)}$ with a guarantee of $\mathcal X(\mathbf
x^{[k]}, \mathbf x)$ being inside $\Omega$ by conservative vertex
advancement. Intuitively, $\{\mathbf y^{(l)}\}$ guide the evolution of
$\{\mathbf x^{(l)}\}$ in the forward step, and in turn $\{\mathbf
x^{(l)}\}$ explore and update the boundary of the feasible region
$\Omega$, which gives feedback to the updates of $\{\mathbf
y^{(l)}\}$. The supplemental video illustrates our two-way
optimization process.

Alg.~\ref{alg:one} outlines the pseudo-code of our method: it keeps
running the two steps alternately, until the termination metric
$\mathbf r^{(l+1)}$ is small enough (\Cref{lst:line:converge}) or it
reaches the maximum number of iterations $L$
(\Cref{lst:line:max_iteration}). We have $\mathbf r^{(l+1)}$ to make
sure that the accumulated moving distance from $\mathbf x^{[k]}$ to
$\mathbf x^{(l+1)}$ should be close enough to the distance from
$\mathbf x^{[k]}$ to $\mathbf y^{[k+1]}$.
Either way, the method can safely report ${\mathbf x}^{(l+1)}$ as its result $\x^{[k+1]}$, no matter if $\mathbf y^{(l+1)}$ stays in $\Omega$ or not.

\subsection{The Proximity Search}
Before diving into details of our method, we first discuss about the
proximity search. Functioning as broad-phase collision culling, the
proximity search serves multi-fold purposes: to form the set of
contact constraints in the backward step (in
Subsection~\ref{sec:contact}), to obtain proximity pair distances for
safe steppings (in Section~\ref{sec:forward}), and to calculate
repulsive forces as a part of dynamics (in
Subsection~\ref{sec:elastic}). In our implementation, the proximity
search is based on the standard grid-based spatial hashing
technique~\cite{Pabst:2010:FSC,Tang:2018:PPS}.

Since the proximity search has non-negligible cost, one challenge is
how to reuse the results as often as possible, rather than redoing the
search in every step. Let $\mathcal P$ be the proximity set in each
step that \textcolor{black}{is} required to be a superset of all possible pairs whose
distances are below a certain bound $D^{\min}$:
\begin{equation}
	\mathcal P \supset \mathcal P_{D^{\min}}= \{\mathcal p  \mid \forall \mathcal p: 	
	\mathtt {dist}_{\mathcal p} \big(\mathbf x^{(l)} \big) < D^{\min}\}.
        \label{eq:relationship_P}
\end{equation}
Assuming that in the $l$-th step, $\mathcal P$ is computed with a
certain bound $D^{\max}$ ($\mathcal P = \mathcal P_{D^{\max}}$,
$D^{(l)} = D^{\max} > D^{\min}$) so that all proximity pairs
satisfying Eq.~\eqref{eq:relationship_P} are collected. To reuse this
computed $\mathcal P$ in the $(l$+1)-th step, we point out that
$\mathcal P$ still contains all of the pairs whose distances are below
$D^{(l+1)}=D^{(l)} - 2 \max\limits_{i} \| \mathbf x_i^{(l+1)} - \mathbf
x_i^{(l)} \| $. If $D^{(l+1)} \geq D^{\min}$, we can reuse $\mathcal
P$ and only need to filter out some elements in $\mathcal P$ with
recomputed distances. Otherwise, $\mathcal P$ is insufficient to
\textcolor{black}{fulfill Eq.~\eqref{eq:relationship_P}}. Thus, we
have to reset $D^{(l+1)}=D^{\max}$ and perform the proximity search to
\textcolor{black}{avoid missing any necessary proximity pair.}

The entire computational cost depends on values of both $D^{\min}$ and
$D^{\max}$. \textcolor{black}{On one hand}, the cost decreases as
$D^{\min}$ decreases, but $D^{\min}$ cannot tend to zero. To build a
full set of contact constraints in the backward step, $\mathcal
P_{D^{\min}}$ needs to collect all pairs whose distances are below a
given activation threshold $\delta$, which should not be too small in
the discrete computing environment as suggested
in~\cite{li2020incremental}. Here we set $\delta = 1$mm and
$D^{\min}\ge \delta$.
\textcolor{black}{On the other hand}, the cost decreases as $D^{\max}$ increases, but
doing so requires more memory cost and distance evaluations as a
result of an increasing number of proximity pairs. In our experiments,
we find that setting $D^{\min} = 2\delta =2$mm and $D^{\max} = 4$mm
usually triggers one proximity search every three steps in average,
which empirically keeps a reasonable balance between memory cost
and computing time.

\begin{figure*}[t]
	\centering
	\subfigure[Falling mats]{\includegraphics[width=1.13in]{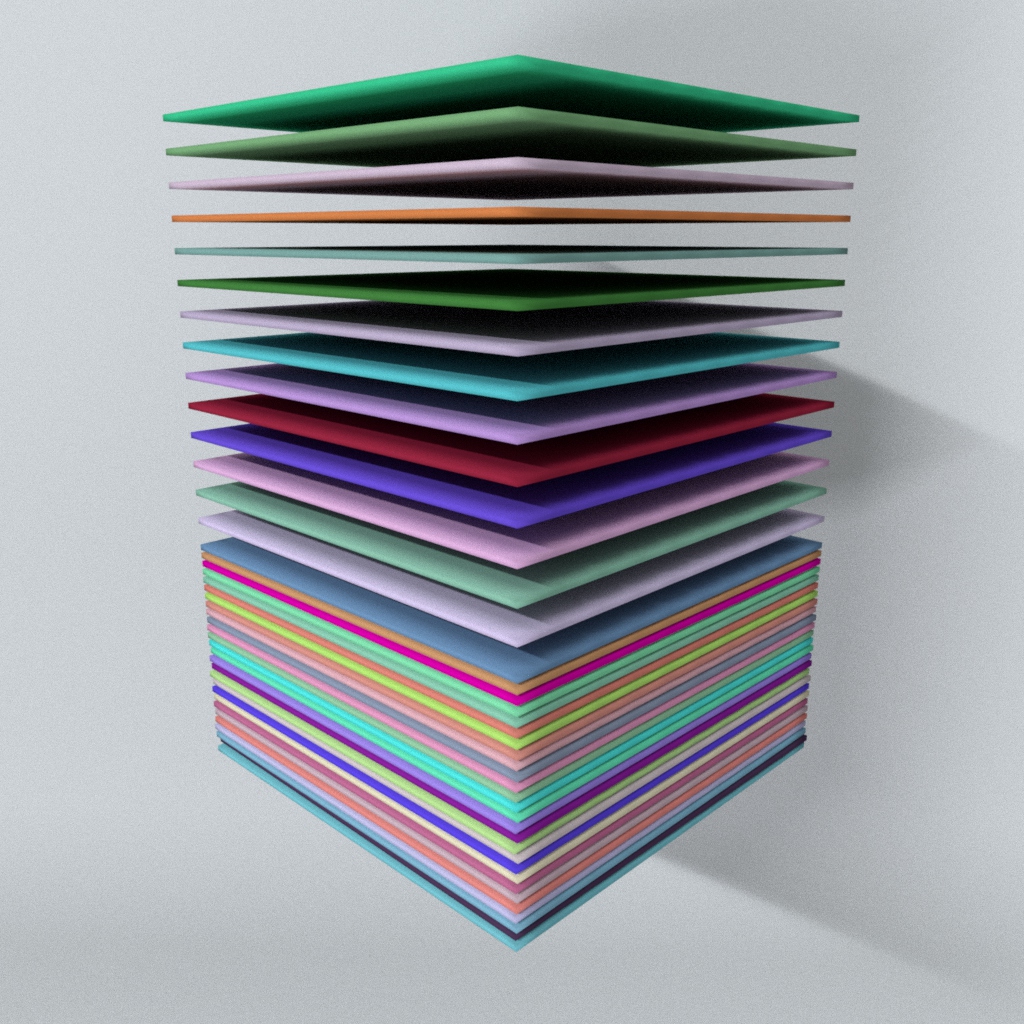}}
	\subfigure[Stacked mats]{\includegraphics[width=1.13in]{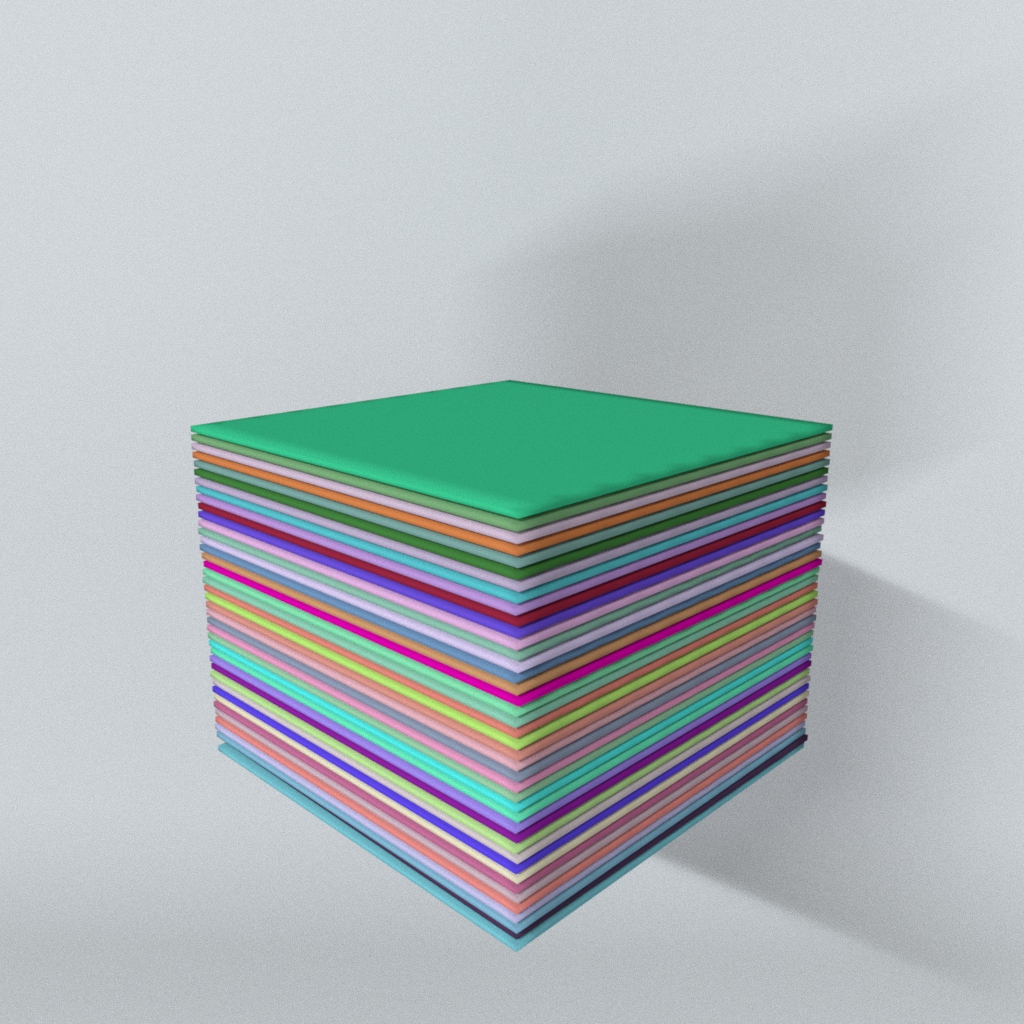}}\hfill
	\subfigure[Twisted hair]{\includegraphics[width=1.13in]{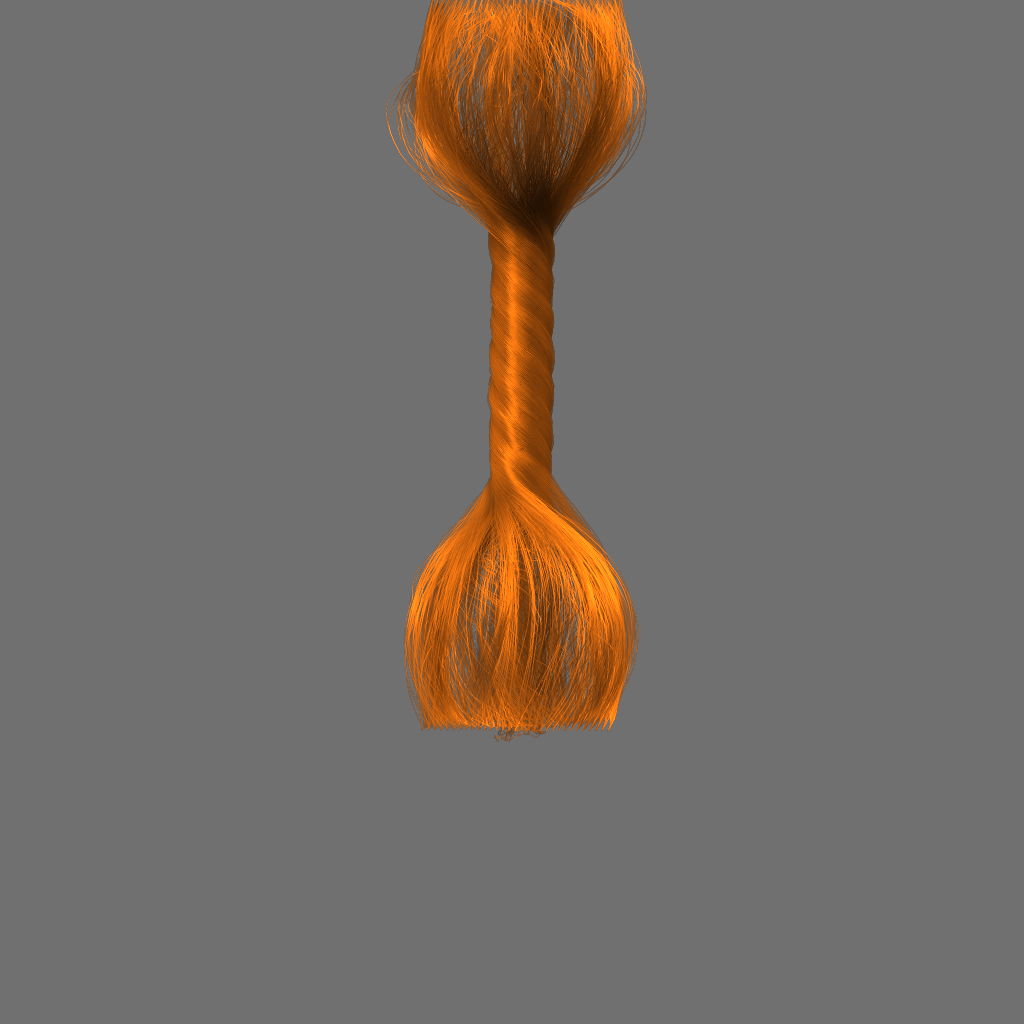}}
	\subfigure[Loose hair]{\includegraphics[width=1.13in]{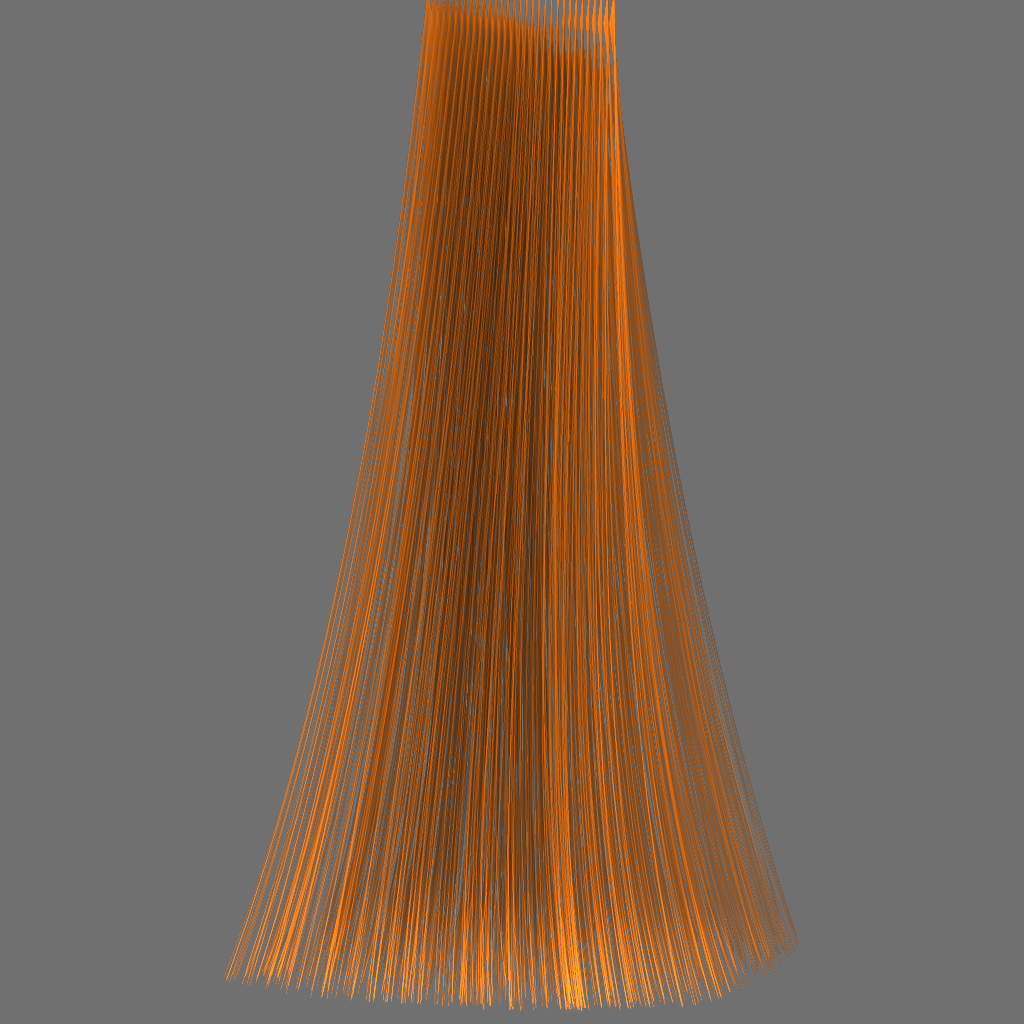}} \hfill
	\subfigure[Falling sand]{\includegraphics[width=1.13in]{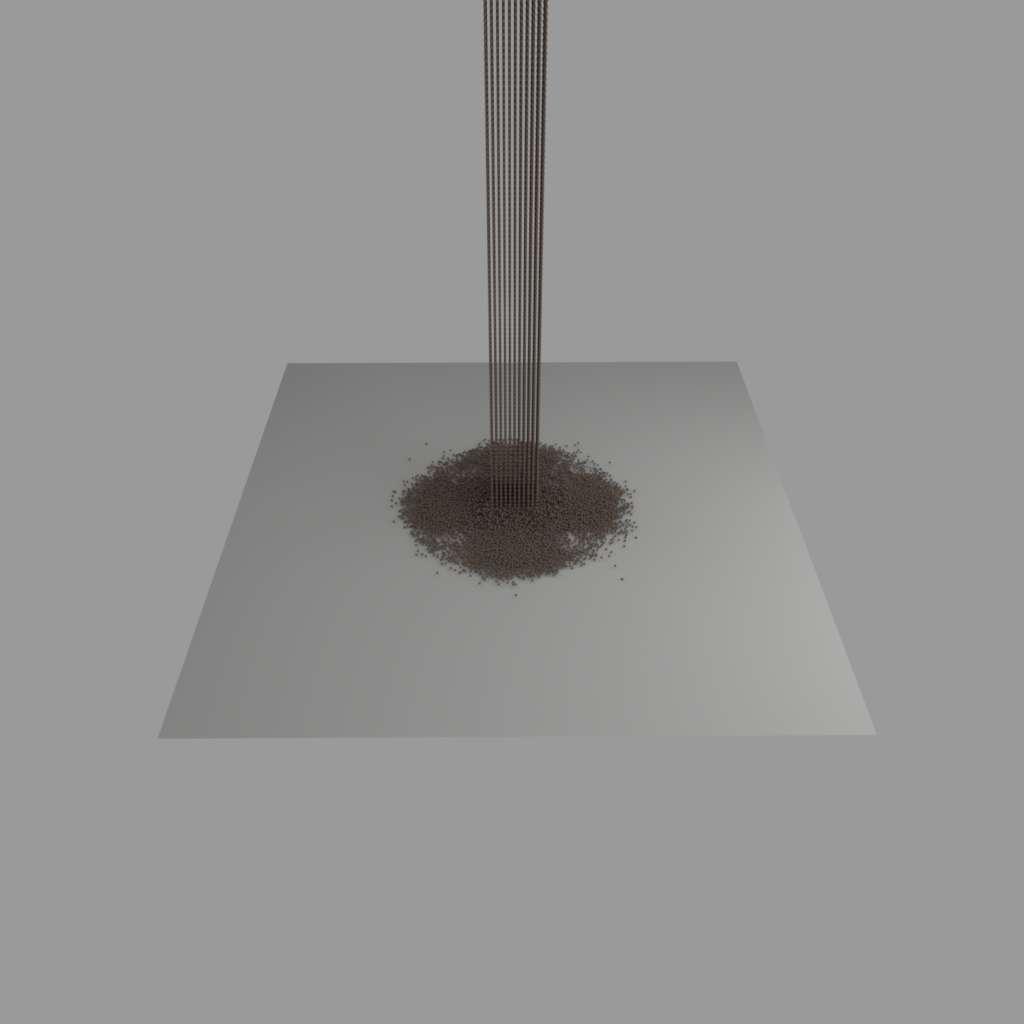}}
	\subfigure[Piled sand]{\includegraphics[width=1.13in]{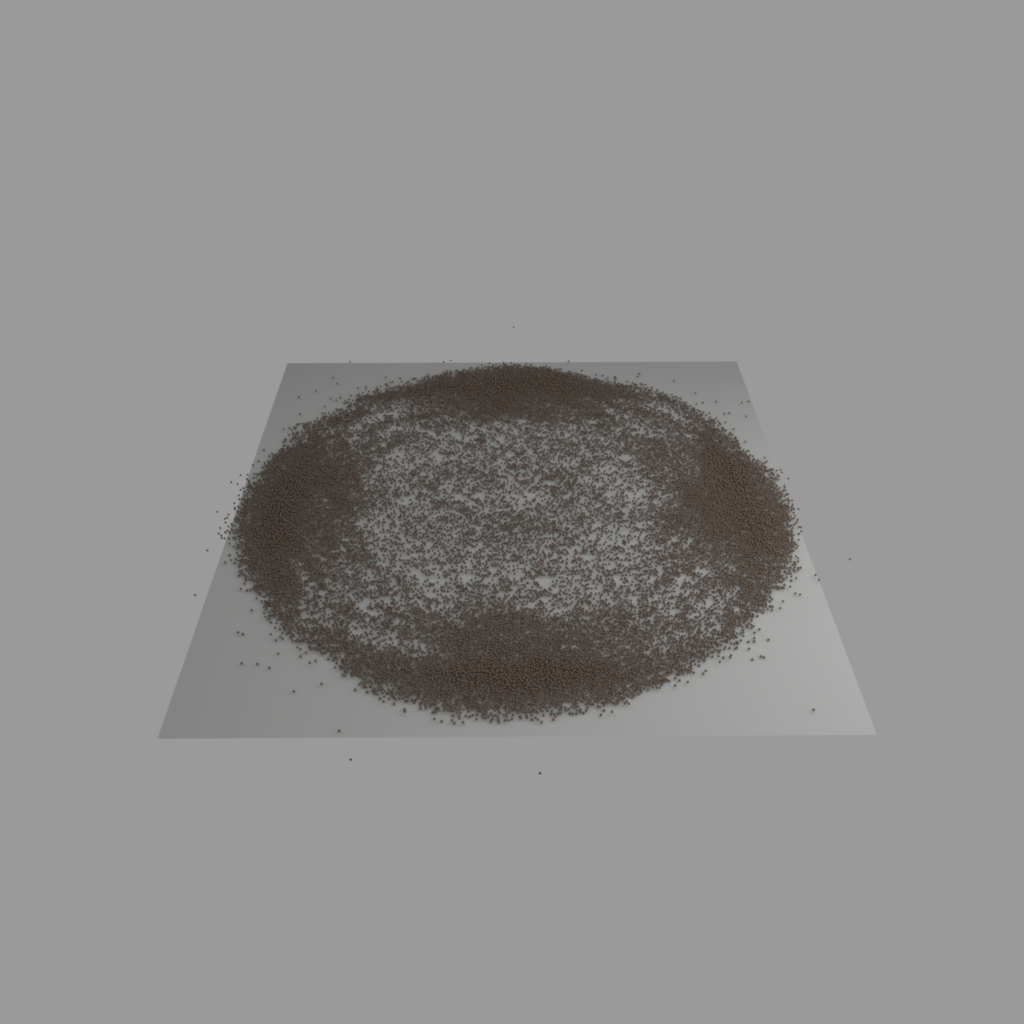}}  
	\vspace*{-0.12in}
	\caption{\textbf{Codimensional deformable bodies.}  Based on the volume enforcement idea, we develop contact constraints for a wide range of primitive proximity pairs.  These contact constraints allow our method to simulate a variety of codimensional examples, including elastic mats in (a) and (b), cloth, hair in (c) and (d), and sand in (e) and (f). }
	\label{fig:cod}
\end{figure*}
      
\section{The Backward Step}
\label{sec:backward}
The backward step in our two-way approach is similar to the impact
zone optimizations
in~\cite{harmon2008robust,narain2012adaptive,tang2016cama,tang2018clotha,Tang:2018:PPS,li2020p}.
Instead of directly performing CCD to find a strictly
intersection-free projection, we roughly optimize an intermediate
target $\y^{(l)}$ almost intersection-free. Note that it is only used
for guiding the vertex advancement in the forward step later, where
the exact \textcolor{black}{intersection-freeness} is imposed as discussed in
Section~\ref{sec:forward}.

In the $l$-th iteration, the goal of the backward step is to project
$\y^{(l)}$ back to $\Omega$ formulated as a constrained optimization:
\begin{equation}
	\y^{(l+1)} = \mathop{\arg\min}\limits_\mathbf{y} \frac{1}{2}\big\|\mathbf{y}- {\mathbf y}^{[k+1]}\big\|^2_{\mathbf M}, \quad
\mathrm{s.t.}\; \mathbf c(\mathbf y) \geq \mathbf 0,
\label{eqn:lsq}
\end{equation}
where $\mathbf c(\mathbf y) \geq \mathbf 0$ contains the contact
constraints and edge length constraints, and $\mathbf{y}$ is
initialized with $\y^{(l)}$.
Let $\mathbf x^{(l)}$ be the current state of the forward step in the $l$-th iteration. 
We linearize the contact constraints at $\mathbf x^{(l)}$: $ \mathbf c (\mathbf{x}^{(l)} ) + \mathbf J^{(l)} (\mathbf {y}-\mathbf{x}^{(l)} ) \geq \mathbf 0$, in which $\mathbf J^{(l)}= {\partial \mathbf  c(\mathbf{x}^{(l)}) }/{\partial \mathbf x} $ is the Jacobian matrix.

{\color{black}
\paragraph{Linearization}
Note that we perform assembly and linearization of
contact constraints at $\x^{(l)}$ rather than $\y^{(l)}$. The reason
is that we not only require $\y^{(l+1)}$ to be close to
intersection-free, but also expect the path from $\x^{(l)}$ to
$\y^{(l+1)}$ to stay inside $\Omega$ as much as possible. Since
$\x^{(l)}$ is inside $\Omega$ in the first place, we are actually
attempting to drive $\y^{(l+1)}$ to cross the boundary of $\Omega$ and
stay on \emph{the same side} as $\x^{(l)}$ rather than $\y^{(l)}$.
Simply projecting the target back to $\Omega$ based on the
constraints assembled around $\y^{(l)}$ can cause the stagnation issue
as shown in Fig.~\ref{fig:linearization_y}, while performing the
linearization at $\x^{(l)}$ effectively avoids such an issue and
reach the convergence with a more reasonable solution as shown in
Fig.~\ref{fig:linearization_x}.

\begin{figure}[t]
	\centering
	\subfigure[$\x^{[k]}$ (above) and $\y^{[k+1]}$ (below)]{\includegraphics[width=1.05in]{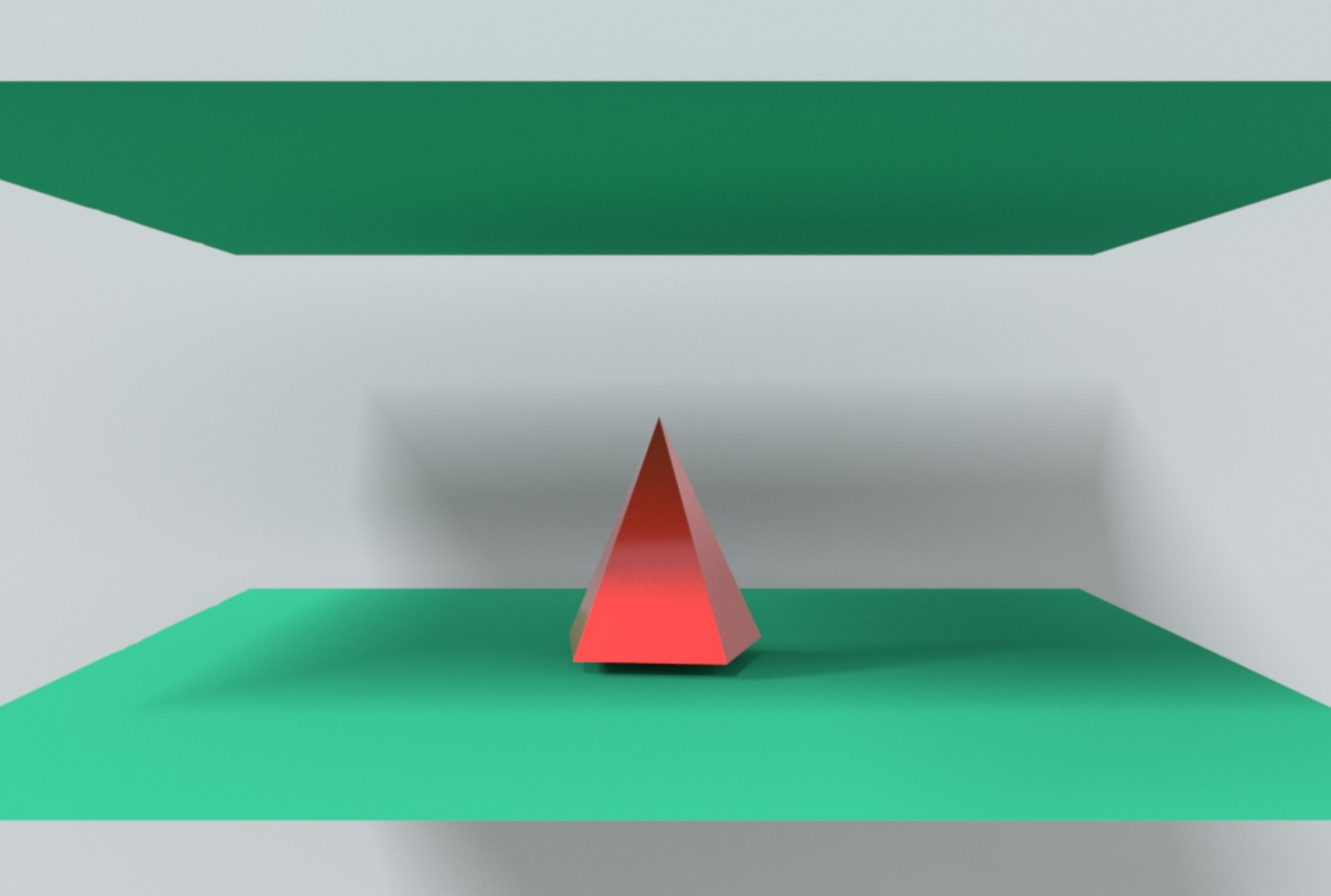}\label{fig:tunneling}}	
	\subfigure[$\x^{[k+1]}$ using the contact constraints assembled around $\y^{(l)}$]{\includegraphics[width=1.05in]{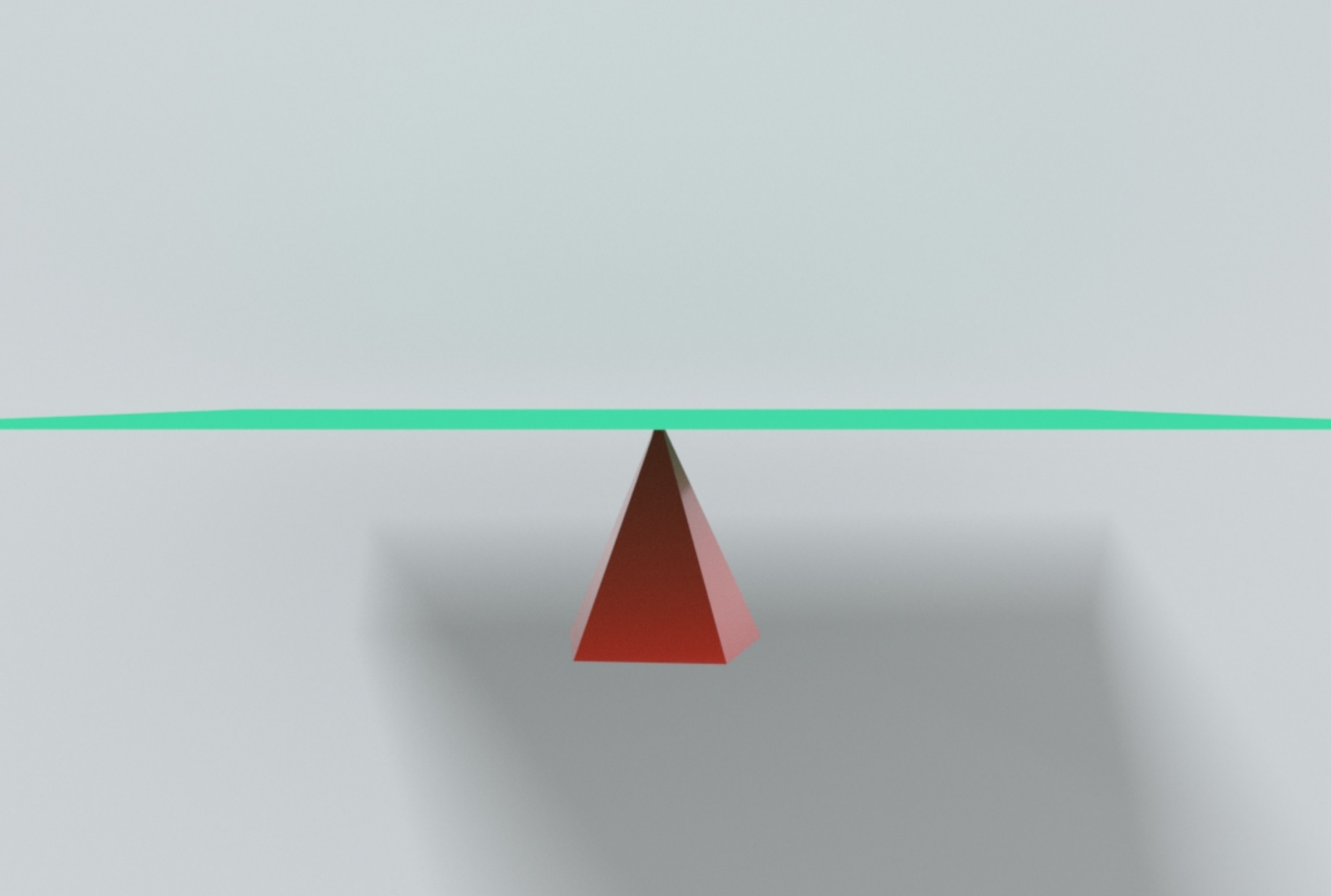} \label{fig:linearization_y} }
	\subfigure[$\x^{[k+1]}$ using the contact constraints assembled around $\x^{(l)}$]{\includegraphics[width=1.05in]{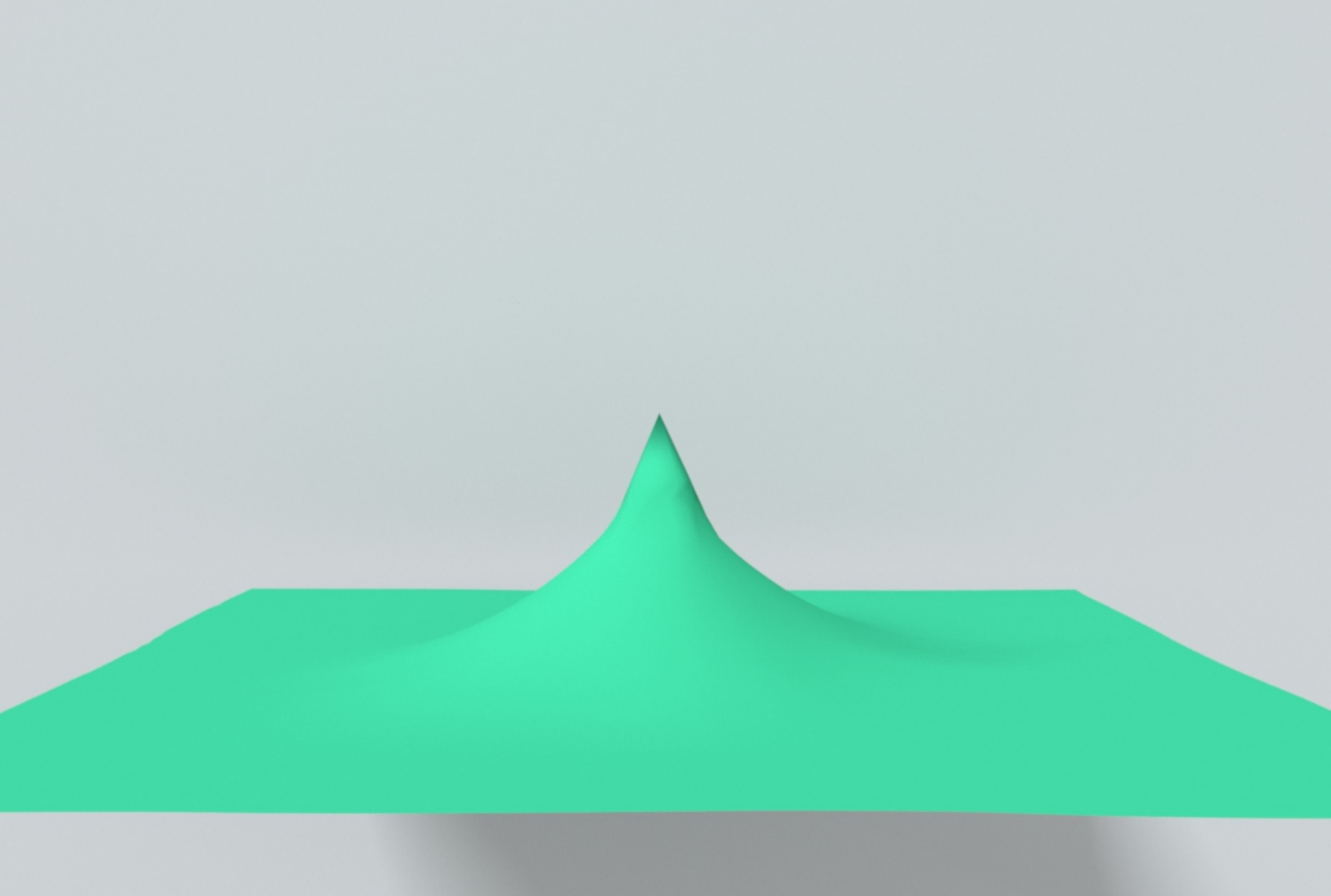} \label{fig:linearization_x} }	
	\vspace{-0.12in}
	\caption{\textbf{Linearization choices.} The tunneling
          artifact appears between two intersection-free states
          $\x^{[k]}$ and $\y^{[k+1]}$ (Fig.~(a)). Projecting
          $\y^{[k+1]}$ back to $\Omega$ based on the constraints
          assembled around $\y^{(l)}$ stagnates $\x^{[k+1]}$
          (Fig.~(b)).  Instead, if we assemble and linearize contact
          constraints around $\x^{(l)}$, $\y^{(l)}$ manages to cross
          the boundary of $\Omega$ and stay on the same side as
          $\x^{(l)}$, eventually advancing $\x^{(l)}$ to a more
          reasonable solution (Fig.~(c)).}
	\label{fig:different_linearization}
	\vspace{-0.12in}
\end{figure}
}

By introducing Lagrangian multipliers, we formulate the following Lagrangian:
\begin{equation}
		\mathcal L(\mathbf{y},\boldsymbol \lambda)= \frac{1}{2}\big\|\mathbf{y}- \mathbf y^{[k+1]} \big\|^2_{\mathbf M} - { \left( \mathbf c\big(\mathbf{x}^{(l)}\big) + \mathbf J^{(l)} \big(\mathbf{y}-\mathbf{x}^{(l)} \big) \right)} ^{\mathsf T}  \boldsymbol \lambda ,
	\label{eqn:lsq_langrangian}
\end{equation}
whose minimizer satisfies the KKT conditions:
\begin{equation}
	\left\{
	\begin{array}{l}
		\nabla_{\mathbf y}\mathcal L=\mathbf M(\mathbf{y}- {\mathbf y}^{[k+1]}) - \big( \mathbf J^{(l)} \big) ^{\mathsf T} \boldsymbol \lambda= \mathbf 0,\\
		\boldsymbol \lambda \geq 0 \perp \mathbf c \big(\mathbf{x}^{(l)} \big) +\mathbf J^{(l)} \big(\mathbf{y}-\mathbf{x}^{(l)} \big) \geq \mathbf 0.
	\end{array}
	\right.
	\label{eqn:KKT_condition}
\end{equation}
Multiplying the first condition in~\Cref{eqn:KKT_condition} with $\mathbf J^{(l)}$, we obtain:
\begin{equation}
	\mathbf J^{(l)}\mathbf{y} = \mathbf J^{(l)} \mathbf M^{-1} \big( \mathbf J^{(l)} \big)^{\mathsf T} \boldsymbol \lambda + \mathbf J^{(l)}{\mathbf y}^{[k+1]}.
\end{equation}
Together with the second condition, we get a  linear complementarity problem (LCP) with only one unknown $\boldsymbol \lambda$:
\begin{equation}
	\boldsymbol \lambda \geq \mathbf 0 \perp \mathbf c \big(\mathbf{x}^{(l)} \big) + \mathbf J^{(l)} \mathbf M^{-1} \big(\mathbf J^{(l)} \big)^{\mathsf T} \boldsymbol \lambda + \mathbf J^{(l)} \big( {\mathbf y}^{[k+1]} -  {\mathbf x}^{(l)} \big) \geq \mathbf 0.
	\label{eqn:LCP}
\end{equation}

Once we solve $\boldsymbol \lambda$, we apply the first condition of Eq.~\eqref{eqn:KKT_condition} to calculate $\y^{(l+1)}$ for the next iteration. Note that $\y^{(l)}$ is the initialization to the problem in Eq.~\eqref{eqn:lsq} and it is calculated from the last $\boldsymbol \lambda$ in the ($l$-1)-th step, so the cumulative effect of the previous $l$-1 iterations to $\y$ is retained.
\subsection{Contact Constraints}   
\label{sec:contact}

An interesting question is how to define the contact constraints $\mathbf c(\mathbf{x}) \geq \mathbf 0$ for a variety of primitive proximity pairs. In our method, we construct our contact constraints in a volume enforcement fashion.

\setlength{\columnsep}{0.16in}
\begin{figure}[htb]
	\centering
        \begin{tikzpicture}
          \node at (0, 0) {
            \includegraphics[width=0.47\textwidth]{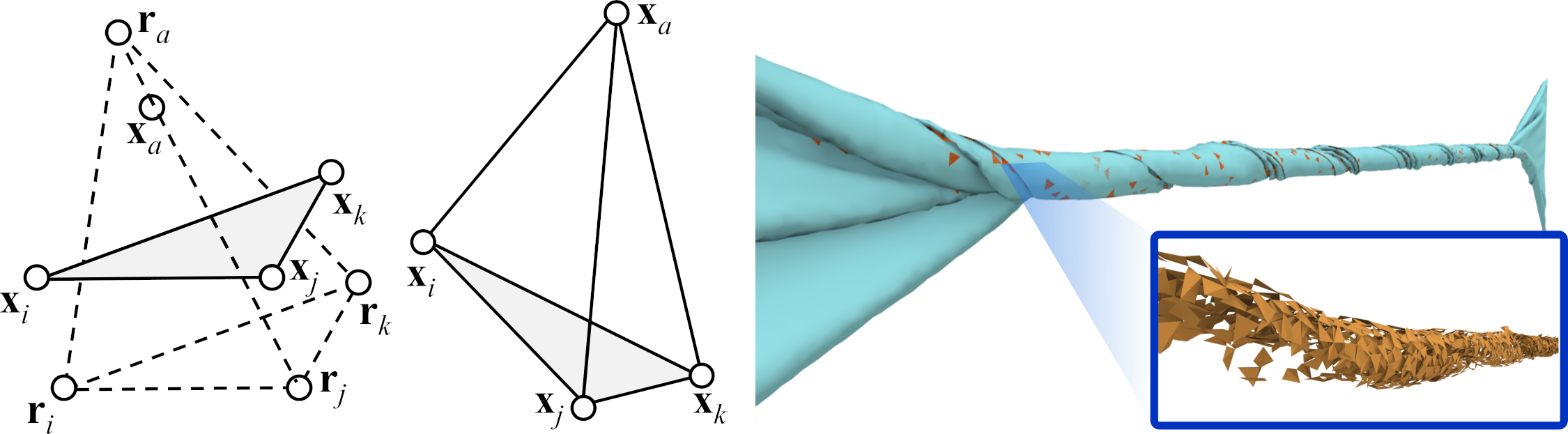}
          };
          \node at (-3.2, -1.4) { (a)};
          \node at (-1.2, -1.4) { (b)};
          \node at (2.2, -1.4) { (c)};                    
        \end{tikzpicture}
	\vspace*{-7mm}
	\caption{\textbf{Vertex-triangle pair.} Our constraint
          enforces the volume of $\mathbf x_a\mathbf x_i \mathbf x_j
          \mathbf x_k$ to be the same as its reference
          $\mathbf{r}_a\mathbf{r}_i\mathbf{r}_j\mathbf{r}_k$
          (Fig.~(a)), in which the vertex and the triangle are
          separated by a threshold distance $\delta$ (Fig.~(b)).
          Tessellated with small tetrahedra in crevices, the simulated
          cloth robustly prevents self-intersections (Fig.~(c)).}
	\vspace*{-.14in}
	\label{fig:vt}
\end{figure}
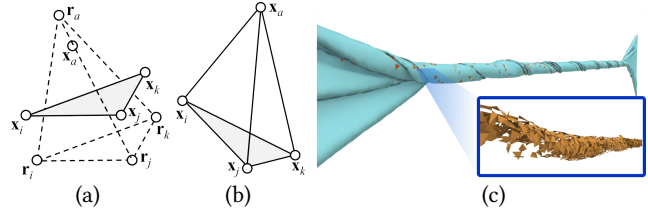
To begin with, we consider a vertex-triangle proximity pair in Fig.~\ref{fig:vt}(a), whose distance is below a certain activation threshold $\delta$.  In our experiment, $\delta$=1mm. Let $\{\mathbf r_a, \mathbf r_i, \mathbf r_j, \mathbf r_k \}$ be its projection, calculated by moving the vertex and the triangle in opposite normal directions until the distance becomes $\delta$:
\begin{equation}
	\left\{
	\begin{array}{l}
		\mathbf r_a \quad= \mathbf x_a \quad + \frac{1}{2} \left( \delta -\left\| \mathbf x_a - b_i \mathbf x_i - b_j \mathbf x_j - b_k\mathbf x_k \right\| \right) \mathbf n,\\
		\mathbf r_{i,j,k}= \mathbf x_{i,j,k} -  \frac{1}{2}  \left( \delta -\left\| \mathbf x_a - b_i \mathbf x_i - b_j \mathbf x_j - b_k\mathbf x_k \right\| \right) \mathbf n,
	\end{array}
	\right.
      \end{equation}          
where $b_i$, $b_j$ and $b_k$ are the barycentric weig\-hts of vertex $a$ on the triangle, and $\mathbf n$ is the constant triangle normal. Under the assumption that the triangle area is constant, we define the contact constraint by requiring the volume of  $\mathbf x_a\mathbf x_i \mathbf x_j \mathbf x_k$ to be greater than or equal to the volume of $\mathbf r_a\mathbf r_i \mathbf r_j \mathbf r_k$:
\begin{equation}
	\quad c(\mathbf x_a, \mathbf x_i, \mathbf x_j, \mathbf x_k) = \mathsf{det}({\partial{\mathbf{x}}}/{\partial{\mathbf{r}}})-1 \geq 0,
	\label{eq:volume_by_det}
\end{equation}
in which $\mathbf r_a$, $\mathbf r_i$, $\mathbf r_j$ and $\mathbf r_k$ are treated as constants, and
$\partial{\mathbf{x}}/{\partial{\mathbf{r}}}$ is the artificial deformation gradient tensor, assuming that the computed $\mathbf r$ is the reference shape.
Note that the contact constraint actually outlines the boundary of the feasible region at $\mathbf x^{(l)}$, so the sign of volume at $\mathbf x^{(l)}$ is positive. If $\mathbf y^{(l)}$ is on the other side of the boundary, the sign of volume at $\mathbf y^{(l)}$ would be negative and the inequality constraint would ``pull'' $\mathbf y^{(l)}$ to the same side as $\mathbf x^{(l)}$.

Based on the same idea, we model the contact constraints for other simplex pairs, including edge-edge pairs, and vertex-vertex pairs.  In the simplest case, the contact constraint for a vertex-vertex pair is:
\begin{equation}
	c(\mathbf x_a, \mathbf x_{i}) = \left\| \mathbf x_i -\mathbf x_a \right\| /\delta -1 \geq 0.
\end{equation}
These constraints enable our method to handle contacts for a variety of codimensional deformable body examples, such as cloth, hair (in Fig.~\ref{fig:cod}(c) and~\ref{fig:cod}(d)) and sand  (in Fig.~\ref{fig:cod}(e) and~\ref{fig:cod}(f)).

\begin{figure}[t]
	\centering
	\subfigure[The number of steps fluctuating over frames]{\includegraphics[width=3.3in]{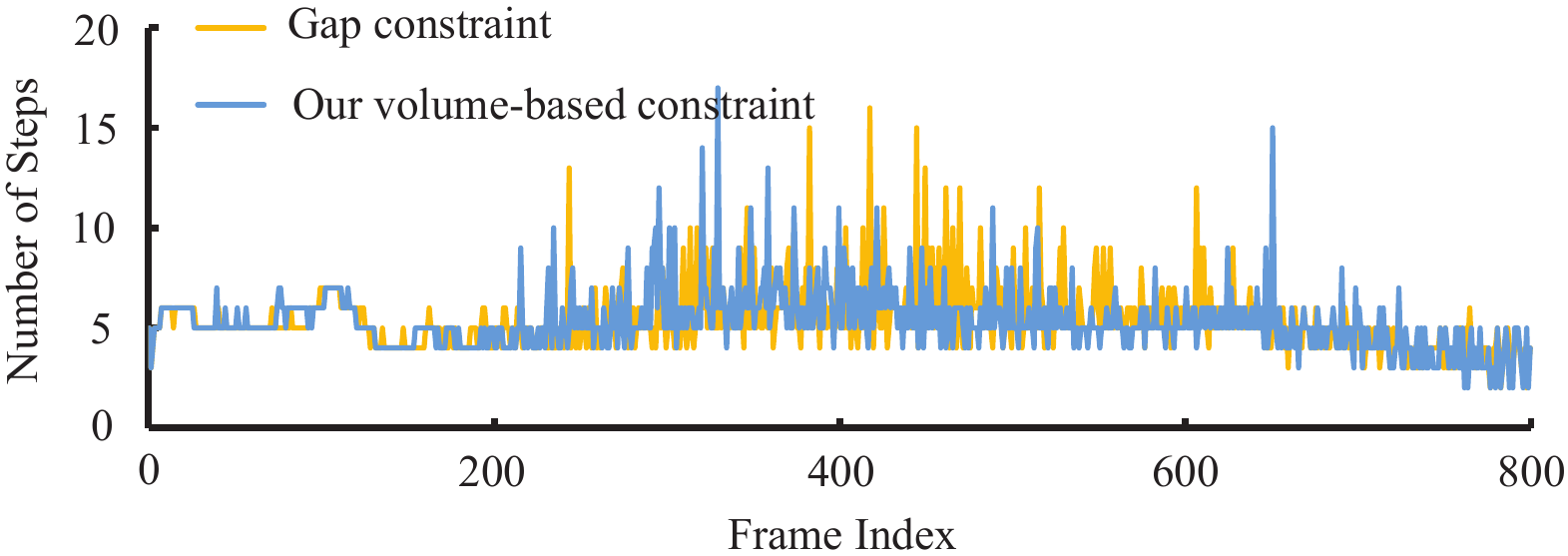}}
	\subfigure[The collision cost fluctuating over frames]{\includegraphics[width=3.3in]{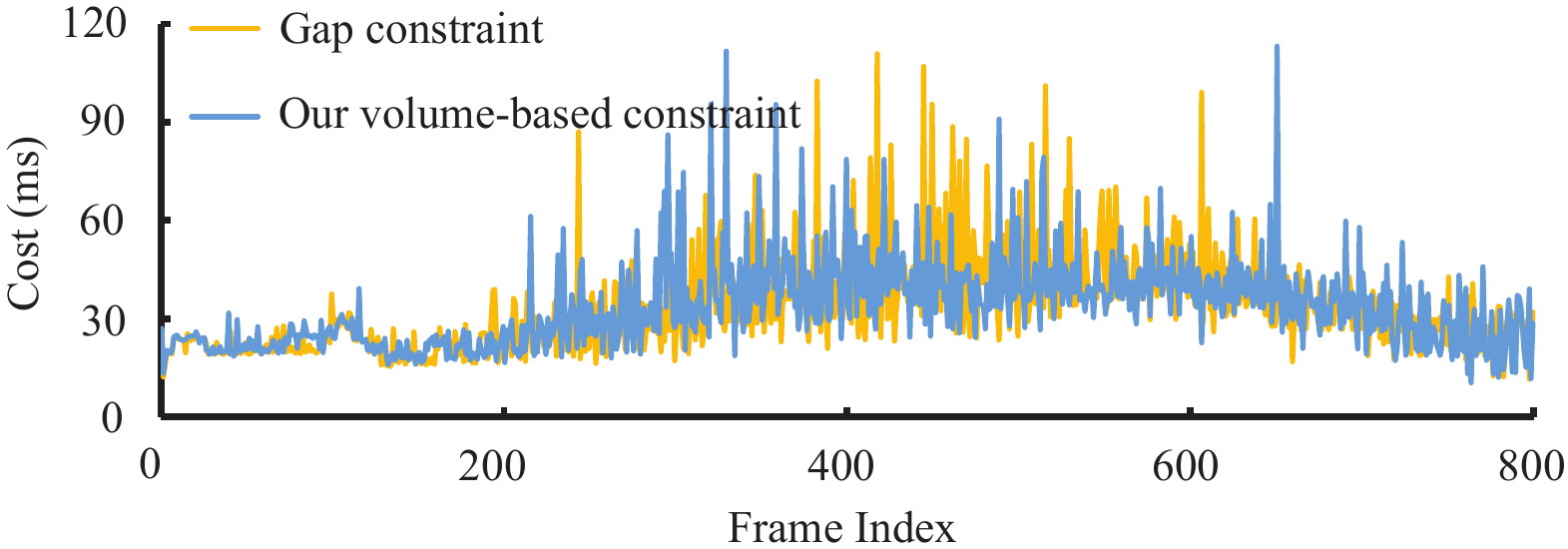}}
	\vspace{-0.12in}
	\caption{\textbf{Comparison with the gap constraints.} {\color{black} We
          count the performances of our method implemented with the gap constraints and the volume-based constraints (as
          ours). Our volume-based constraints have comparable performance 
          compared with the gap constraints and do not have a negative influence on performance. By default, we use the bow knot example for evaluations in this paper.}}
	\label{fig:gap_or_volume}
\end{figure}

{\color{black}We find such volume-based constraints do not suffer from locking artifacts, which usually come from the edge-edge pairs. The possible reason is that if we renew the reference volume in every intermediate step as our method does, it does not cause the resistance of shearing or twisting, whereas if the reference shapes are constant, these artifacts would be observed as discussed in ~\cite{muller2015air}.
  A potential problem with volumetric constraints is that they may increase the triangle area or edge length, which may influence the simulation quality or performance negatively.
  However, this issue is barely problematic in our experiments: we test the popular gap constraints~\cite{andrews2022contact} as a replacement, and using both kinds of constraints has comparable performances as Fig.~\ref{fig:gap_or_volume} shows, and our examples show that volumetric constraints can work well without obvious artifacts.
}

Sifakis et al.~\shortcite{sifakis2008globally} explored a similar idea, but they chose to preserve the volume, rather than enforce the volume to a desired value. In comparison, our constraints can keep pairs well separated, so that fewer collisions can occur in later updates.

\subsection{Edge length Constraints} \hspace{0.12in}\
\label{edge_length_constraint}
When the target $\y^{[k+1]}$ is close to $\x^{[k]}$, minimizing the
mass-weighted Euclidean distance keeps a minimal change of kinetic
energy after collision handling and the change of potential should not
be significant either. However, as $\y^{[k+1]}$ starts to get far away
from $\x^{[k]}$, simply using this kinetic energy norm to measure the
projection distance could be inappropriate. It can cause spuriously
large, local deformations
(see~\Cref{fig:cmp_arcsim,fig:cmp_with_impact_zone}) and even
oscillations over time, as a result of a sharp change of potential
before and after collision handling.  We can incorporate additional
deformation resistance terms into the objective function, but it
breaks a simple LCP formulation and thus is more numerically
involved. Given the fact that potentials are basically penalizing
non-rigid deformations, we therefore attempt to preserve the shape of
each element by preserving its edge lengths, so that each element
mostly undergoes a rigid transformation and the change of potential
stays at a low level after collision handling.

For keeping it as a simple LCP formulation where 
many fast iterative techniques can be used, we follow the strategy
in~\cite{macklin2021constraint} to suppress spurious distortions by
incorporating constraints without complicating the objective function,
and linearize it at $\y^{[k+1]}$:
\begin{equation}
  c(\mathbf x_i, \mathbf x_j)= \sigma - {\big\|\mathbf x_i -\mathbf x_j \big\|
\mathord{\left/
			{\vphantom {\big\|\mathbf x_i -\mathbf x_j \big\| \big\|\mathbf y_i^{[k+1]} -\mathbf y_j^{[k+1]} \big\|}} \right.
			\kern-\nulldelimiterspace}
    \big\|\mathbf y_i^{[k+1]} -\mathbf y_j^{[k+1]} \big\|} \geq 0,
\end{equation}
in which $i$ and $j$ are the two vertex indices of one edge and
$\sigma$ is the maximum violation ratio. Here we relax the strict bilateral constraint
to the soft unilateral one and permit a certain level of
violation for performance consideration.

\begin{figure}[t]
	\centering
	\subfigure[The number of steps fluctuating over frames]{\includegraphics[width=3.3in]{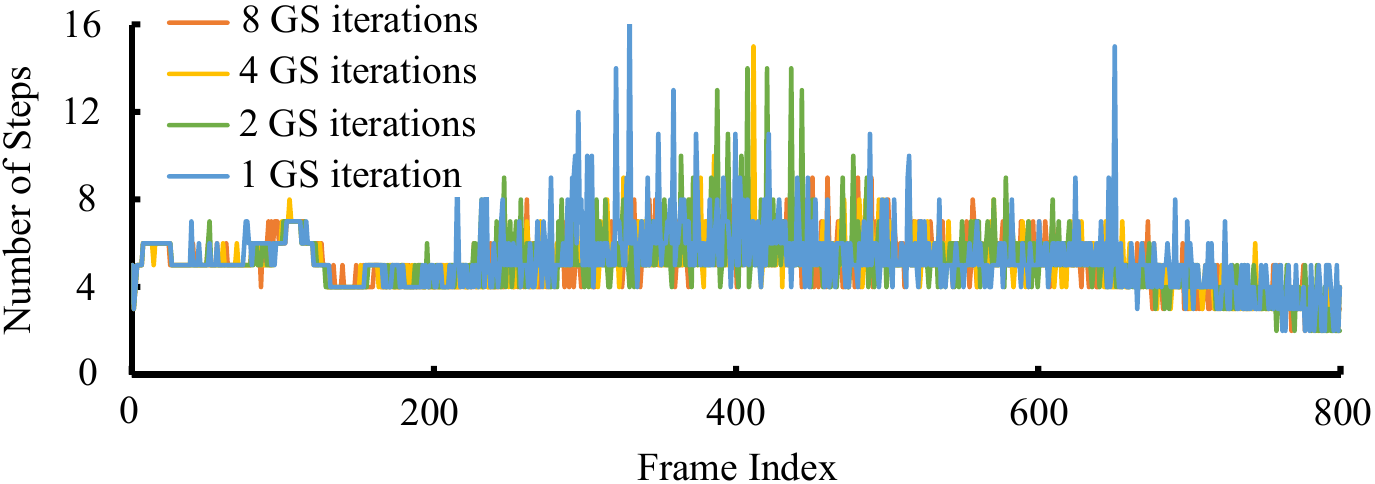}}
	\subfigure[The collision cost fluctuating over frames]{\includegraphics[width=3.3in]{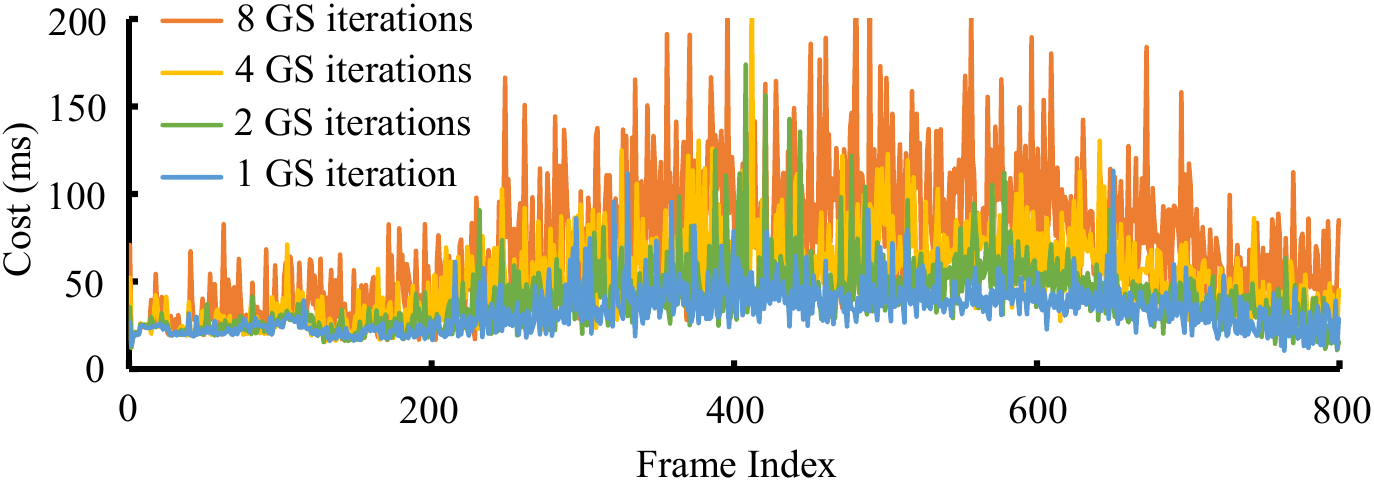}}
	\vspace{-0.12in}
	\caption{\textbf{Performance w.r.t. Gauss-Seidel iterations.}
          We count the performances of our method with different
          numbers of Gauss-Seidel iterations per backward
          step. By applying more Gauss-Seidel iterations, we can solve the LCP
          problem more accurately in every backward step, but it does not necessarily reduce
          the number of steps (Fig.~(a)). Thus, we recommend using
          only one Gauss-Seidel iteration per step for the efficiency
          purpose (Fig.~(b)).
        }
	\label{fig:different_iteration_number_of_SOR}
\end{figure}
\subsection{An Inexact GPU-Based Optimizer}
\label{sec:gpulcp}

A popular way of solving the LCP problem in Eq.~\eqref{eqn:LCP} is to apply a projected iterative method~\cite{erleben2013numerical}, which enforces $\boldsymbol \lambda \geq \mathbf 0$ after every iteration solving the linear system. In our simulator, we adopt standard multi-color Gauss-Seidel as our solver. {\color{black}We use the randomized graph coloring method in ~\cite{fratarcangeli2016vivace} while we assign color (graph node) to constraints, not  vertices. This is different from the strategy in \cite{fratarcangeli2016vivace}.}
Please refer to~\cite{fratarcangeli2016vivace} for more details.

\begin{figure}[t]
	\centering
	\subfigure[The number of steps fluctuating over frames]{\includegraphics[width=3.3in]{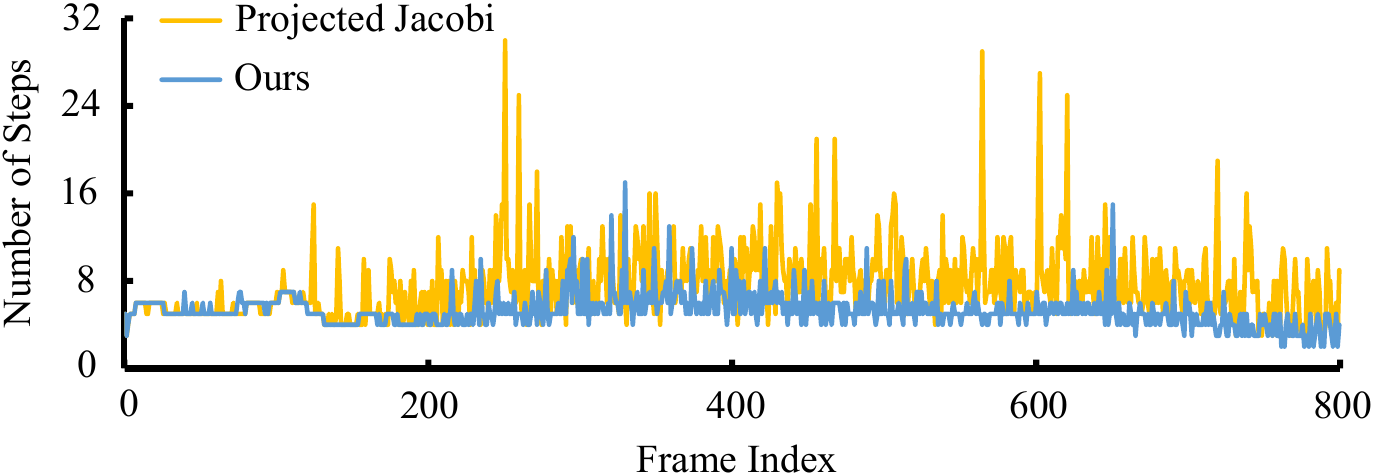}}
	\subfigure[The collision cost fluctuating over frames]{\includegraphics[width=3.3in]{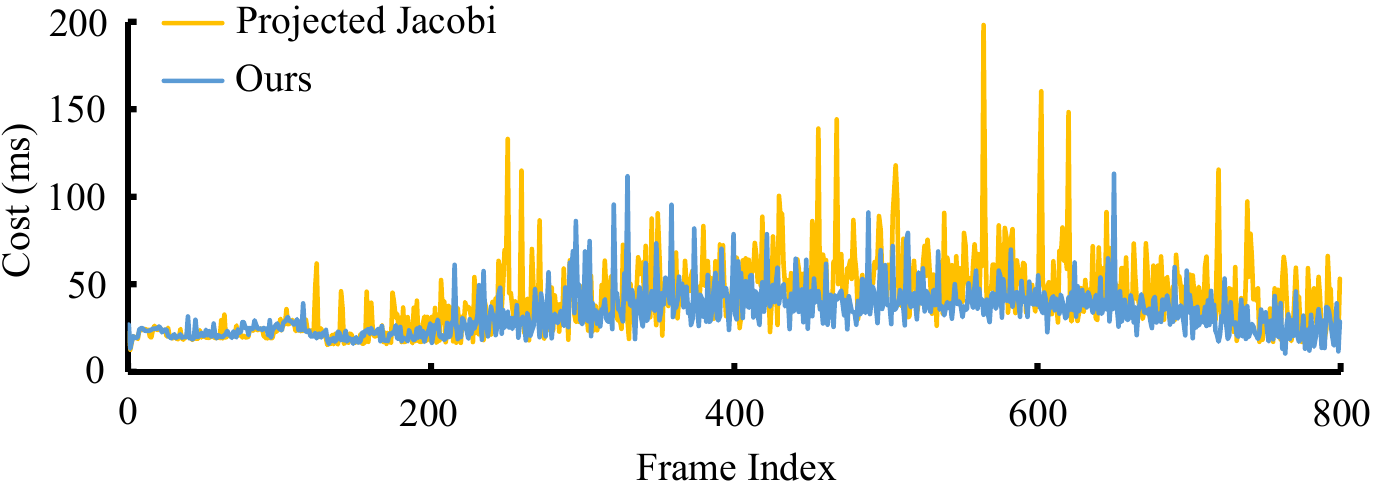}}
	\vspace{-0.12in}
	\caption{\textbf{Comparison with projected Jacobi method.} We
          count the performances of our method implemented with the projected Jacobi
          method and the projected Gauss-Seidel method (as
          ours). Compared with ours, using projected Jacobi needs more
          steps (Fig.~(a)) to conditionally converge under metric
          $\mathbf r$ and thus more computational costs (Fig.~(b)).}
	\label{fig:jacobi}
\end{figure}

\subsubsection{Inexactness} \label{sec:inexactness}\hspace{0.12in} One interesting question is how many iterations should we spend on solving the LCP problem?  The more iterations we use, the more exactly we get the problem solved. But given the fact that we face a new LCP problem in the next backward step, it will be a waste if we spend too much computational cost on a single problem. Ultimately, our choice should be based on  the total collision cost, fundamentally determined by two factors: the total number of steps for reaching convergence and the costs associated with backward and forward steps. According to Fig.~\ref{fig:different_iteration_number_of_SOR}, increasing the number of Gauss-Seidel iterations negatively affects the overall performance. Therefore, we choose to use a single iteration per backward step by default.

\subsubsection{A projected Jacobi implementation} \hspace{0.12in} The analysis in Subsection~\ref{sec:inexactness} motivates us to consider an even more inexact implementation, i.e., replacing one projected Gauss-Seidel iteration by one projected Jacobi iteration.  After testing this implementation, we conclude that it is not a suitable choice for two reasons. First, projected Jacobi needs an under-relaxation factor to ensure its convergence, which further lowers the convergence rate. Second, the nonsmooth nature of our method makes Chebyshev acceleration~\cite{Wang:2015:ACS} ineffective across steps. Overall, the method with projected Jacobi needs more steps and more costs for collisions as Fig.~\ref{fig:jacobi} shows. 


\subsection{Comparison to An Augmented Lagrangian Optimizer} 
\begin{figure}[t]
	\centering
	\subfigure[The number of steps fluctuating over frames]{\includegraphics[width=3.3in]{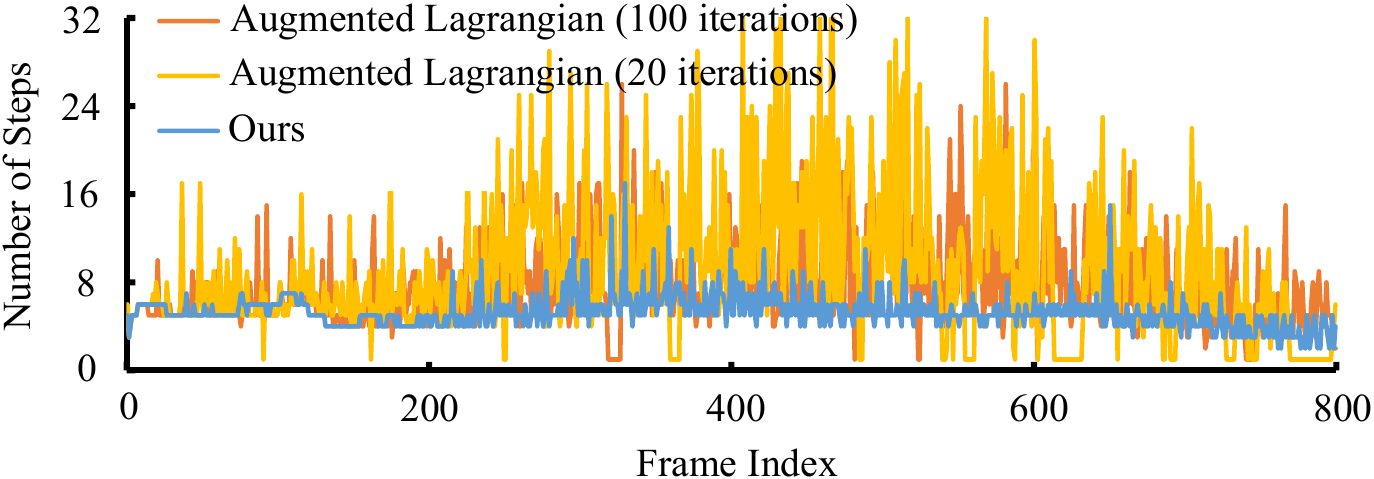}}
	\subfigure[The collision cost fluctuating over frames]{\includegraphics[width=3.3in]{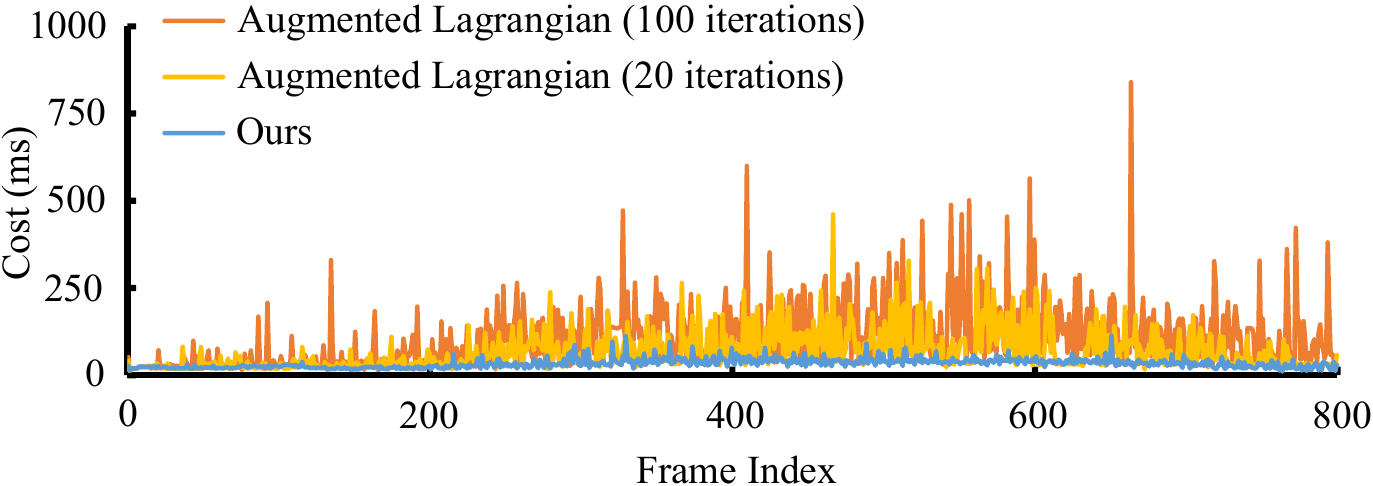}}
	\vspace{-0.12in}
	\caption{\textbf{Comparison with augmented Lagrangian method.}
          We compare the performance of our own optimizer and the
          augmented Lagrangian optimizer
          in~\cite{tang2018clotha}. Since their optimizer converges
          significantly slower than ours, it has to run multiple
          iterations per backward step to reduce the total number of steps.
          However, the improved convergence does not pay off
          the extra cost, which makes their method less efficient.}
	\label{fig:cmp_tang_2018}
        \vspace{-0.12in}
\end{figure}

\hspace{0.12in}
We can adopt other constrained optimization techniques to perform the task in our backward step as well.
Specifically, we would like to evaluate the performance of an augmented Lagrangian optimizer with the gradient descent method, advocated by Tang et al.~\shortcite{tang2018clotha}. 
The strength of their optimizer is its simplicity: it does not need to solve any system for primal or dual variables, and the only major computational components are gradient and constraint evaluations. But since their optimizer converges considerably slower, the method with their optimizer must run multiple iterations per backward step to reduce the total number of steps.  In our experiment, we implement their optimizer with two options: one running 20 iterations per step and one running 100 iterations per step.
{\color{black} We find when we use fewer iterations of their optimizer, its ability to change $\mathbf r$ degenerates very quickly. To avoid iterations with no revenue for their optimizer, if we find after one step, the change of $\big\| \mathbf r \big\|_{\infty}$ is tiny ($\big\| \mathbf r^{(l)} \big\|_{\infty} - \big\| \mathbf r^{(l+1)} \big\|_{\infty} $<FLT\_EPSILON), the collision handling module terminates immediately.
  Fig.~\ref{fig:cmp_tang_2018} shows that when using 20 iterations per step, the two-way method with their optimizer needs a large number of steps and often early terminates, especially when the knot is tight.
  When increasing to 100 iterations per step, the early termination could be alleviated, but the two-way method with their optimizer still needs a large number of steps compared with ours which is always terminated under metric $\mathbf r$ in a small time consumption. 
}
Overall our optimizer is a better choice.

\section{The Forward Step}
\label{sec:forward}
In each forward step, we move the vertices from the current state $\mathbf x^{(l)}$ toward ${\mathbf y}^{(l+1)}$ asynchronously:
\begin{equation}
	{\mathbf x}_i^{(l+1)} = {\mathbf x}_i^{(l)} + \alpha_i^{(l+1)} \big({\mathbf y}_i^{(l+1)} -  {\mathbf x}_i^{(l)} \big).
	\label{eq:forward}
\end{equation}
The key question is how to find the safe step size $\alpha_i^{(l+1)}$ for every vertex $i$, so that $ \mathcal X(\mathbf x^{(l)}, \mathbf x^{(l+1)}) \subset \Omega$. One way of obtaining a safe step size is to use continuous collision detection (CCD).  CCD tests calculate exact moments when proximity pairs intersect, using which we then determine how far $\mathbf x_i$ can travel. {\color{black}However, previous works show that CCD tests could be prone to errors ~\cite{brochu2012efficient,wang2014defending,tang2014fast} and require considerable efforts to get accelerated on GPUs~\cite{tang2011collision,tang2018clotha,tang2016cama}}. 

In our method, we propose to use an inexpensive yet reliable scheme for calculating the safe step size.  Our key idea is based on the simple fact that a proximity pair cannot intersect, if none of its vertices moves more than half of its distance. To use this idea, the method needs a set of proximity pairs $\mathcal P$, in which each pair contains two non-adjacent simplices with its distance below a global threshold $D^{(l)}$. Given $\mathcal P$, the method calculates $D_i$, the shortest distance of the proximity pairs involving vertex $i$:
\begin{equation}
	D_i=\min\limits_{ \{ a, \, b\} \in  \mathcal P } \mathtt {dist} \big(  \mathbf x_a^{(l)}, \mathbf x_b^{(l)} \big) \le D^{(l)},\, \forall a, b:  a \ne b \; \mathrm{and}\; i \in a \cup b.
\end{equation}
in which $a$ and $b$ are the two simplices and $\mathtt {dist} \big(  \mathbf x_a^{(l)}, \mathbf x_b^{(l)} \big)$ is their distance.
We treat $D_i/2$ as an upper bound on the displacement of vertex $i$ to ensure the intersection-free condition:
\begin{equation}
	\forall i: \; {\big\| \mathbf x_i^{(l)} - \mathbf x_i^{(l+1)} \big\|}  < D_i/2
	\Rightarrow (1-t)\mathbf x^{(l)} + t \mathbf x^{(l+1)} \in \Omega,
\end{equation}
for any $t\in [0, 1]$. 
Therefore we formulate our forward step by updating the position of vertex $i$ as:
\begin{equation}
	\left\{
        \begin{array}{l}
          \alpha_i^{(l+1)}=\min \left( {{{0.5 \gamma {D_i}} \mathord{\left/
				{\vphantom {{0.5 \gamma {D_i}} {\big\| {\mathbf y_i^{(l + 1)} - \mathbf x_i^{(l)}} \big\|}}} \right.
          \kern-\nulldelimiterspace} {\left\| {\mathbf y_i^{(l + 1)} - \mathbf x_i^{(l)}} \right\|}}, 1} \right), \\
          \mathbf x_i^{(l+1)} = \mathbf x_i^{(l)} + \alpha_i^{(l+1)}\big( {\mathbf y_i^{(l + 1)} - \mathbf x_i^{(l)}} \big),\\
          r_i^{(l+1)} = r_i^{(l)}\big(1-\alpha_i^{(l+1)}\big)   ,       
          \end{array}
      \right.
	\label{eq:forward2}
\end{equation}
in which $\gamma$ is a damping factor preventing proximity pairs from getting too close in a single forward step.  Using the step size calculated for every vertex, we ensure that $\mathbf x^{(l+1)}$ is an acceptable intermediate state, regardless of the search direction. 

\begin{figure}[t]
	\centering
	\subfigure[$\x^{[k]}$ (above) and $\y^{[k+1]}$ (below)]{\includegraphics[width=1.66in]{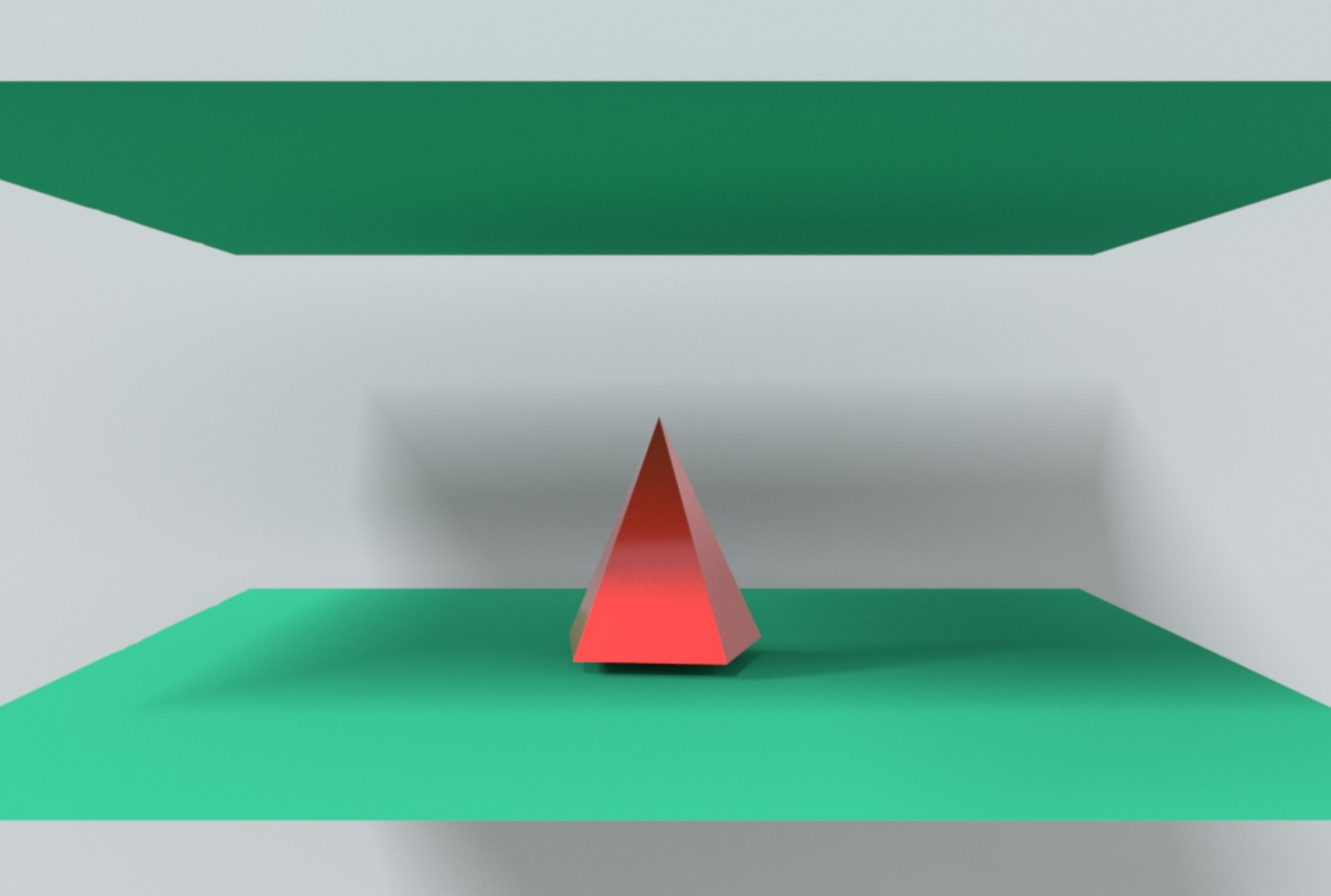}}	
	\subfigure[stack view of $\x^{[k+1]}$ with various $\epsilon$]{\includegraphics[width=1.66in]{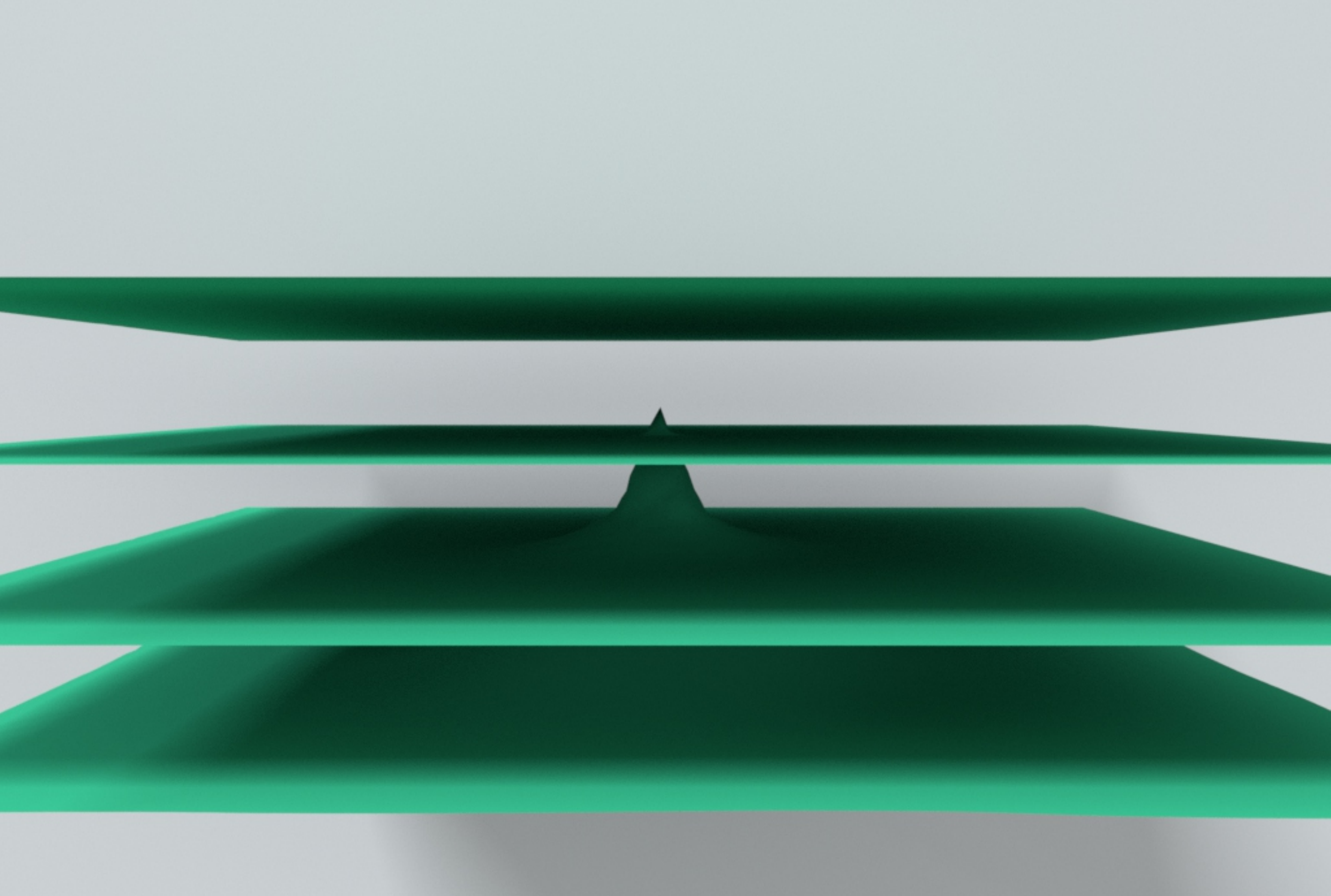}}
	\subfigure[$\x^{[k+1]}$ with $\epsilon=0.75$]{\includegraphics[width=0.802in]{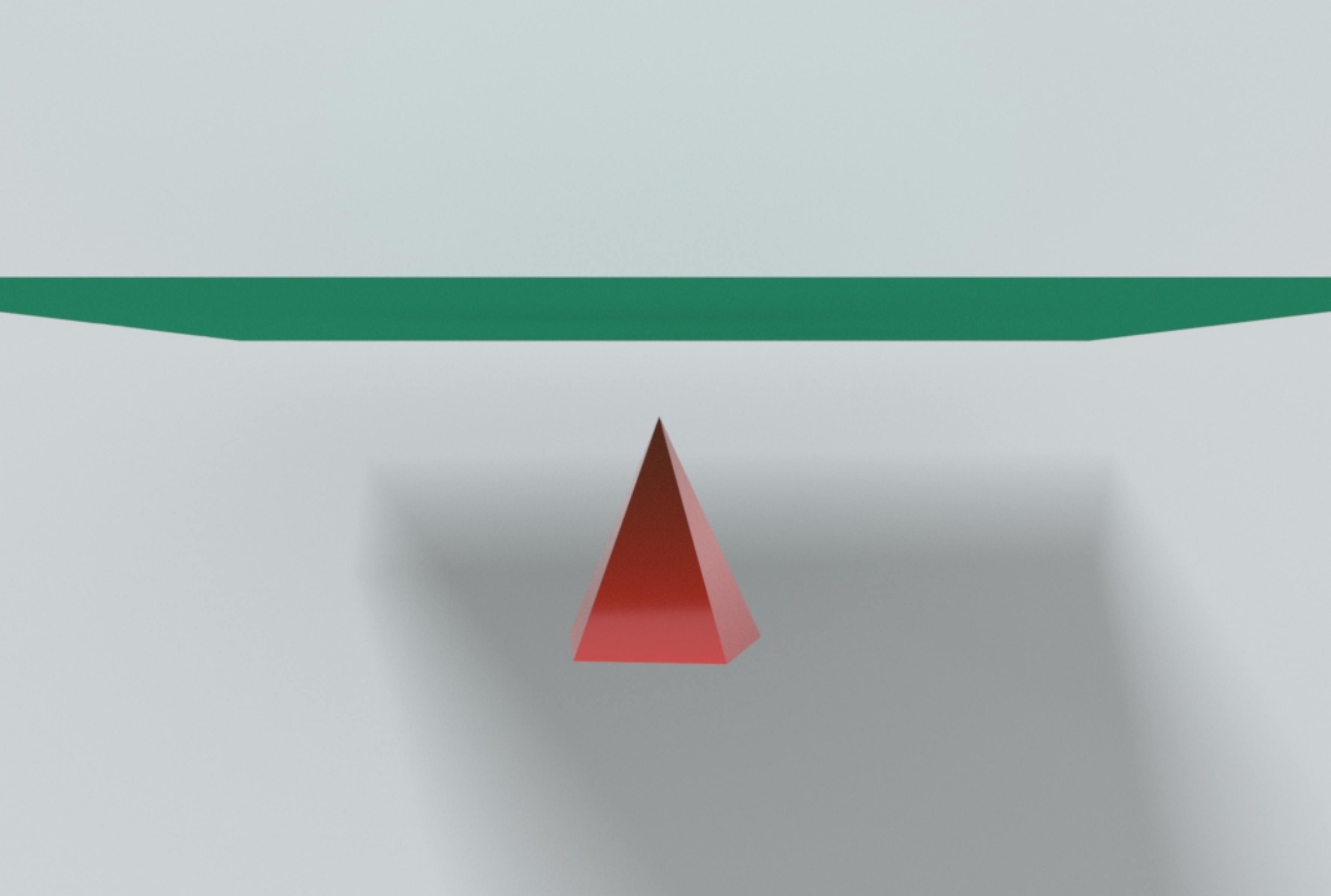}}
	\subfigure[$\x^{[k+1]}$ with $\epsilon=0.50$]{\includegraphics[width=0.802in]{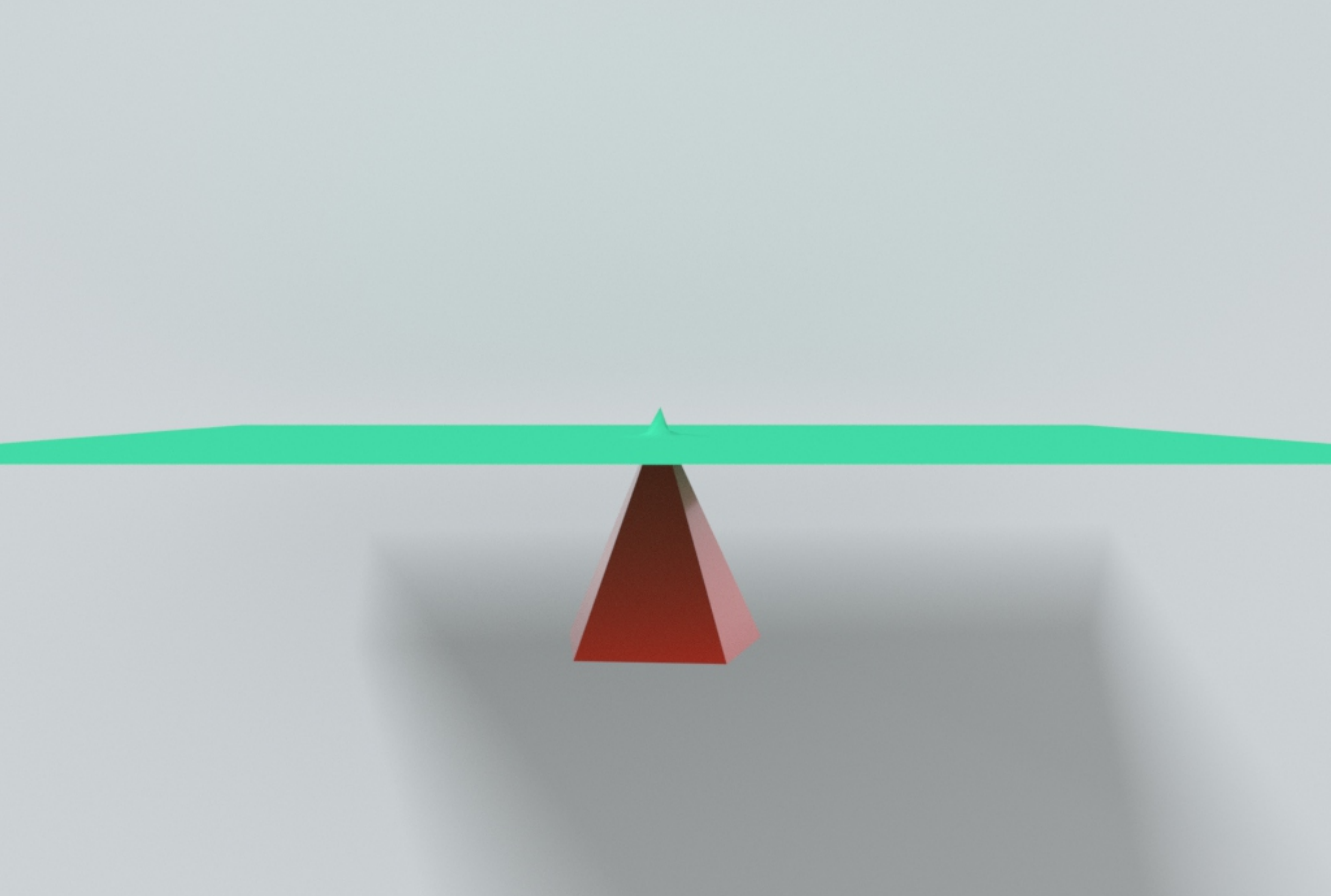}}
	\subfigure[$\x^{[k+1]}$ with $\epsilon=0.25$]{\includegraphics[width=0.802in]{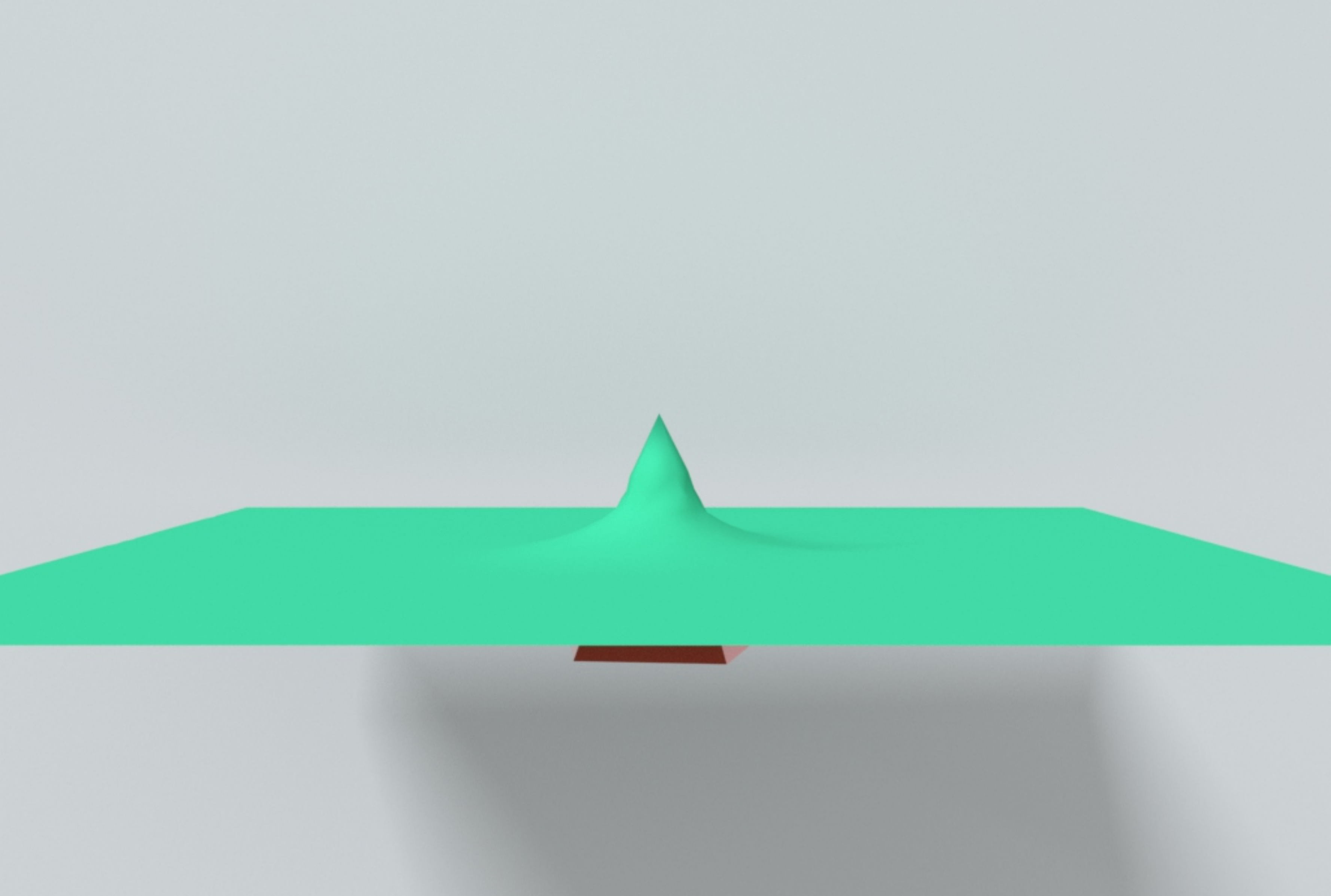}}
	\subfigure[$\x^{[k+1]}$ with $\epsilon=0.0001$]{\includegraphics[width=0.802in]{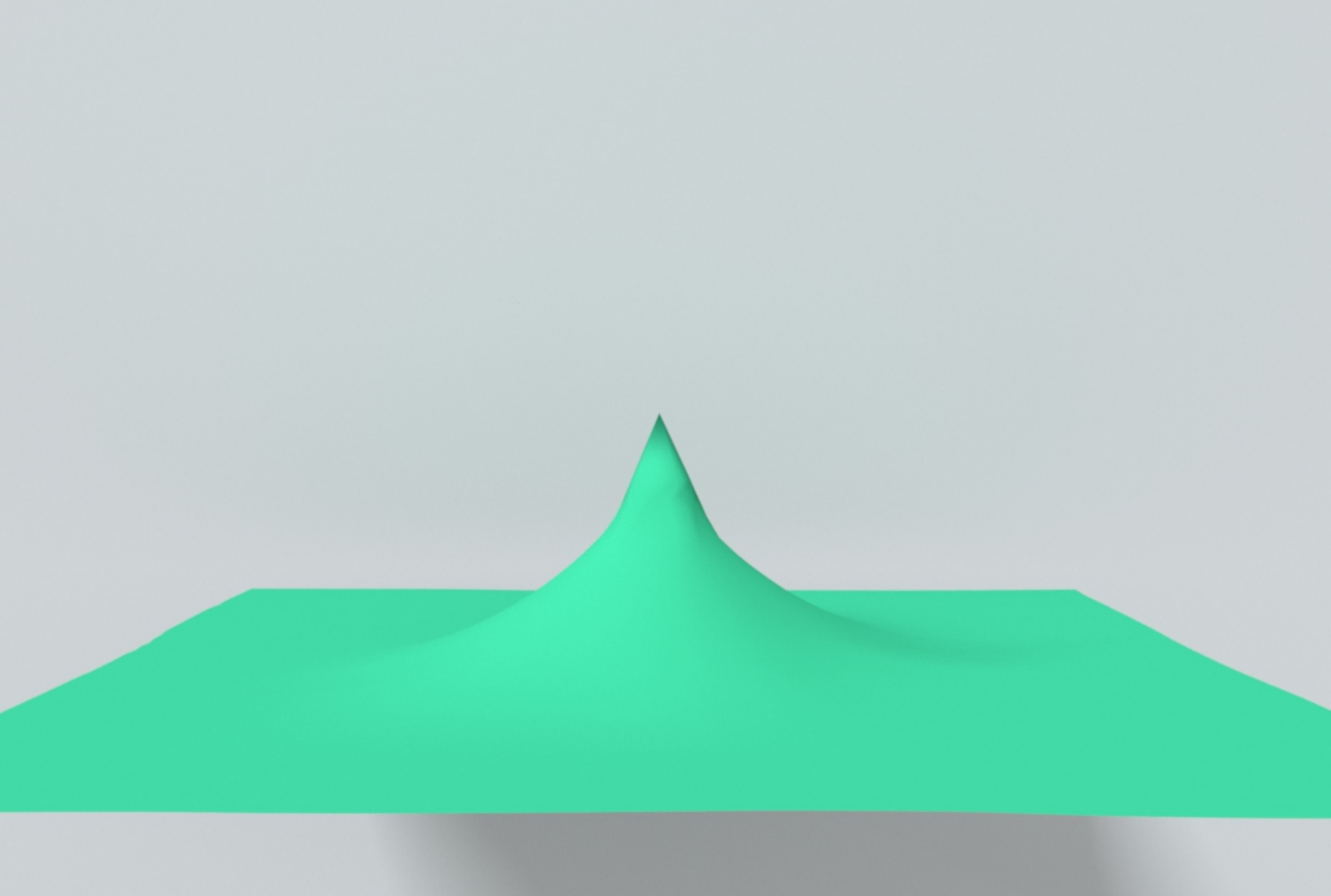}  \label{fig:early_termination_our_epsilon} } 	
	\vspace{-0.12in}
	\caption{\textbf{Termination condition and threshold
            $\boldsymbol{\epsilon}$.} With the same $\x^{[k]}$ and
          $\y^{[k+1]}$, the accumulated moving distance from
          $\x^{[k]}$ to $\x^{[k+1]}$ is getting close to the distance
          from $\x^{[k]}$ to $\y^{[k+1]}$ as $\epsilon$ approaching
          zero, reducing the risk of early termination.}
	\label{fig:early_termination}
\end{figure}
      
A special feature we would like to mention in Eq.~\eqref{eq:forward2} is the termination metric $r_i$.  The non-smooth nature of our optimization method makes it difficult to define the termination condition by the step size $\alpha_i^{(l+1)}$ directly, without potential early termination risks.  To address this issue, we come up with a termination metric $r_i$ in an accumulated fashion.  Intuitively, it keeps track of the remaining step size needed for vertex $i$ to reach its target and we terminate the method once $\| \mathbf r^{(l+1)} \|_{\infty}$ drops below a certain threshold $\epsilon$. As the threshold $\epsilon$ tends to zero, the accumulated moving distance from $\mathbf x^{[k]}$ to $\mathbf x^{[k+1]}$ tends to be close enough to the distance from $\mathbf x^{[k]}$ to $\mathbf y^{[k+1]}$, so the early termination risks of the optimization could be eliminated as shown in Fig.~\ref{fig:early_termination_our_epsilon}.

Compared with CCD tests, distance evaluations used by our scheme are computationally inexpensive, reliable against floating-point errors and easy to parallelize on GPUs.

\subsection{Comparison to CCD-Based Schemes}
\setlength{\columnsep}{0.15in}
\begin{wrapfigure}{R}{0.27\textwidth}
	\centering
	\vspace*{-.08in} 
	\includegraphics[width=1.65in]{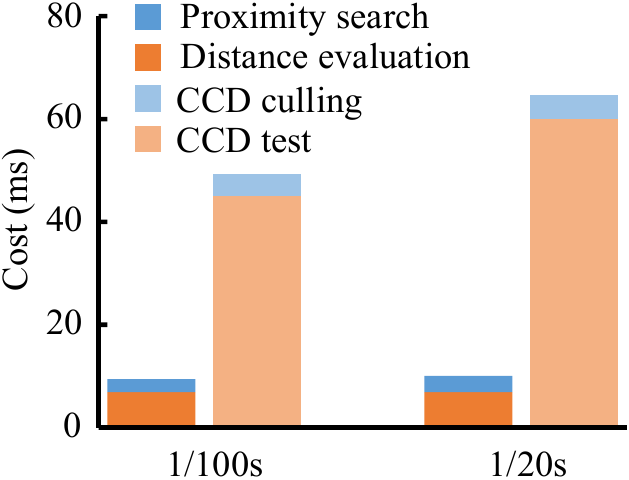}
	\vspace*{-.06in}  
	\caption{\textbf{Comparison with CCD.} Overall, the total cost
          of CCD-free step size schemes by the collision handling process
          is five to seven times faster than
          CCD-based schemes in a time step.}
	\vspace*{-.1in}
	\label{fig:bid}
\end{wrapfigure}
To compare our CCD-free step size scheme with CCD-based schemes, we implement a CCD-based scheme by the tests provided by I-Cloth~\cite{tang2018clotha}, which is one of the fastest CCD implementations on a GPU. We also adjust both schemes to use the same proximity search tool provided by I-Cloth, so that we can eliminate the difference in the implementations of proximity search. On the same NVIDIA GeForce GTX 2080 Ti GPU, we evaluate both schemes in a number of examples with two large time steps: $\Delta t$=1/100s and $\Delta t$=1/20s. 

In the examples with $\Delta t$=1/100s, the experiment shows our proximity search cost is about 60 percent of the broad-phase CCD culling cost, while the distance evaluation cost is about 15 percent of the narrow-phase CCD test cost. Together our CCD-free scheme is five times faster than the CCD-based scheme.

In the examples with $\Delta t$=1/20s, the narrow-phase CCD test cost increases significantly while  the distance evaluation cost increases marginally. As a result, our CCD-free scheme is about seven times faster than the CCD-based scheme.

We note that the computational costs of the two step size schemes alone do not provide the full picture of their difference. In general, our CCD-free scheme provides smaller step sizes and causes 10 to 20 percent more steps needed for convergence. This further increases the cost of the backward step by 10 to 20 percent. But overall, it is still beneficial to use the CCD-free scheme, given the large computational cost needed for CCD tests.

\paragraph{The Special CCD-Based Scheme in~\cite{Wu:2020:ASF}}
\textcolor{black}{The step size scheme adopted by the hard phase
  in~\cite{Wu:2020:ASF} is also based on CCD
  tests. Compared with other CCD-based schemes,
    their approach restricts all of the edge lengths to be less than a constant
    upper bound, and accordingly derives a series of sufficient
  conditions to prevent intersections, which can be achieved by
  handling vertex-vertex contacts only.  As a result,
  their CCD tests are less expensive, making
  their method suitable for fast simulation of
  virtual garments that are almost inextensible.}

\textcolor{black}{ However, such a benefit comes at the cost of limited
  applicability. If we apply their scheme to simulate
  general shell deformations where this length
  restriction can not be adopted, e.g., the sustaining inflation or
  contraction of membrane in the normal flow example
  (Fig.~\ref{fig:normal_flow}), then CCD tests have to consider all
  necessary vertex-triangle and edge-edge contacts again. In such
  cases, their method would be less efficient.
}

\section{ Implementation Details}
\label{sec:implement}
In this section, we discuss the implementation details of our two-way method in a GPU-based deformable body simulator.

\subsection{Dynamics Solvers}
Since our method works as a standalone module for collision handling,
it is naturally compatible with most of the dynamics solvers. In our
simulator, \textcolor{black}{we follow the pipeline in Eq.~\eqref{eq:step_and_project} to
solve the nonlinear optimization problem stemming from deformable body dynamics.}
In detail, $Q_k$ is a quadratic
proxy of energies written as \textcolor{black}{below:}
 \begin{equation}
    Q_k={{E}\big(\mathbf x^{[k]},\mathbf x^{t},\mathbf v^{t}\big) } + {\mathbf b^{[k]}}^\intercal (\mathbf y - \x^{[k]}) + \frac{1}{2} (\mathbf y- \x^{[k]})^\intercal \mathbf G^{[k]} (\mathbf y- \x^{[k]}), \quad\;      
    \label{eq:quadratic_problem}
  \end{equation}
in which $\mathbf b^{[k]}$ is the gradient and $\mathbf G^{[k]}$ is the modified Hessian of $\left.E(\mathbf x,\mathbf x^{t},\mathbf v^{t} )\right|_{\mathbf x=\mathbf x^{[k]}}$ after positive semi-deﬁnite projection.
  We apply the conjugate gradient method with a block Jacobi preconditioner to solve the linear system emerged from the quadratic model in every Newton iteration.
  {\color{black}We treat the
  Newton iteration as an update and run our two-way 
  method for collision handling right afterwards.}
Currently, our simulator uses CUDA 11.2 and the CUB library for reduction and sorting operations.

In our implementation, we fix the number of Newton iterations as a
constant.  Ideally, this number is related to the time step: the
solver should run more Newton iterations as the time step increases
for more accurate results. {\color{black}But even if we choose to run a single
Newton iteration for a large time step, i.e., $\Delta t$=1/10s, our
method still can guarantee intersection-free with much less artifacts compared with existing collision handling algorithms as Fig.~\ref{fig:cmp_with_impact_zone} shows.}

\begin{figure}[t]
	\centering
	\subfigure[The initial state]{\includegraphics[width=1.65in]{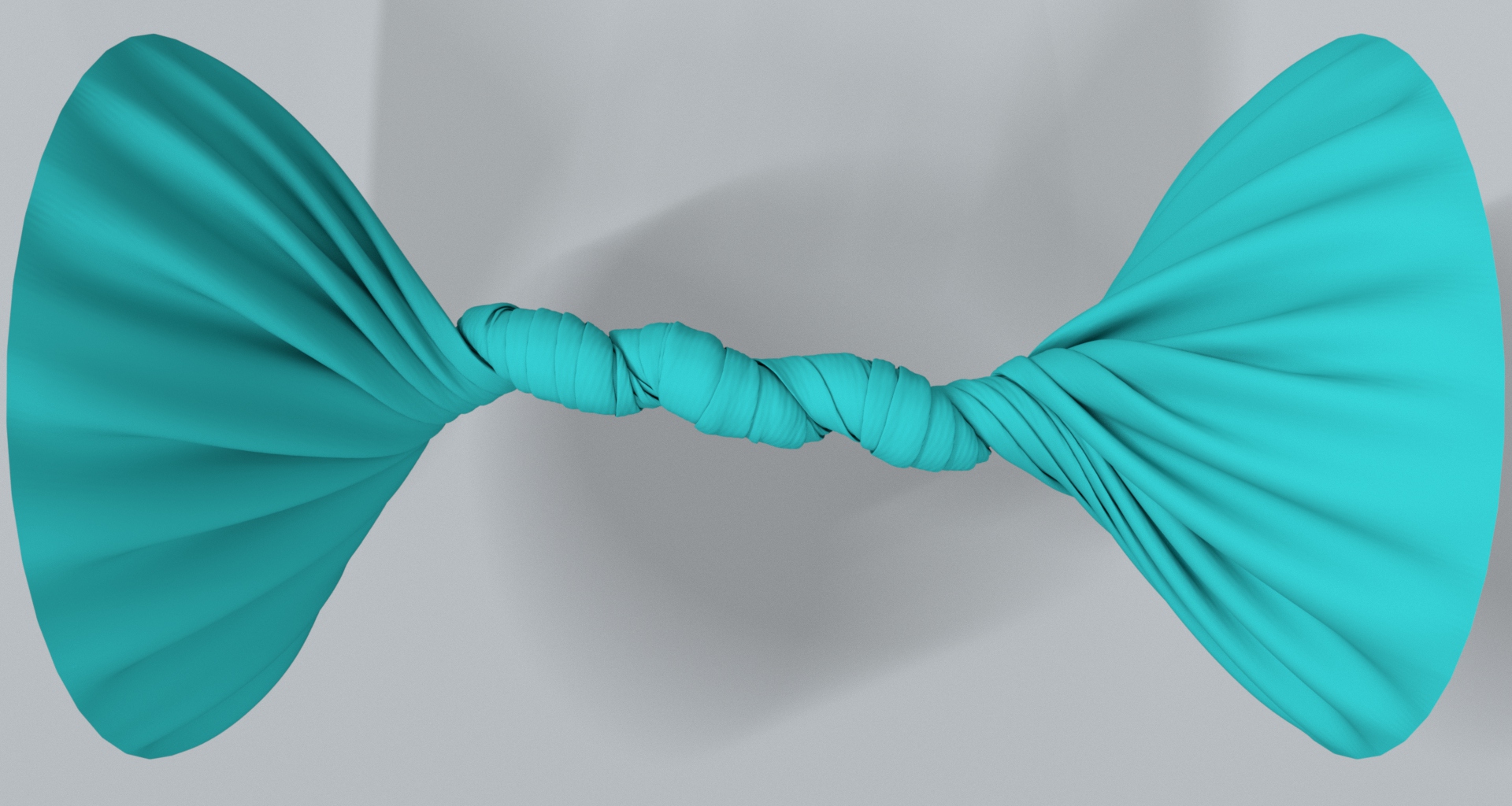}}\hfill
	\subfigure[The twisted state]{\includegraphics[width=1.65in]{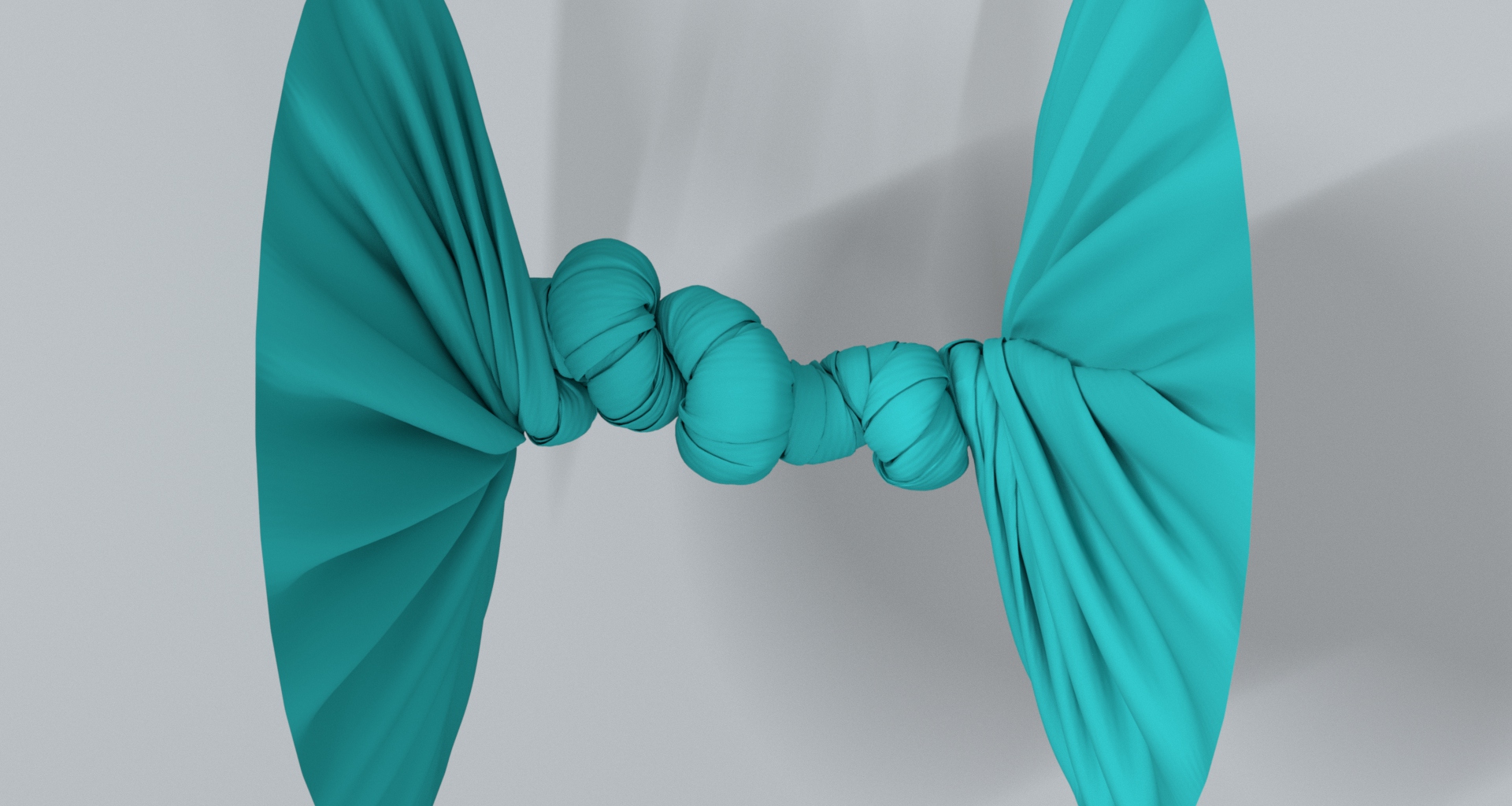}}\hfill
	\vspace*{-0.12in}
	\caption{\textbf{Twisting tube.} When a compliant cloth tube
          is severely squeezed, it experiences intense and frequent
          collisions in the middle part as shown in (b).  Our method
          robustly handles these collisions in this example taking
          $\Delta t$=1/100s as the time step.}
	\label{fig:twist_cylinder_with_bulk_effects}
\end{figure}

\subsection{Elastic and Repulsive Models}
\label{sec:elastic}
To simulate codimensional deformable body examples shown in
Fig.~\ref{fig:cod}, we provide a number of elastic models in our
dynamics solver, including the St. Venant-Kirchhoff model for
tetrahedral meshes, the co-rotational linear model and the quadratic
bending model~\cite{Bergou:2006:QBM} for triangular meshes, and the
mass-spring model for hair strands.
{\color{black}The eigensystems of the Hessian matrix for these energy models have been analyzed~\cite{Choi_2002,etzmuss2003fast,teran2005robust,smith2019analytic} and at most only $3\times3$ singular value decomposition (SVD) is needed for positive semi-deﬁnite projection, which can be solved efficiently on GPU~\cite{Gao:2018:GPU_MPM} or the Hessian matrix is natively guaranteed to be positive semi-definite~\cite{Bergou:2006:QBM}. Some other energy models can also be incorporated into this framework, such as the discrete shell model~\cite{grinspun2003discrete} where the positive semi-definite projection involving larger matrix eigendecomposition should be done on the CPU like~\cite{li2020codimensional,chen2022unified}.} 
Similar to many other
simulators~\cite{narain2012adaptive,tang2016cama,tang2018clotha,li2020p},
our simulator incorporates a quadratic-energy-based repulsive model
into deformable body dynamics, to reduce the collision complexity in
simulation.

\begin{figure}[t]
	\centering
	\subfigure[The initial state]{\includegraphics[width=1.1in]{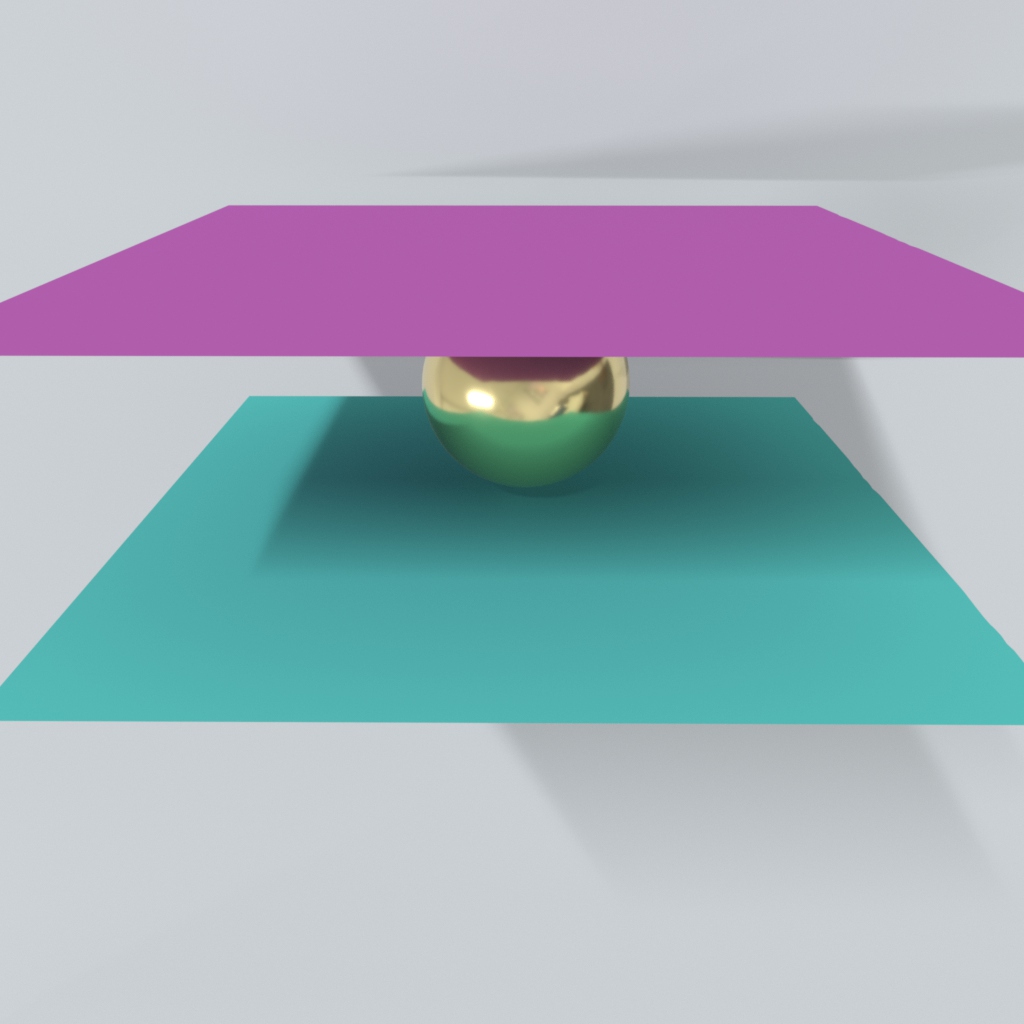}}
	\subfigure[The falling state]{\includegraphics[width=1.1in]{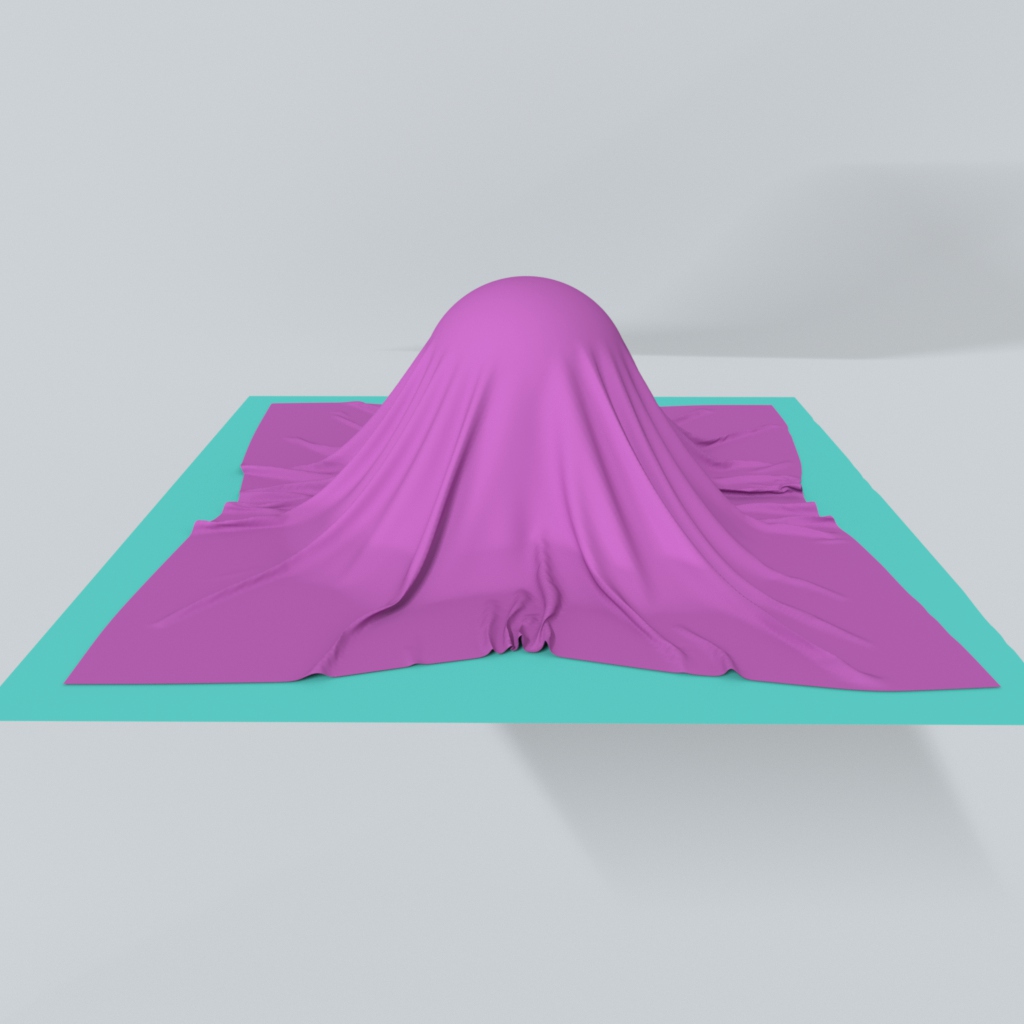}}
	\subfigure[The sliding state]{\includegraphics[width=1.1in]{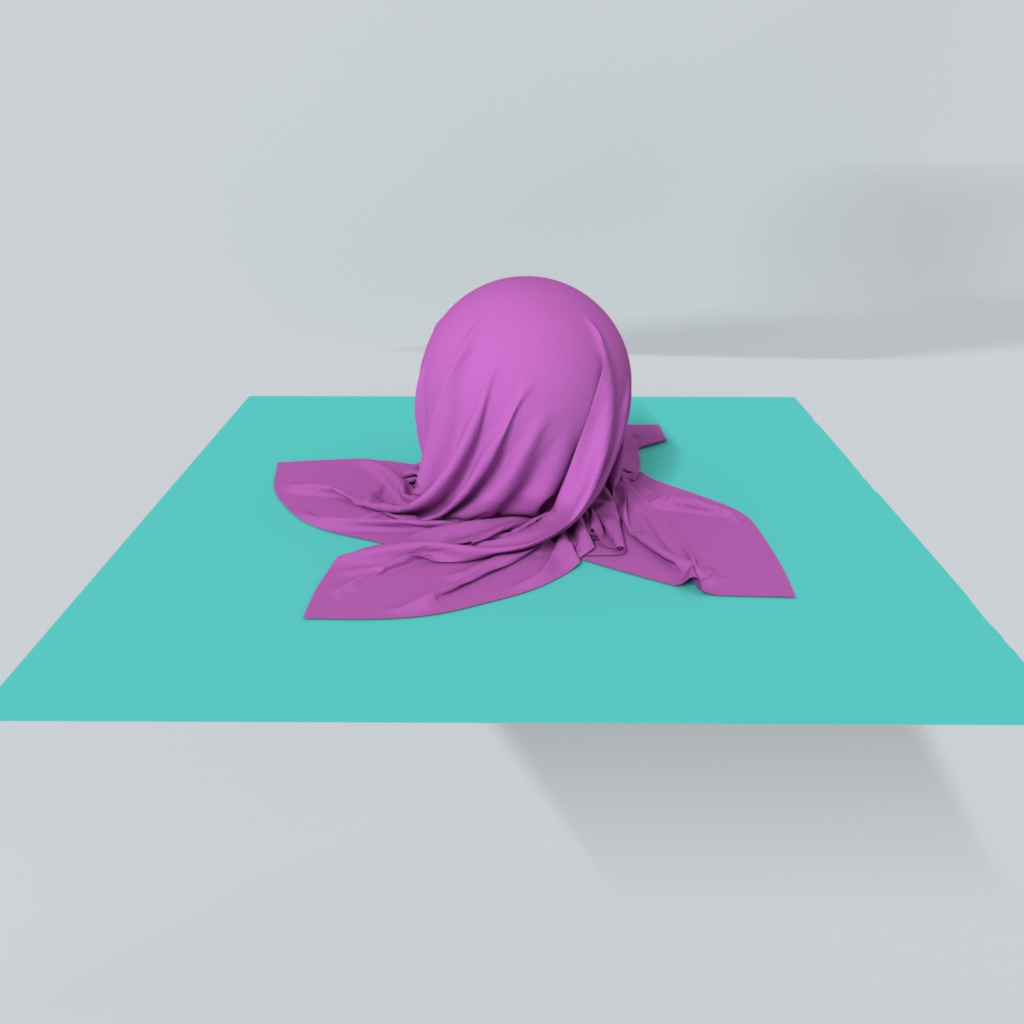}}
	\vspace{-0.12in}
	\caption{\textbf{Rotating sphere.} When a square cloth patch falls onto a rotating sphere, it forms multiple folds and wrinkles due to its frictional contacts with the sphere and the ground floor.}
	\label{fig:sphere_contact}
\end{figure}

\subsection{Frictional Contacts}
We adopt the velocity filtering approach proposed in~\cite{bridson2002robust} to handle frictional contacts.  First we run the dynamics solver to compute the target state $\mathbf y^{[k+1]}$ with no friction. \textcolor{black}{We then calculate penetration depths to estimate collision impulses and use them to determine associated frictional impulses by Coulomb's law as well.  Finally, we update $\mathbf y^{[k+1]}$ by both impulses and treat it as the new target state for collision handling.}  We note that the velocity filtering strategy is more suitable in deformable-rigid body contacts, than in self body contacts, since penetration depth estimations are inaccurate and irrelevant to the actual collision handling process, especially if the time step is large. Fig.~\ref{fig:sphere_contact} demonstrates \textcolor{black}{our current} frictional effects \textcolor{black}{achieved} by the velocity filtering.

\begin{table}[t]
	\caption{\textbf{Statistics and performances.} This table lists the time step size, the maximum and mean numbers of steps and time cost spent by our two-way method in one time step on various examples.}
	\label{tab:examples}
	\vspace{-0.12in}
	{
		\small 
		\begin{center}
			\begin{tabular}{  c|c|c|c}
				Name  & Time & Avg. (Max.)  & Avg. (Max.)\\
				{\small (\#verts., \#edg./\#tri./\#tet., ref.)} & step (s) & \# of steps & cost (s)\\
				\hline
				{ Needle (10k, 20k, Fig.~\ref{fig:early_termination_our_epsilon})} & 1/20& 46 (1304) & 0.036 (1.942)\\
				{ Blade (59k, 116k, Fig.~\ref{fig:blade})} &1/20& 55 (779) & 0.267 (4.463)\\
				\hline                                 
				{ Funnel (49K, 98K, Fig.~\ref{fig:funnel}a)} &1/20 & 27 (675) & 0.087 (2.161) \\
                                { Funnel (49K, 98K, Fig.~\ref{fig:funnel}b)} &1/40 & 23 (437) & 0.055 (1.198) \\
                                { Funnel (49K, 98K, Fig.~\ref{fig:funnel}c)} &1/80 & 12 (165) & 0.040 (0.776) \\
				{ Funnel (49K, 98K, Fig.~\ref{fig:funnel}d)} &1/160 & 10 (95) & 0.029 (0.296) \\
                                \hline
				{ Sphere (50k, 100k, Fig.~\ref{fig:sphere_contact})} & 1/100 & 19 (219) & 0.044 (0.475)\\
				\hline
				{ Dress  (30K, 60K, Fig.~\ref{fig:multi_layer_dress}a)}  & 1/100 & 39 (142) & 0.133 (0.511)\\
				{ Gown (27K, 51K, Fig.~\ref{fig:multi_layer_dress}c)}& 1/100 & 32 (150) & 0.092 (0.443)\\		
				\hline
				{ Bow knot (71K, 142K, Fig.~\ref{fig:teaser}a)}  & 1/100 & 5.4 (17) & 0.034 (0.113)\\
				{ Reef knot (37K, 71K, Fig.~\ref{fig:teaser}e)}  & 1/100 & 7 (23) & 0.024 (0.097)\\
				{ Tube (25k, 51K, Fig.~\ref{fig:twist_cylinder_with_bulk_effects})}  & 1/100 & 19 (54) & 0.308 (1.500)\\
				\hline								
                                { Mat (33k, 88K, Fig.~\ref{fig:cod}a)} &  1/100 & 9 (128) & 0.020 (0.303) \\
                                { Hair (63K, 62k, Fig.~\ref{fig:cod}c)} &  1/100 & 16 (503) & 0.252 (15.821)\\
                                { Sand (30K, - , Fig.~\ref{fig:cod}e)} & 1/100 & 51 (160) & 0.223 (1.550) \\
			\end{tabular}
		\end{center}
	}
\end{table}
\begin{figure}[t]
	\centering    
	\subfigure{\includegraphics[width=3.3in]{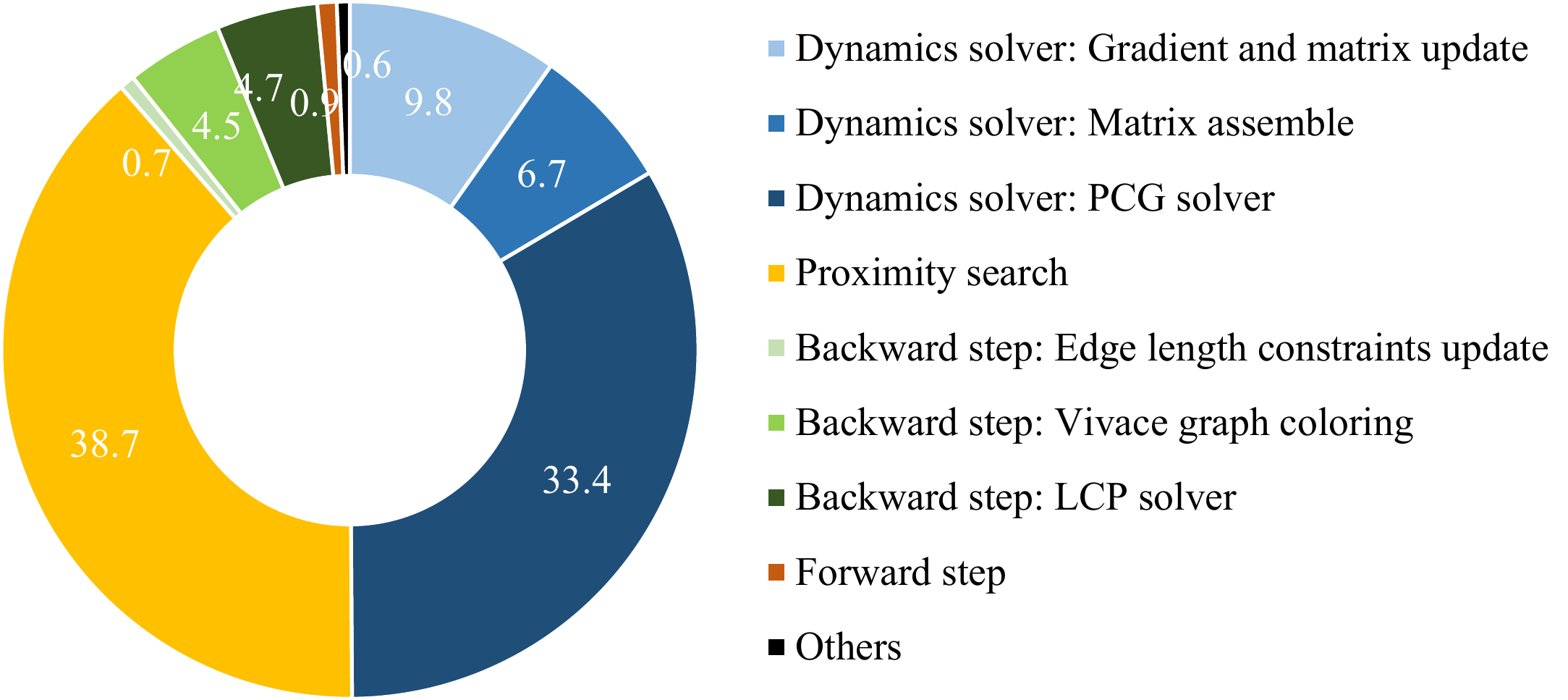}}
	\vspace*{-0.12in}
	\caption{\textbf{Breakdown analysis.} We visualize a typical
          breakdown of the cost spent in each stage on the bow knot example.
          As shown in this figure, the PCG solves
          and the proximity search are the two most expensive parts
          in our simulator.  }
	\label{fig:bowknot_breakdown}
\end{figure}

\section{Results and Discussions}
We test our simulator on an Intel Core i5-7500 3.4GHz CPU and an NVIDIA GeForce GTX 2080 Ti GPU. Table~\ref{tab:examples} summarizes the statistics and performances of our examples, and Table~\ref{tab:notations} provides major parameters and their default values used in our algorithm. 
{\color{black}Most results are computed with the default $\delta=1$mm. Specifically, $\delta=0.5$mm is used for the knotting examples in Fig.~\ref{fig:teaser} and $\delta=0.2$mm is used for the hair example in Fig.~\ref{fig:cod}.}
Our examples include both scenes of \textcolor{black}{common} benchmarks with complex collisions (Figs.~\ref{fig:teaser} and~\ref{fig:twist_cylinder_with_bulk_effects}), and virtual garments of multiple layers dressed on human bodies (Fig.~\ref{fig:multi_layer_dress}). Related animation \textcolor{black}{sequences} are provided in the supplemental video. In general, the cost for handling contacts depends on the number of colliding elements and the number of steps, which are in essence determined by the time step and \textcolor{black}{the shape} complexity.  By default, the step number limit $L$ is set to $512$ when $\Delta t=1/100$s,
and $2,048$ when $\Delta t=1/20$s. In practice, our method typically converges in condition of $\|\mathbf{r}\|_\infty$ less than 64 steps, far before reaching the step number limit $L$ as Table~\ref{tab:examples} shows.
\begin{table}[t]
	\caption{\textbf{Parameters and their default values.}}
	\label{tab:notations}
	\vspace{-0.12in}
	\begin{center}
		\begin{tabular}{  c|c|c}
			Symbol & Meaning & Value \\
			\hline
			$L$  & The limit on the number of steps & 512 to 2,048 \\
			$\epsilon$ & The termination threshold & 0.0001 \\
			$D^{\min}$ & The proximity search lower bound & 2mm\\
			$D^{\max}$ & The proximity search upper bound & 4mm\\
			$\delta$  & The constraint activation threshold & 1mm \\
			$\sigma$ & The violation ratio limit & 1.1 \\
			$\gamma$ & The damping factor on movement  & 0.9  \\
			\hline
		\end{tabular}
	\end{center}
\end{table}

\subsection{Breakdown Analysis}
Fig.~\ref{fig:bowknot_breakdown} provides a breakdown of the computational cost spent in the bow knot example (Fig.~\ref{fig:teaser}). It shows that PCG solves and proximity search are the two most expensive components in our simulator.  In comparison, the actual cost spent by the forward step and backward step is much lower, i.e., occupying only 10.8 percent of the total cost. It suggests that our simulator still has a large space for performance improvement and it should benefit greatly from faster linear solvers and proximity search algorithms in the future.

\begin{figure}[t]
	\centering
	\subfigure[$\Delta t$=1/20s]{\includegraphics[width=1.65in]{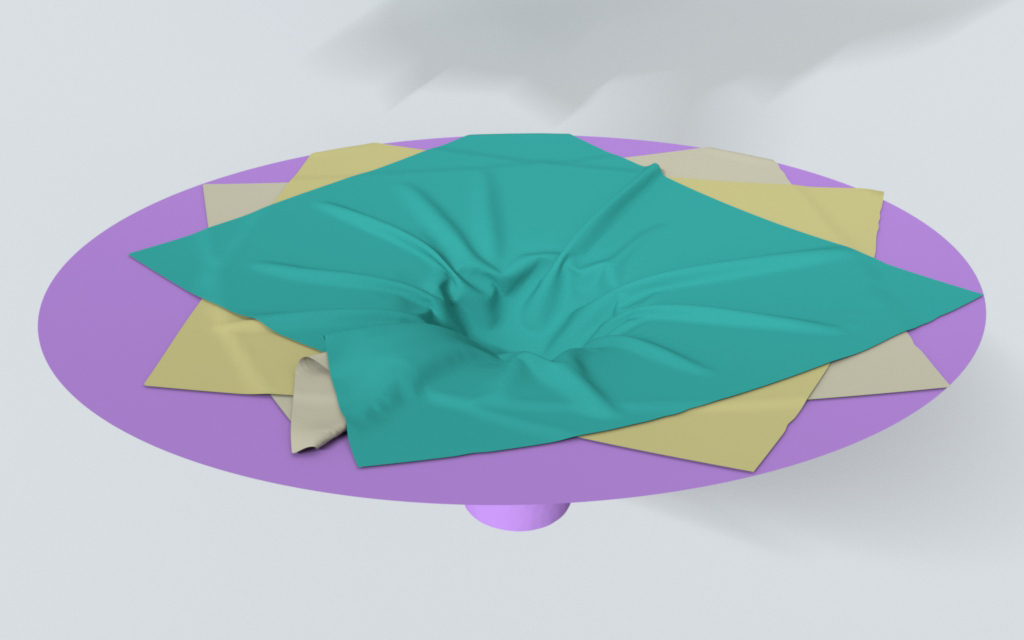}}
	\subfigure[$\Delta t$=1/40s]{\includegraphics[width=1.65in]{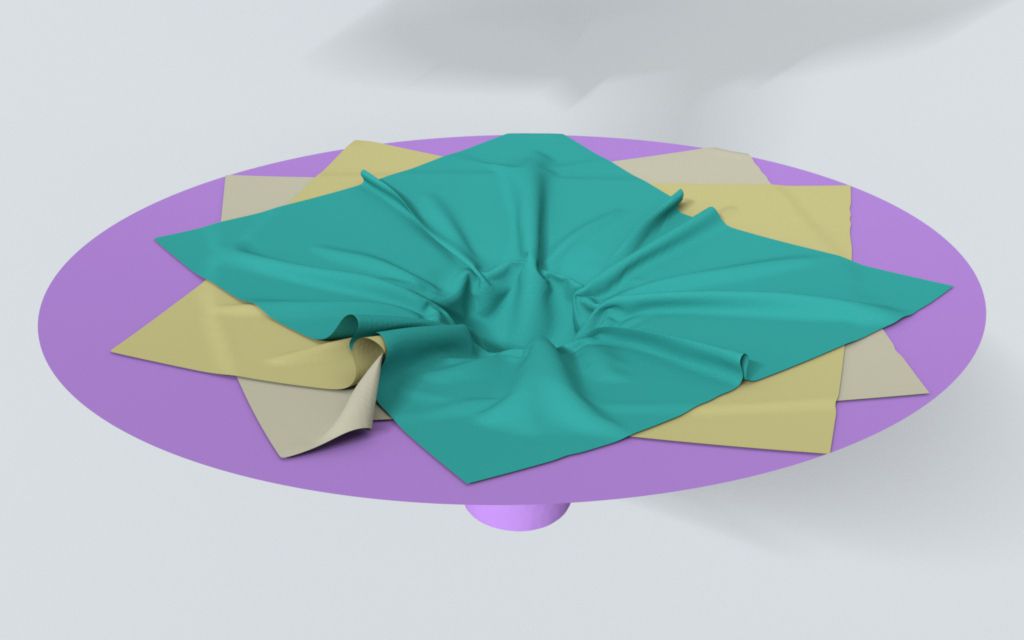}}
	\subfigure[$\Delta t$=1/80s]{\includegraphics[width=1.65in]{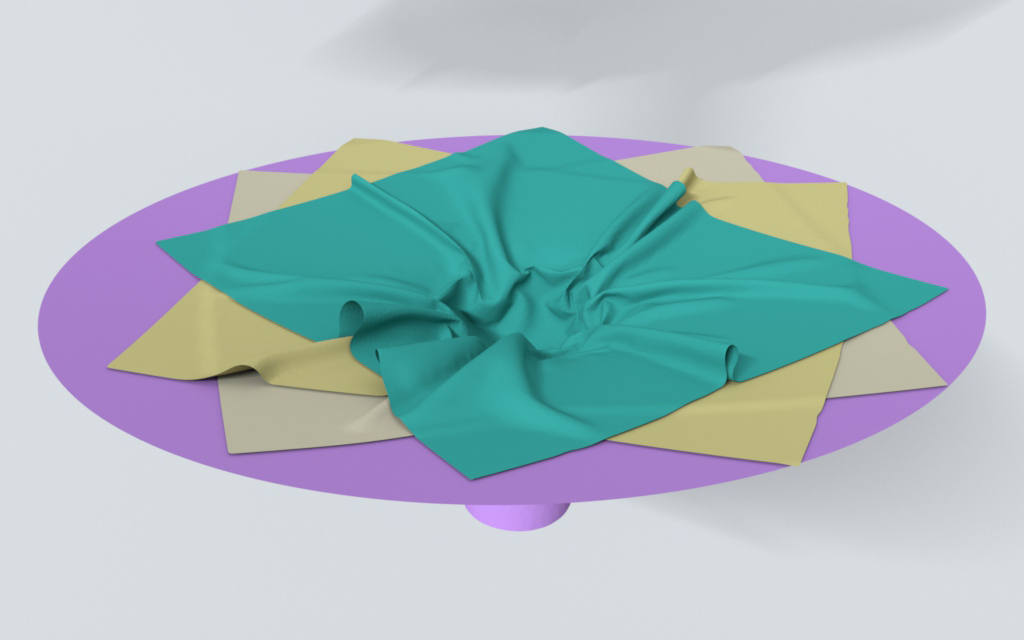}}
	\subfigure[$\Delta t$=1/160s]{\includegraphics[width=1.65in]{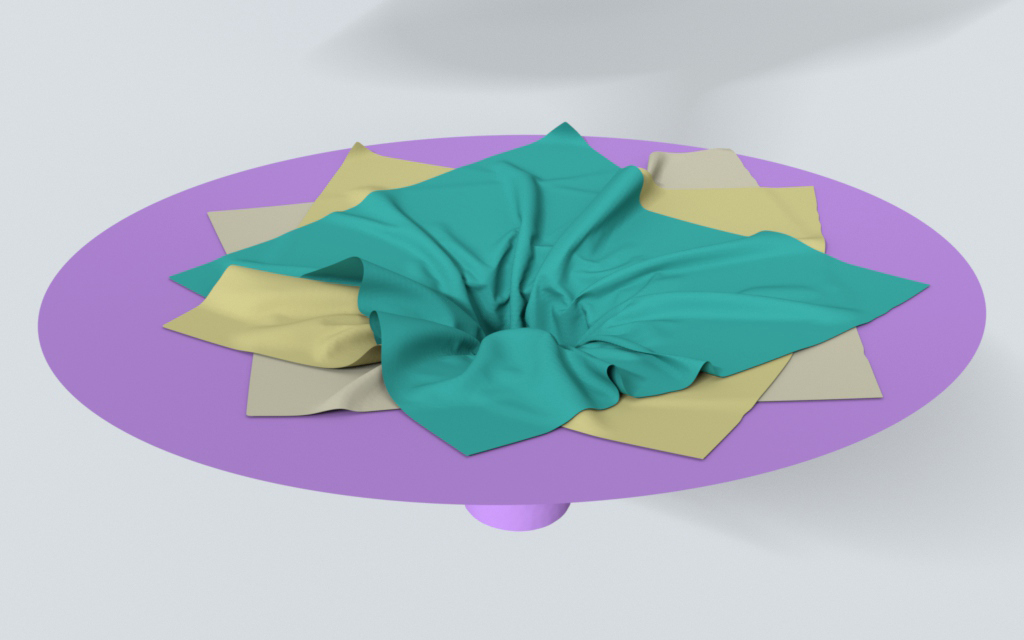}}            
	\vspace*{-0.12in}
	\caption{\textbf{Funnel.} We simulate this example with
          different time step sizes. While our method can reliably
          process collisions at a very large time step, i.e., $\Delta
          t=1/20$s, we suggest using a smaller time step in practice for
          more accurate results with less artificial damping.}
	\label{fig:funnel}
\end{figure}
\subsection{Sensitivity to Time Steps}
While our method is safe and robust regardless of time step size, its
performance per step drops as the time step increases.  \textcolor{black}{Fundamentally,
this is due to more severe collision cases and more steps needed for the method to converge.}
In our experiment, we test the
funnel example with four time steps: $\Delta t$=1/160s, $\Delta
t$=1/80s, $\Delta t$=1/40s and $\Delta t$=1/20s, and we intentionally
set $L$=65,536 \textcolor{black}{so that we can know how many steps are needed for convergence.} Fig.~\ref{fig:funnel}
compares our simulation results in this example and
Table~\ref{tab:examples} provides their performances. From the
performance perspective only, it is beneficial to use a larger time
step to reduce the computational overhead associated with every time
step. However, due to the existence of artificial damping,
we should avoid very large time steps in actual applications.

\subsection{Comparison to Existing Collision Handling Methods}
\label{sec:impact}
\begin{figure}[t]
	\centering
	\subfigure[$\x^{[k]}$ and $\y^{[k+1]}$ when $\theta=0^{\circ}$]{\includegraphics[width=0.78in]{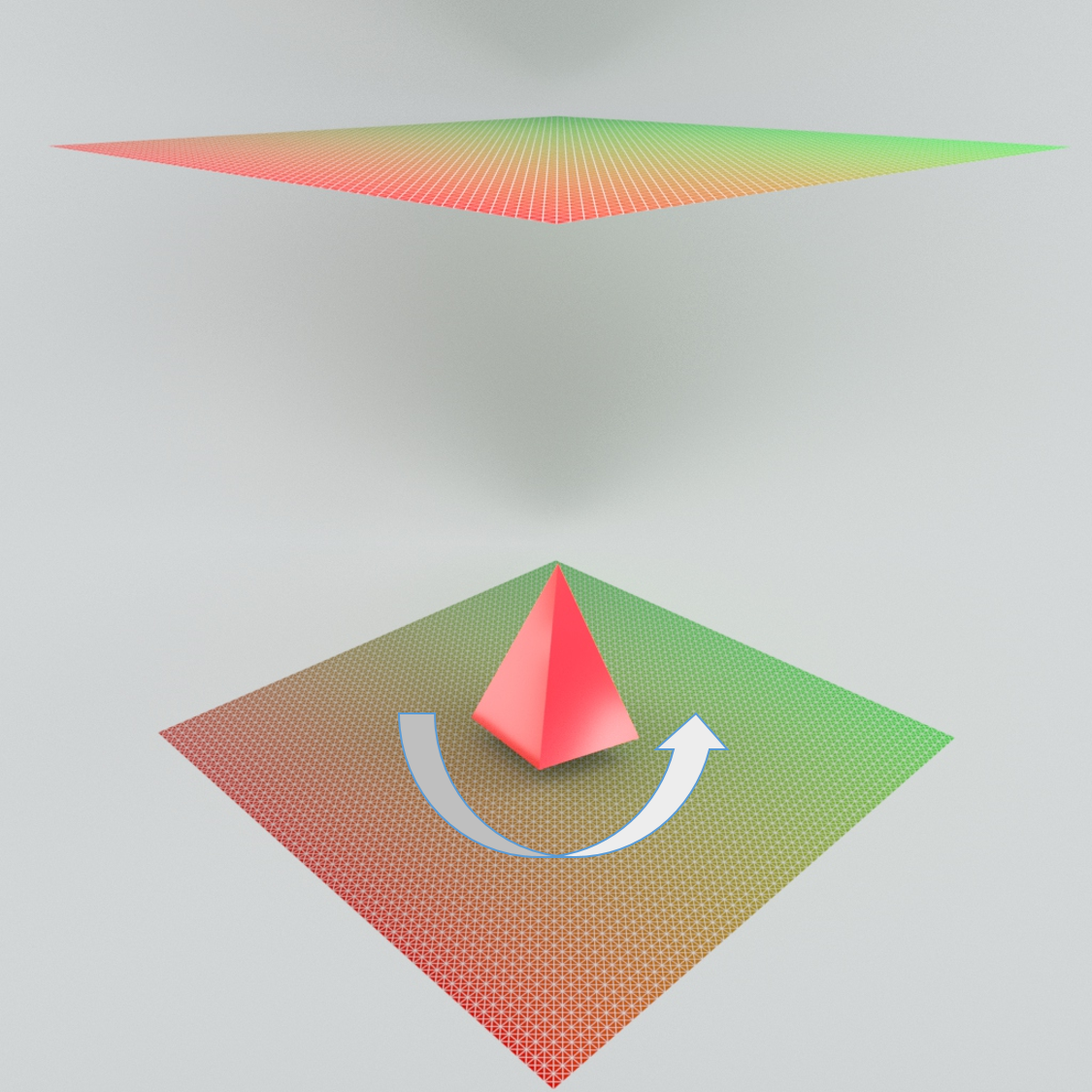} \label{fig:setting_a}} 
	\subfigure[$\x^{[k]}$ and $\y^{[k+1]}$ when $\theta=45^{\circ}$]{\includegraphics[width=0.78in]{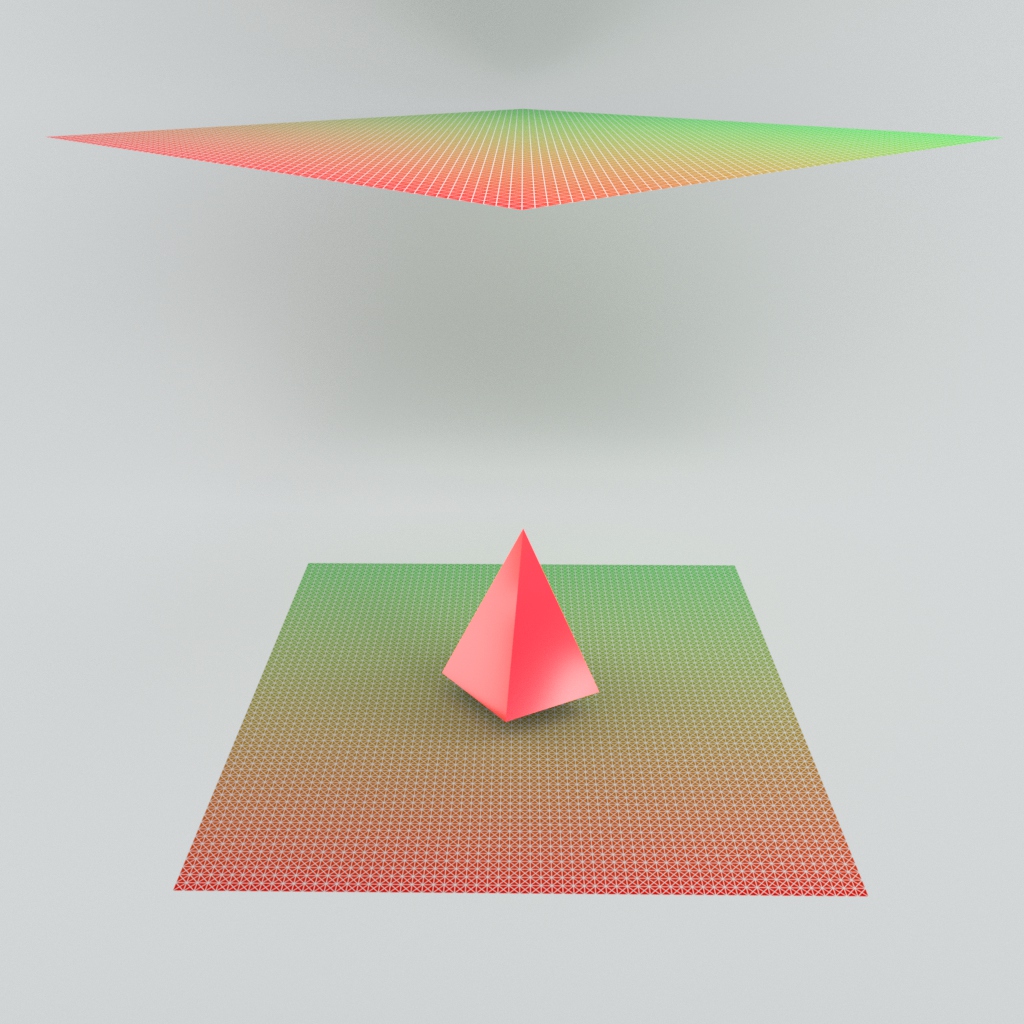} \label{setting_b}} 
	\subfigure[$\x^{[k]}$ and $\y^{[k+1]}$ when $\theta=90^{\circ}$]{\includegraphics[width=0.78in]{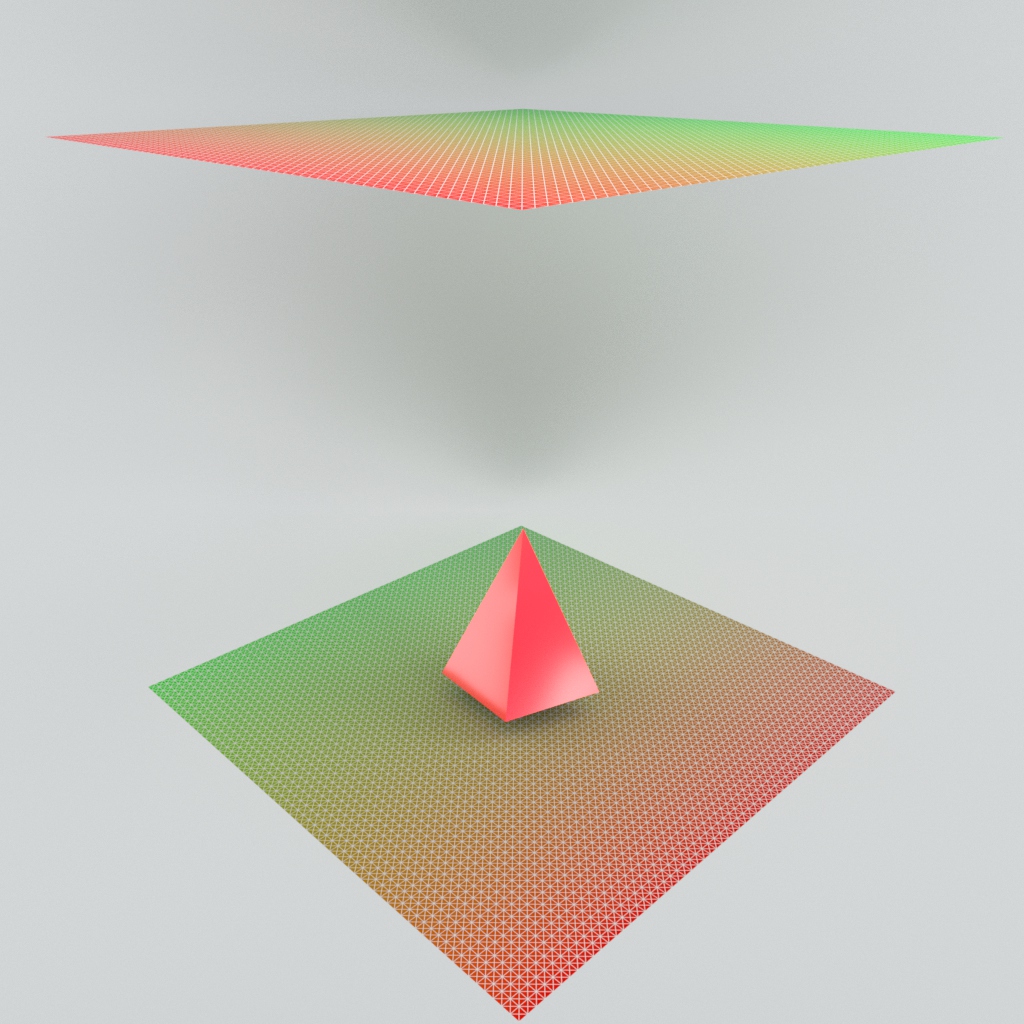} \label{setting_c}} 
	\subfigure[$\x^{[k]}$ and $\y^{[k+1]}$ when $\theta=135^{\circ}$]{\includegraphics[width=0.78in]{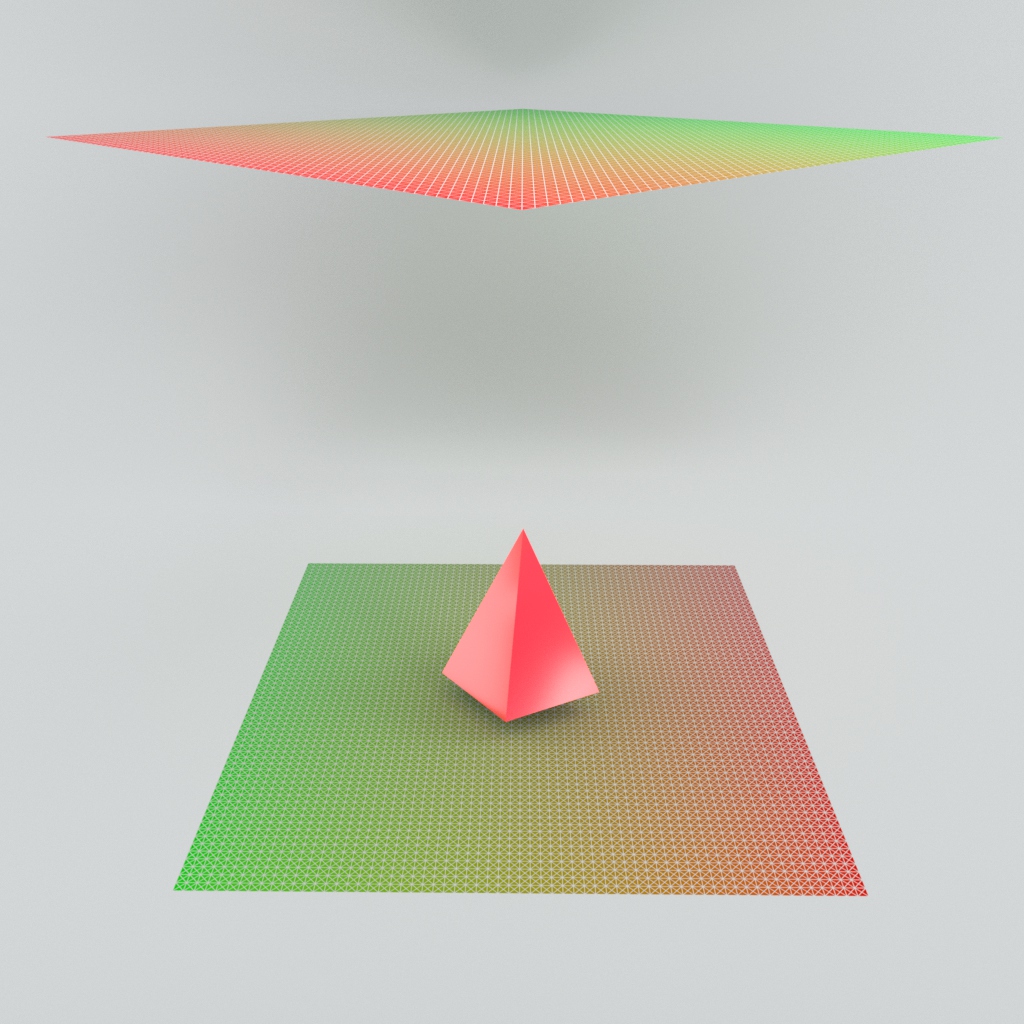} \label{setting_d}} 
	\subfigure[Time cost comparison with ARCSim~\cite{narain2012adaptive}]{\includegraphics[width=3.3in]{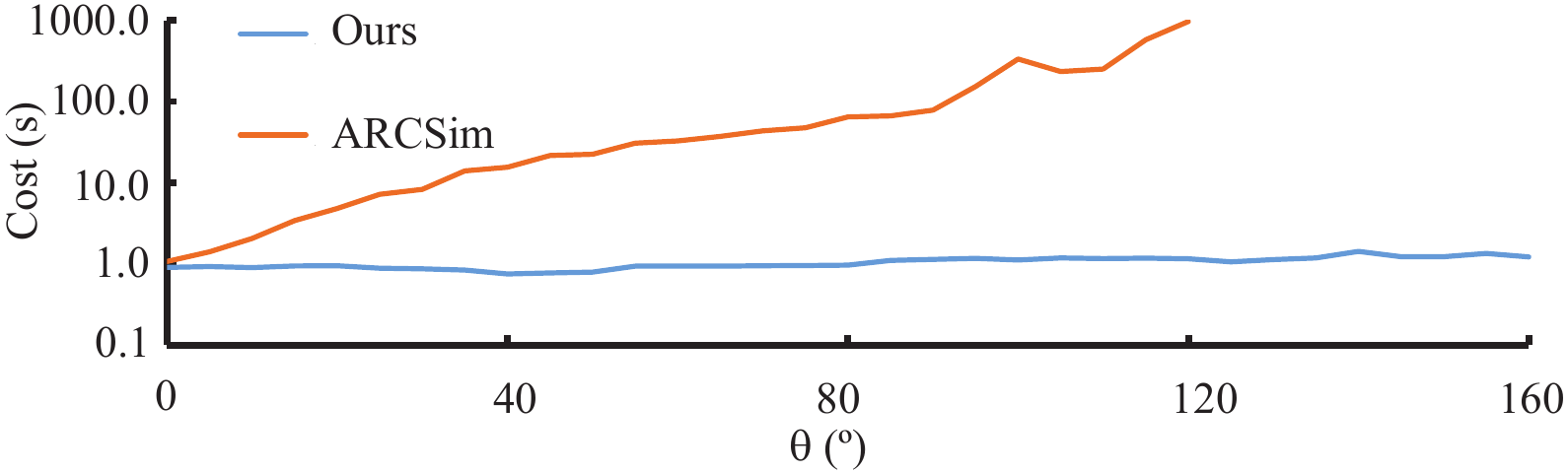} \label{fig:timing_cmp_arcsim}}
	\vspace{-0.12in}
	\caption{\textbf{Spinning cloth patch.}
          \textcolor{black}{From (a) to (d), the spin angle $\theta$
            between $\x^{[k]}$ and $\y^{[k+1]}$ is increasing, and we
            use the color ramp to illustrate region correspondence. As
            $\theta$ increases, ARCSim is getting harder to find a
            valid path, while our method is almost free of such
            drastic growth in time cost (Fig.~(e)).}}
	\label{fig:cmp_arcsim_setting}
\end{figure}

As we argued above, the major problems of existing collision handling
algorithms~\cite{harmon2008robust,narain2012adaptive,tang2016cama,tang2018clotha,Tang:2018:PPS,li2020p}
for the step-and-project method are their linear path assumption and
inappropriate measurement of distance. Such issues are exposed in a
simple example as shown in Fig.~\ref{fig:cmp_arcsim_setting}:
\textcolor{black}{we initialize the angular velocity of various
  magnitudes in plane, leading to different targets $\y^{[k+1]}$ with
  increasing spin angles after the dynamics step.}
As the spin angle increases, the actual trajectory of every vertex
should be close to a helix passing through the {\color{black} five-faces-spike} obstacle surface under
gravity. Therefore, restricting to a linear path \textcolor{black}{can
  make the algorithm difficult to find a valid path not colliding with
  the obstacle, which might become impossible for complex
  shapes.} Fig.~\ref{fig:timing_cmp_arcsim} demonstrates such an
increasing difficulty: as $\theta$ increases, the inelastic impact
zone method~\cite{harmon2008robust} implemented in
ARCSim~\cite{narain2012adaptive} requires drastically increasing cost
to find a valid path. We note that when $\theta$ reaches
$125^{\circ}$, the program cannot converge in two hours no matter how
we tune the parameters.

\begin{figure}[t]
	\centering
	\subfigure[$\x^{[k]}$ and $\x^{[k+1]}$ by our method when \quad \quad $\theta=45^{\circ}$]{\includegraphics[width=1.6in]{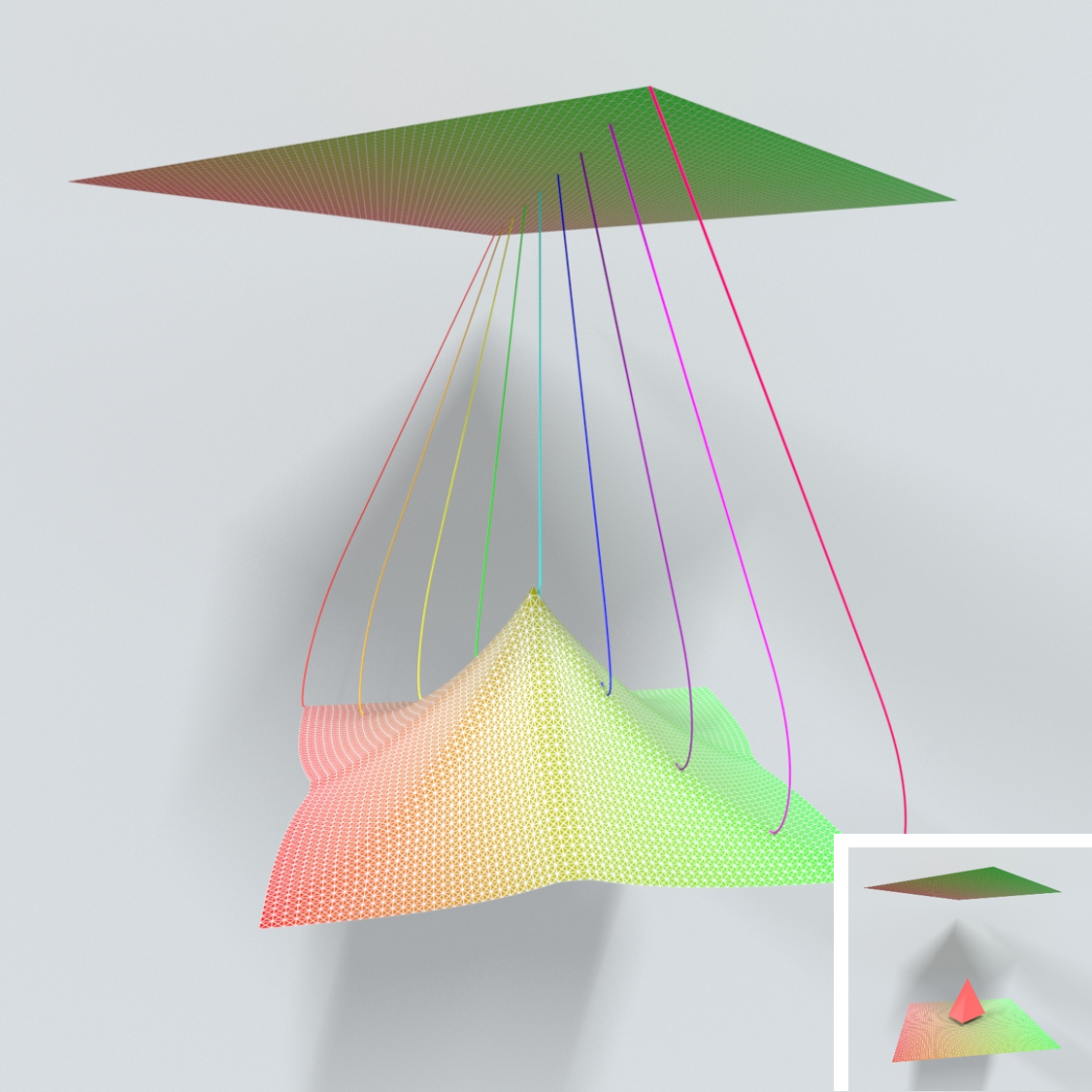}} 
	\subfigure[$\x^{[k]}$ and $\x^{[k+1]}$ by ARCSim when \quad \quad $\theta=45^{\circ}$]{\includegraphics[width=1.6in]{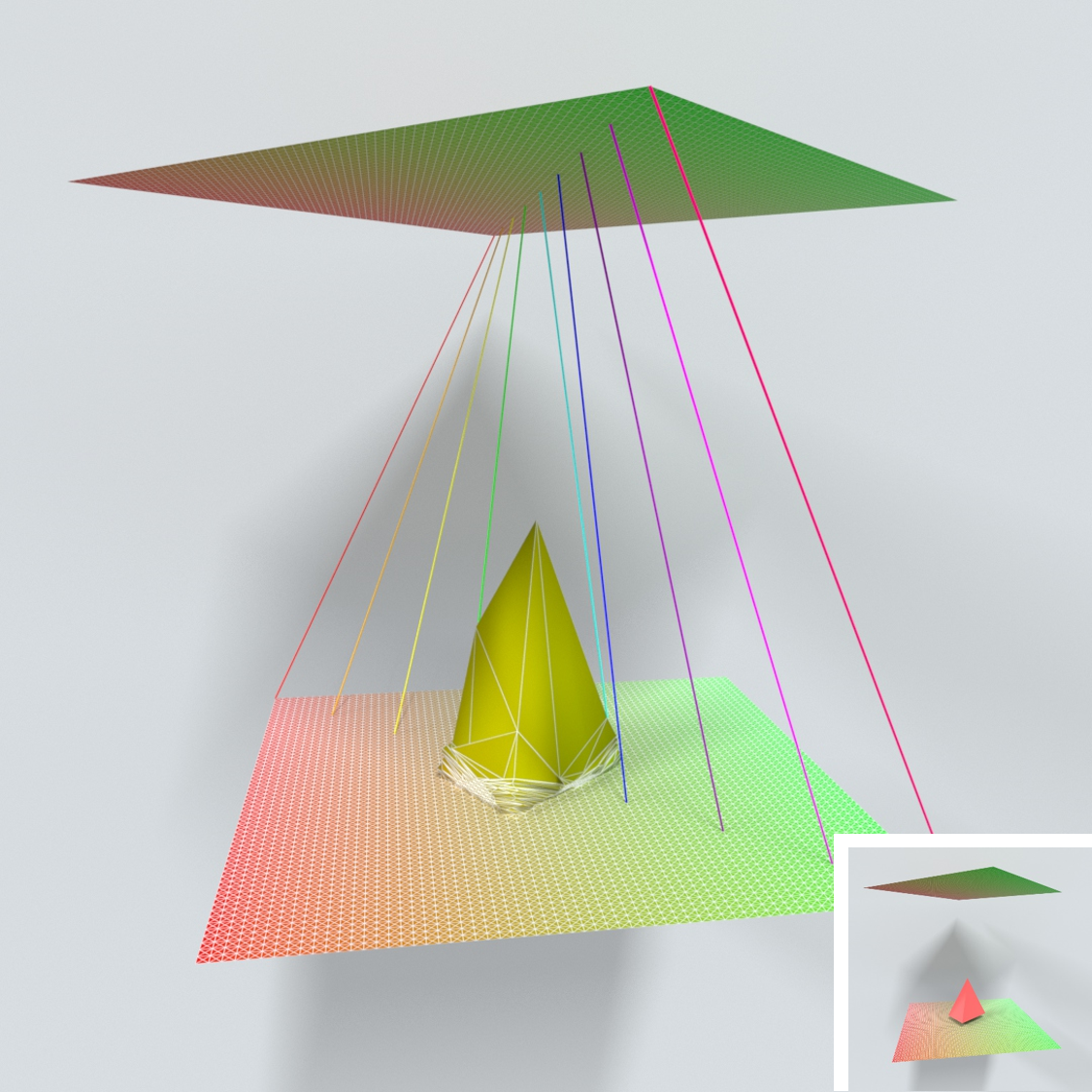} \label{fig:arcsim_a}} 
	\subfigure[$\x^{[k]}$ and $\x^{[k+1]}$ by our method when \quad \quad  $\theta=90^{\circ}$]{\includegraphics[width=1.6in]{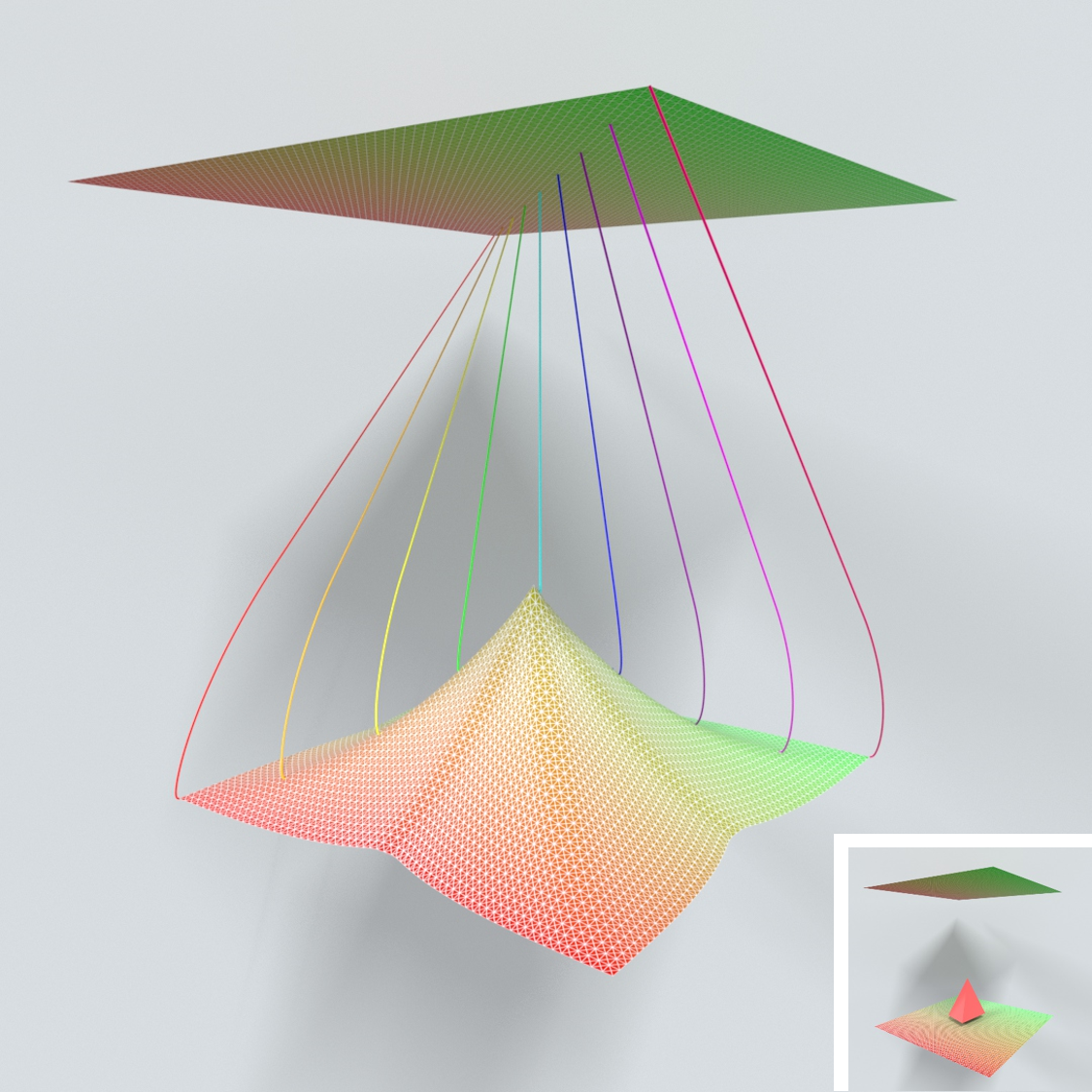}} 
	\subfigure[$\x^{[k]}$ and $\x^{[k+1]}$ by ARCSim when \quad \quad $\theta=90^{\circ}$]{\includegraphics[width=1.6in]{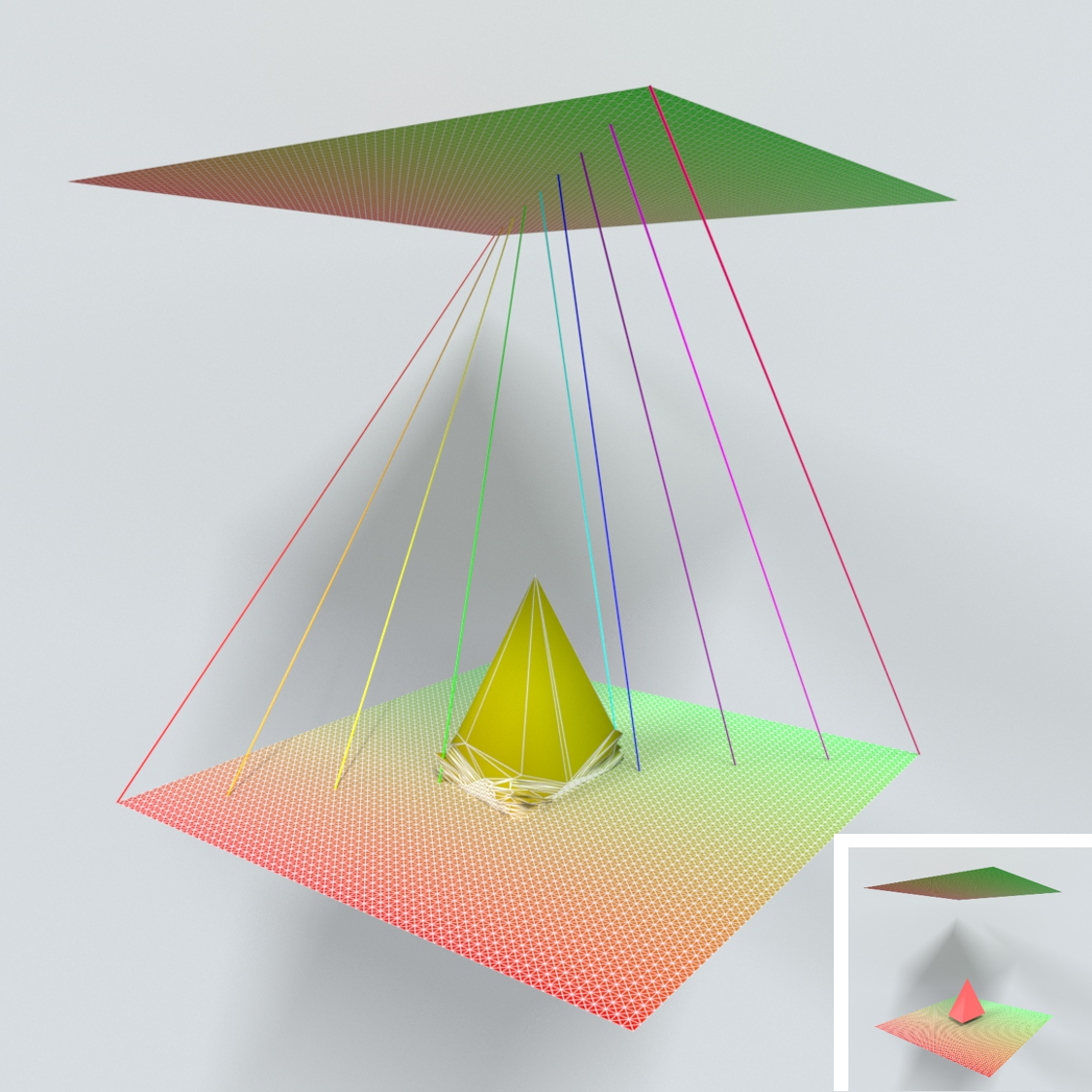} \label{fig:arcsim_b}} 
		
	\vspace{-0.12in}
	\caption{\textbf{Trajectories avoiding collisions.} Here we
          track the trajectories of nine vertices on one diagonal of
          the cloth patch, which are progressively generated by our
          two-way approach. Compared to the impact zone
          optimization implemented by ARCSim~\cite{narain2012adaptive}, our method allows
          a curved path from $\x^{[k]}$ to $\x^{[k+1]}$ to resolve
          collisions where neither intersections nor spuriously
          stretched elements can appear, while restricting to a linear
          path may fail to find a faithful solution at a reasonable
          cost.}
	\label{fig:cmp_arcsim}
	\vspace{-0.12in}
\end{figure}

Instead, our method runs inexpensive forward steps to
\textcolor{black}{form a piecewise linear path safely.  In fact, this
  not only avoids expensive CCD tests, but also greatly increases the
  chances to find a valid path by exploring a larger space compared to
  a single line segment}. Furthermore, with the help of the additional
edge length constraints, the over-stretched artifacts arising from
long-distance projection are pleasingly
removed. \textcolor{black}{Consequently}, our two-way method can
robustly remove all intersections along the path and
reach a plausible and collision-free result (see Fig.~\ref{fig:cmp_arcsim}
and the supplemental video) at a low cost
(Fig.~\ref{fig:timing_cmp_arcsim}).

\begin{figure}[t]
	\centering
	\subfigure[CAMA~\cite{tang2016cama}]{\includegraphics[width=1.1in]{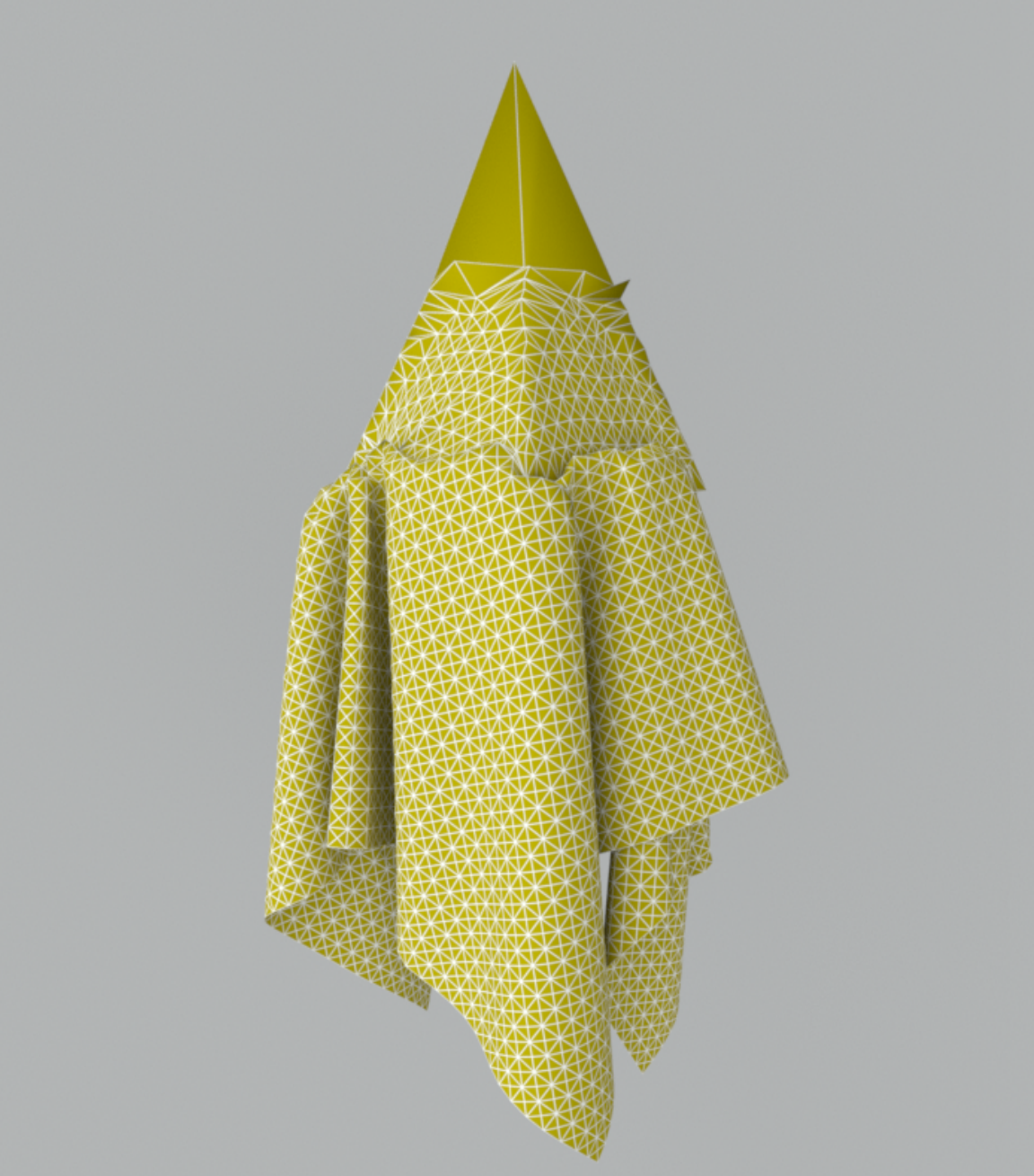}}	
	\subfigure[ARCSim~\cite{narain2012adaptive}]{\includegraphics[width=1.1in]{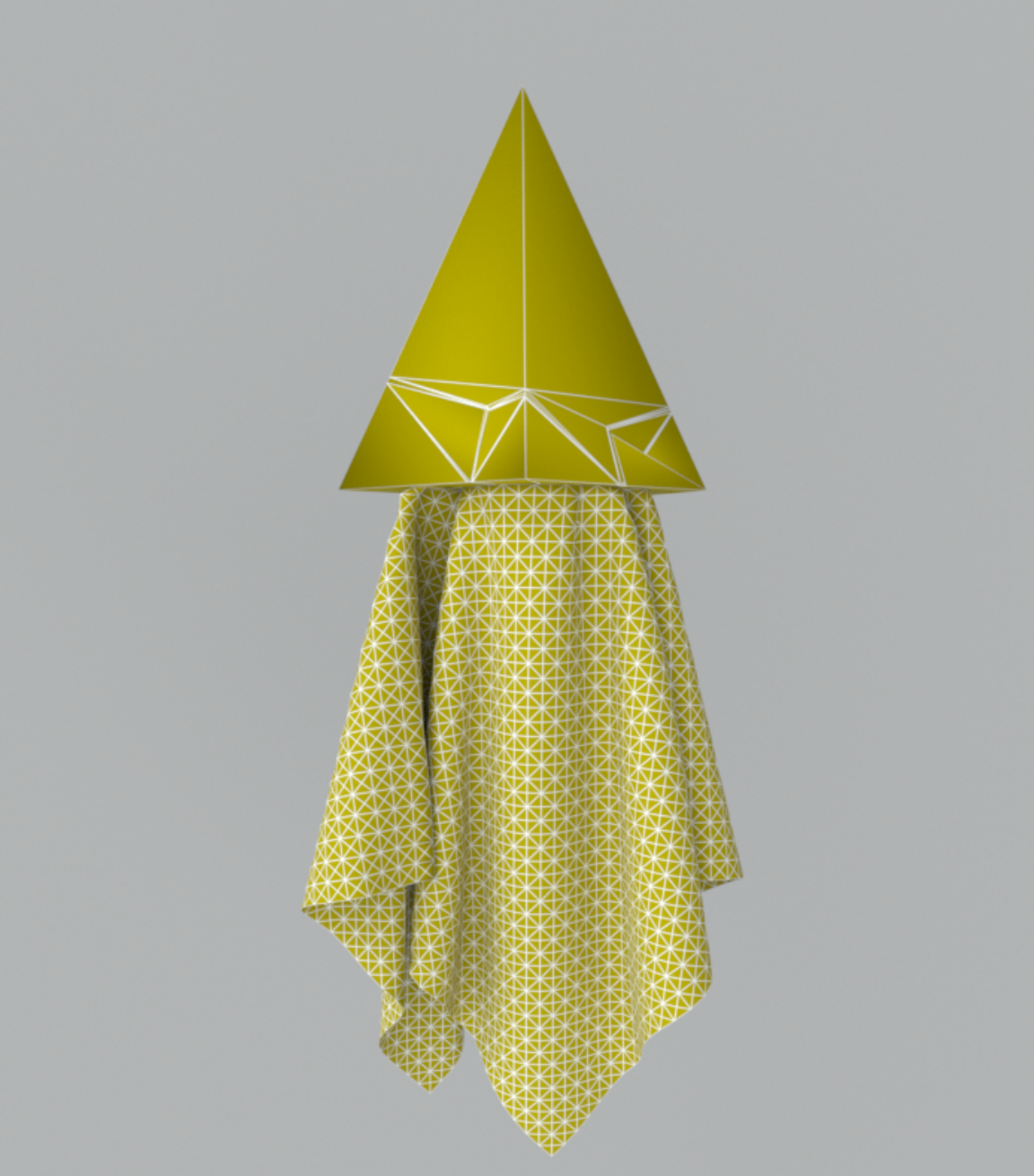}}
	\subfigure[Our method]{\includegraphics[width=1.1in]{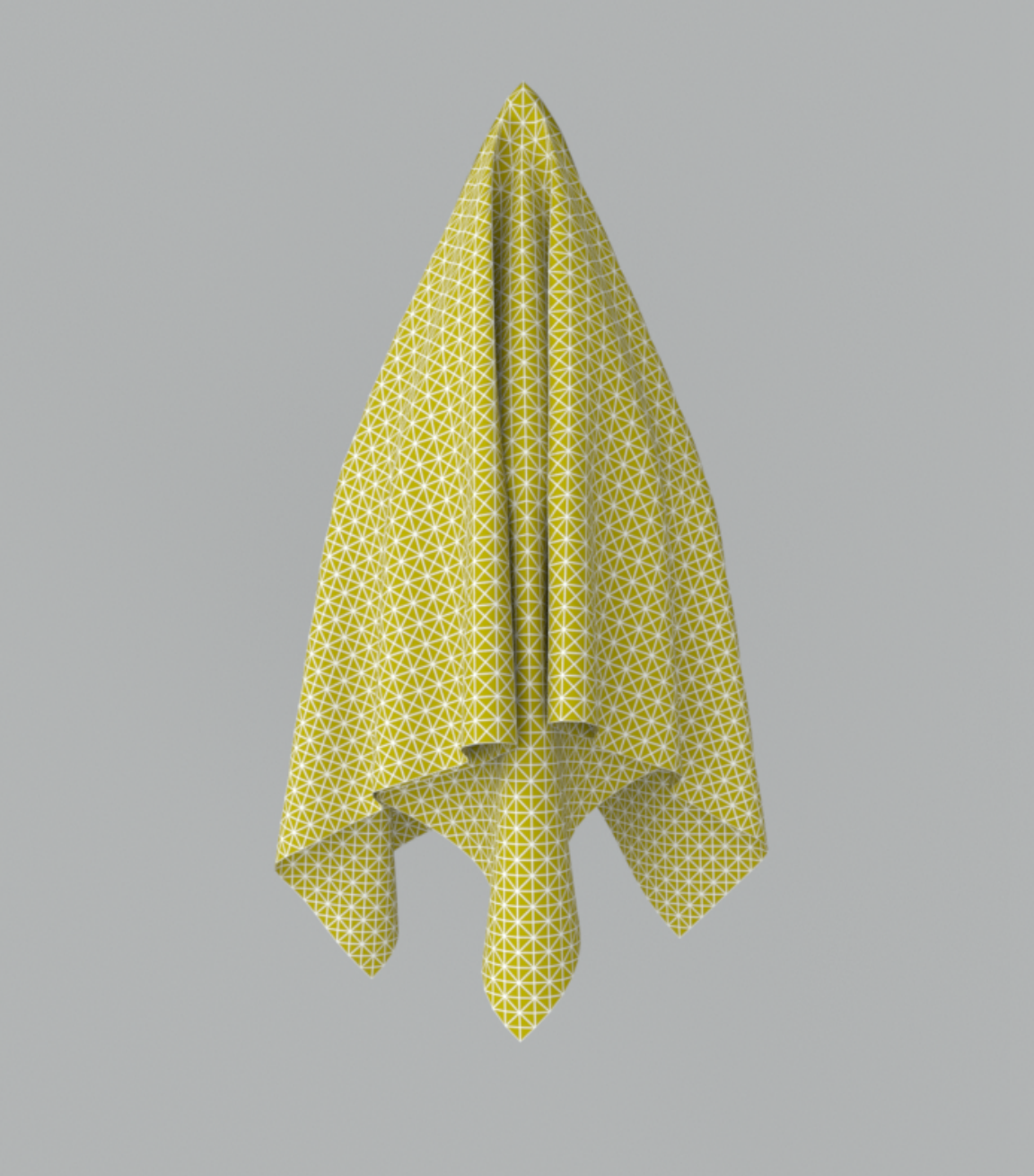}}
        \subfigure[Time cost comparison]{\includegraphics[width=3.3in]{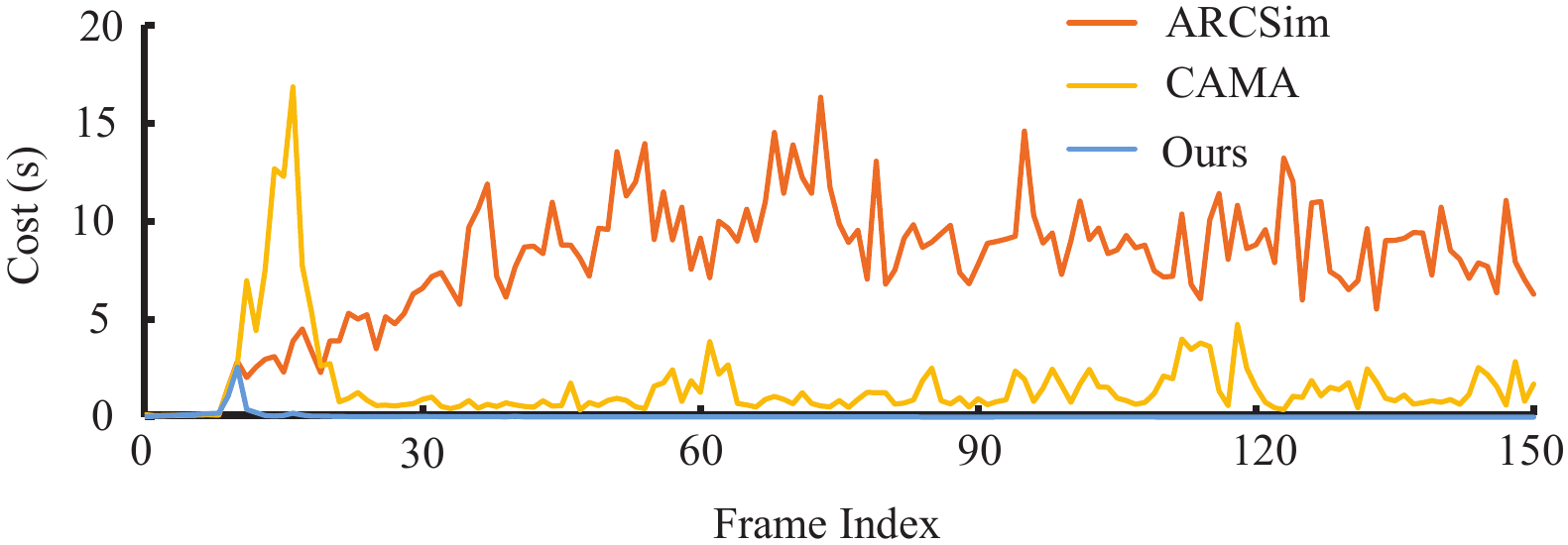}}	
	\vspace{-0.12in}
	\caption{\textbf{Robustness to large time steps.} A piece of
          cloth falling on a sharp cone is simulated with a large
          time step ($\Delta t=1/10$s). Compared with existing impact
          zone methods such as~\cite{tang2016cama,narain2012adaptive},
          our result has much less visual artifacts over time, {\color{black} even with much less computational cost.}
	}
	\label{fig:cmp_with_impact_zone}
\end{figure}

\textcolor{black} {Due to the high
  quality of the outputs in our two-way collision handling module,}
the entire simulation also suffers from much fewer nonphysical
artifacts than the existing methods as
Fig.~\ref{fig:cmp_with_impact_zone} and the supplemental video
demonstrate. {\color{black}Specifically, ARCSim~\cite{narain2012adaptive} and CAMA~\cite{tang2016cama}, these two implementations of non-rigid impact zone optimization~\cite{harmon2008robust} are involved for evaluations, where obvious nonphysical artifacts are observed. Note that other implementations, such as~\cite{harmon2008robust,tang2018clotha,li2020p,Tang:2018:PPS}, should suffer from similar artifacts due to the inheritance of methodology as we introduced above, though with different performances. In detail, to control the comparison targets, starting with the same $\x^{0}$, we run one Newton iteration by the same implementation and parameter setting to get $\y^{[1]}$ from a starting point $\x^{[0]}$ (the same as $\x^{t}$) in the dynamics step, and
then we save the $\x^{[0]}$\textemdash$\y^{[1]}$ pair as OBJ files and ``feed''  this pair to different collision handling methods right afterwards to get different $\x^{[1]}$ (the same as $\x^{t+1}$), and then we save $\x^{[1]}$ as a OBJ file to ``feed'' back to the dynamics step, update velocity as $\mathbf{v}^{t+1}=\frac{\x^{t+1}-\x^{t}}{\Delta t}=\frac{\x^{[1]}-\x^{[0]}}{\Delta t}$ (we only have $k=0$ because we only run one Newton iteration) and start a new time step. This helps us keep the entire  pipeline with all components the same, except for swapping in different projection methods to be compared. Note that the input and output of these methods have the same form as ours, a given $\x^{[k]}$\textemdash$\y^{[k+1]}$ pair and a resulted $\x^{[k+1]}$.}

{\color{black}Theoretically, we note that there exists one class of inequality-constrained nonlinear optimization methods working in a step-and-project manner, the projected gradient (Newton) method \quad~\cite{conn1988global,lin1999newton,jorge2006numerical}
  . Given a gradient direction or Newton direction (a linear path of high dimension), the projection operator progressively projects the state into feasible domain and forms a piecewise linear path of high dimension. The piecewise linear property is important for incorporating appropriate global convergence techniques into it, such as computing the generalized Cauchy point. Compared with existing collision handling methods in the step-and-project pipeline, our method works in a similar way to it and the formed constrained path shares the common pieciewise-linear property with it. Therefore, we leave it to extend our method into an intact projected gradient (Newton) method for collision handling as future work.}

	\begin{figure*}[t]
	\centering
	\subfigure[The initial state]{\includegraphics[height=0.77in]{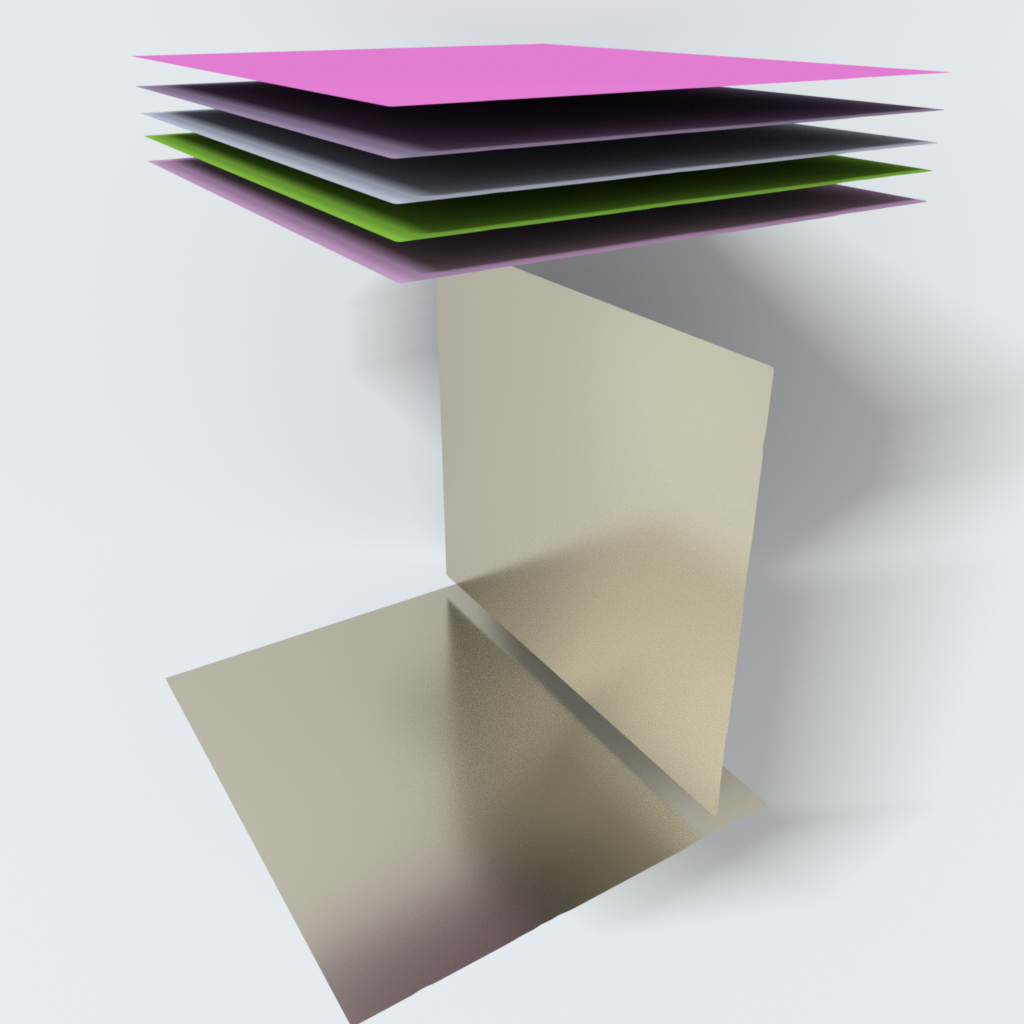}}
	\subfigure[Before repair]{\includegraphics[height=0.77in]{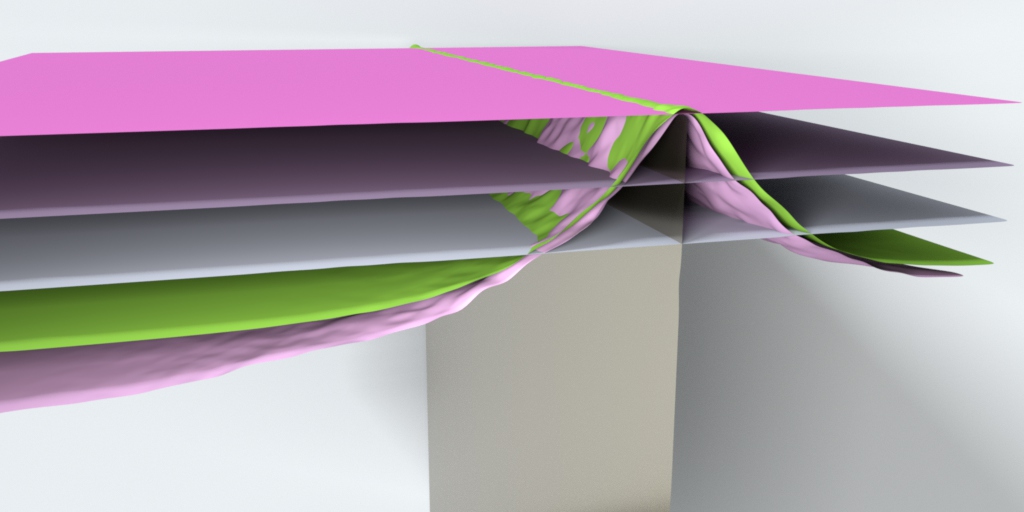}}
	\subfigure[After repair]{\includegraphics[height=0.77in]{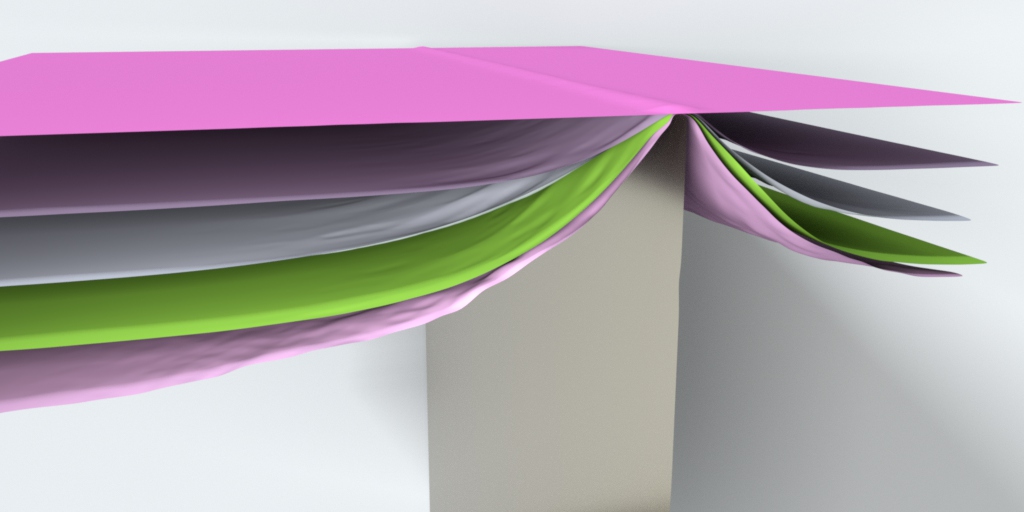}}
	\subfigure[Before repair]{\includegraphics[height=0.77in]{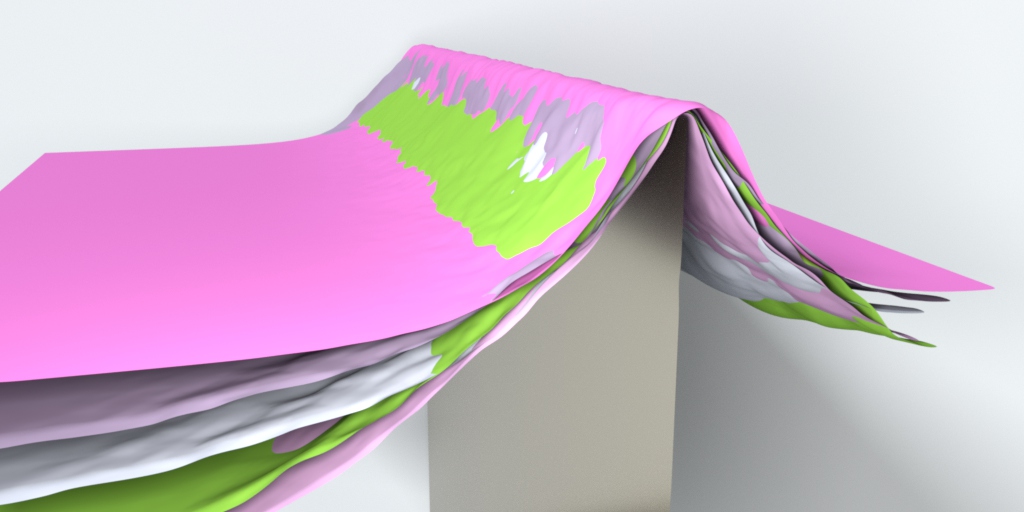}}
	\subfigure[After repair]{\includegraphics[height=0.77in]{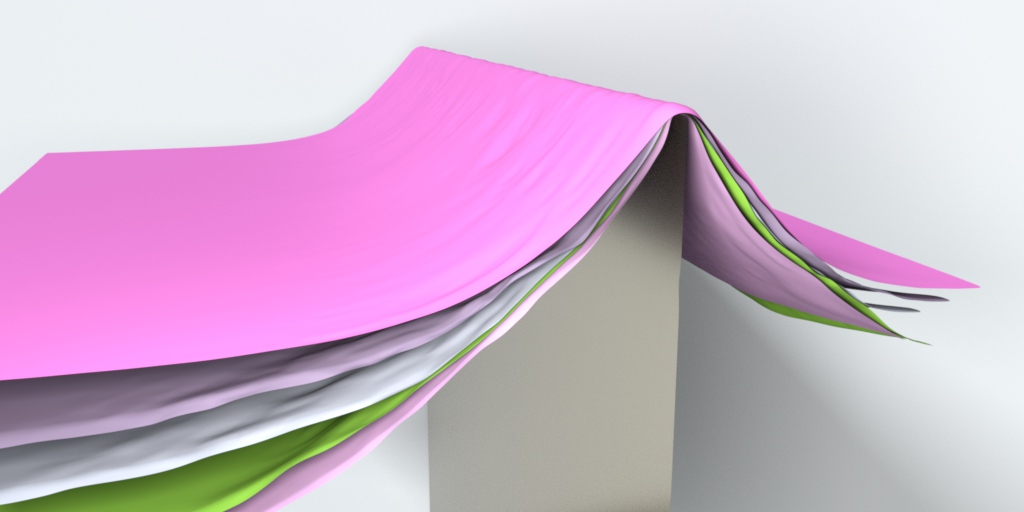}}            
	\vspace{-0.12in}
	\caption{\textbf{Cloth and blade colliding.} In this example, five square cloth patches drop onto a sharp metal blade, then drape, slide and stack under gravity. Our method can robustly project seriously penetrated frames ((b) and (d)) to intersection-free states ((c) and (e)).}
	\label{fig:blade}
\end{figure*}

\begin{figure}[t]
	\centering
	\subfigure[The initial state]{\includegraphics[width=1.1in]{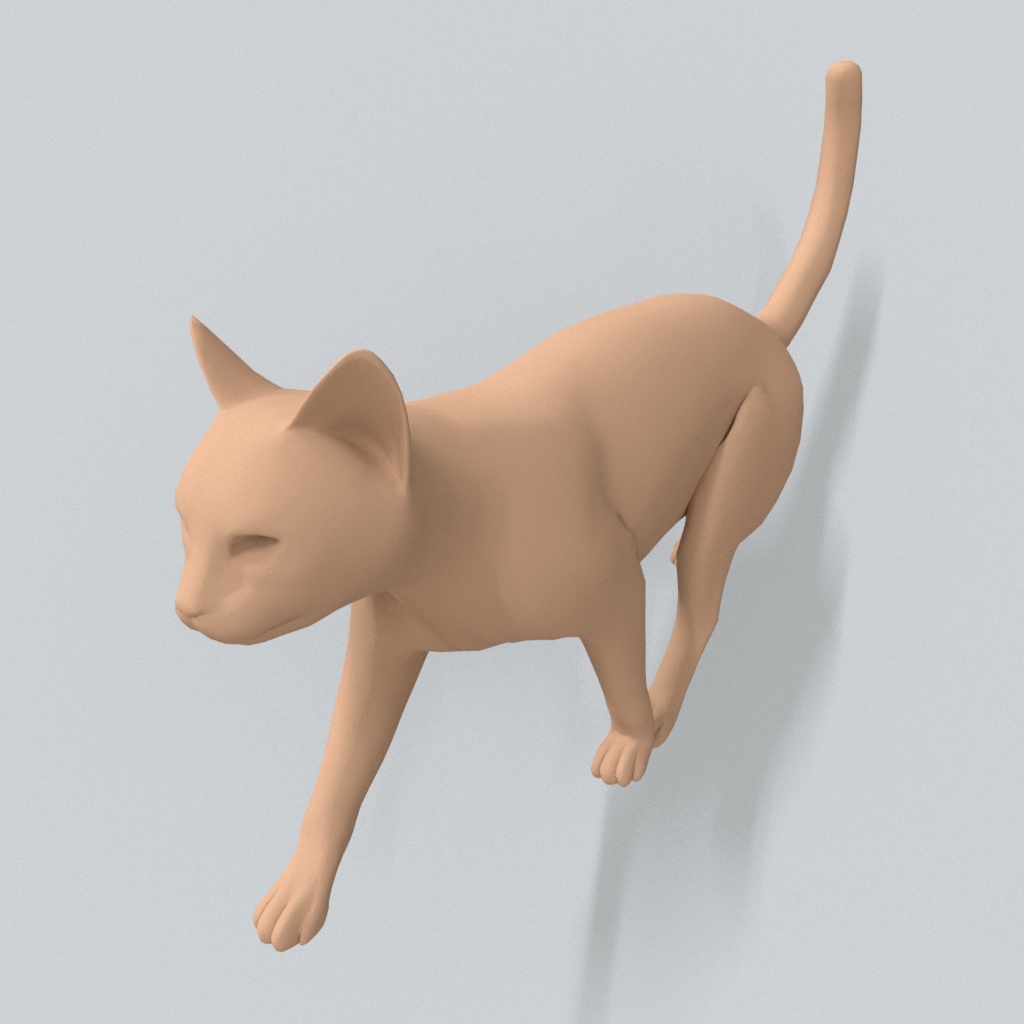}}	
	\subfigure[The positive flow]{\includegraphics[width=1.1in]{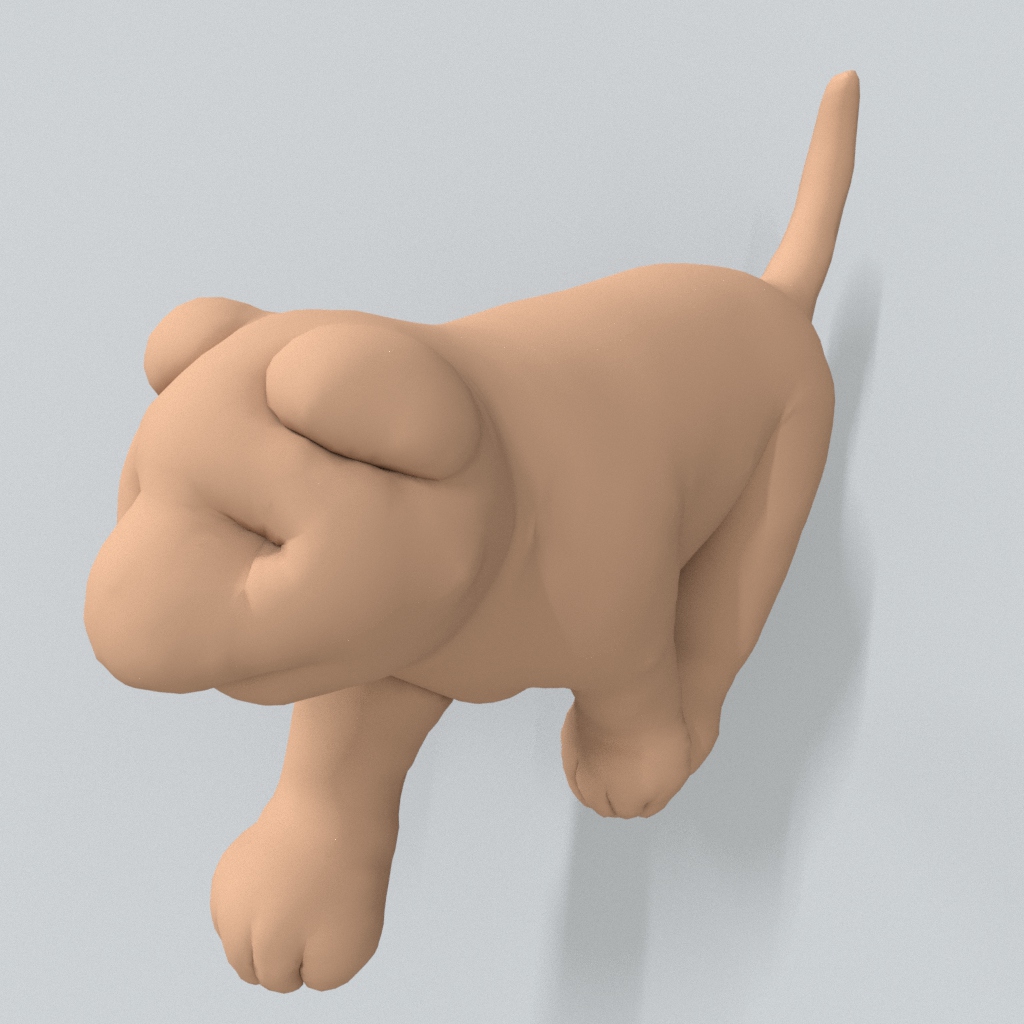}}
	\subfigure[The negative flow]{\includegraphics[width=1.1in]{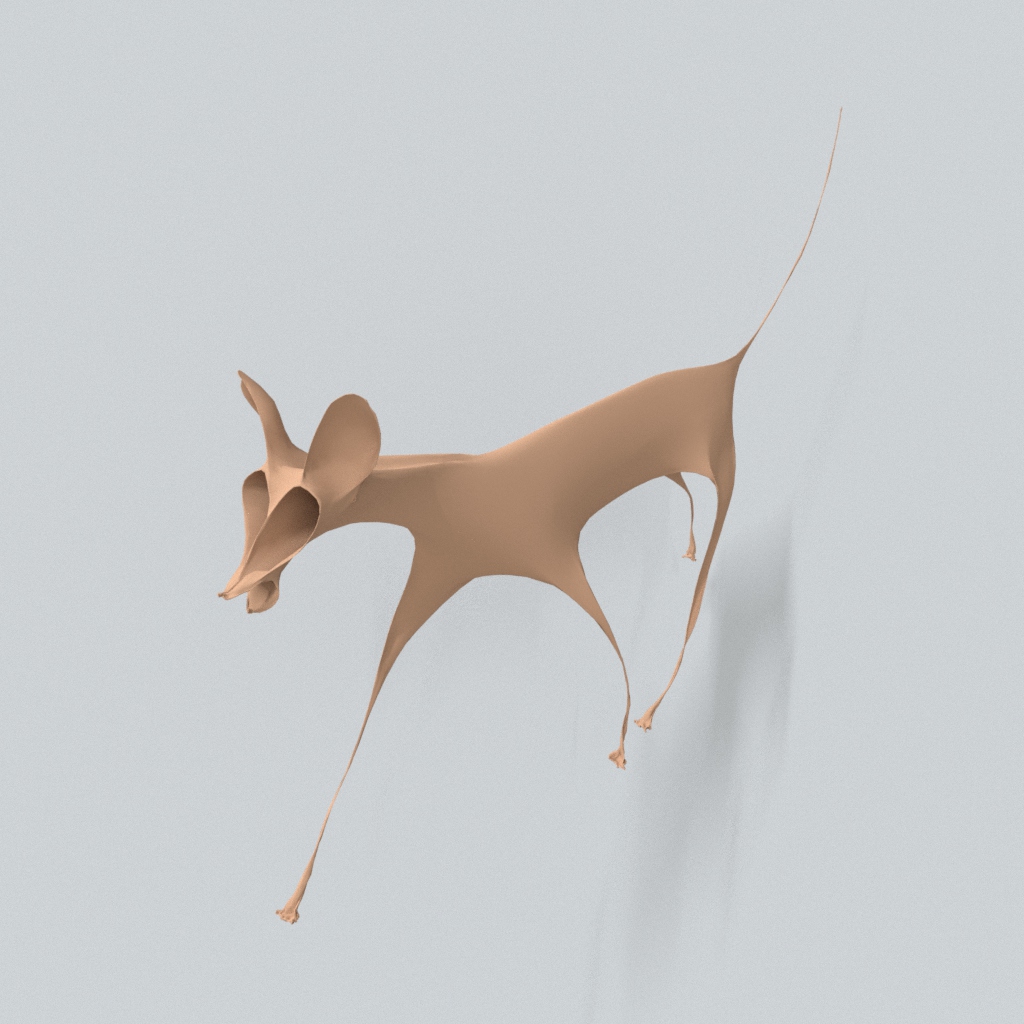}}
	
	\subfigure[The initial state]{\includegraphics[width=1.1in]{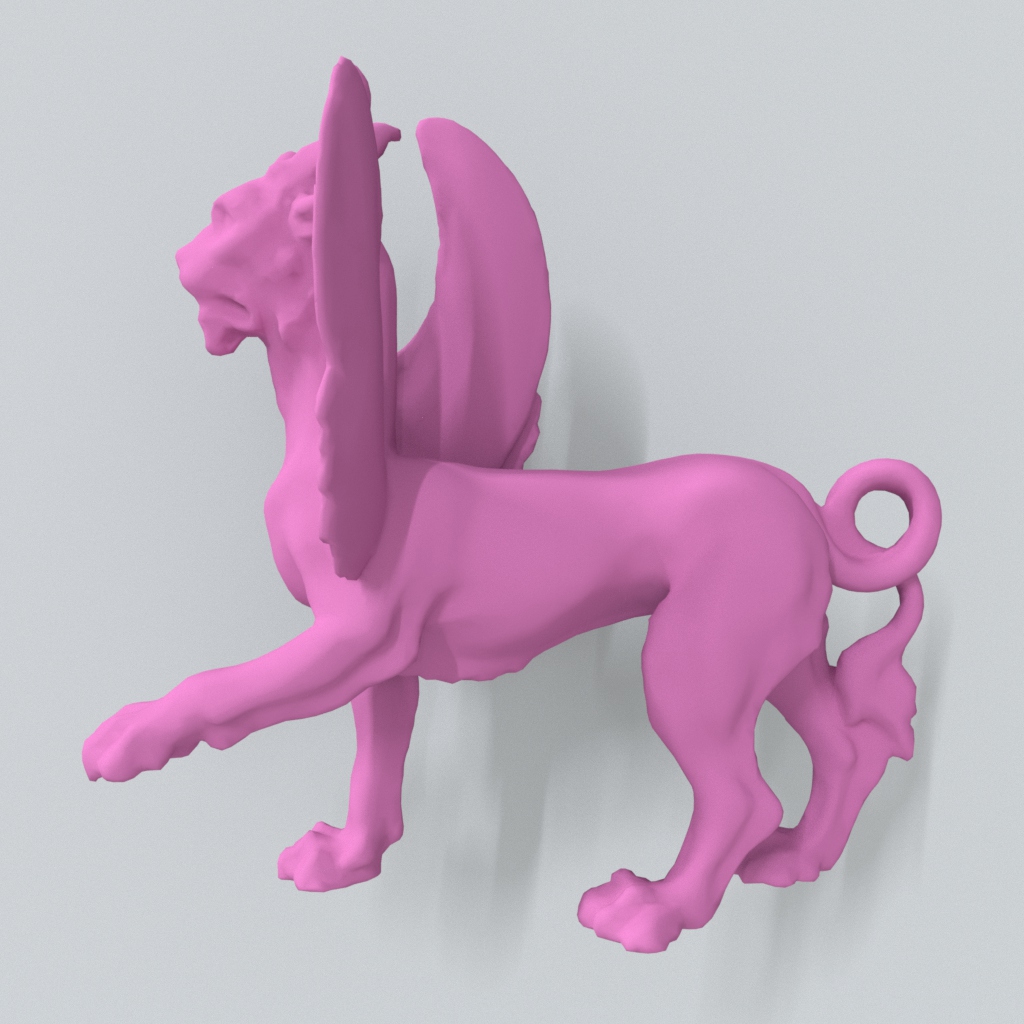}}	
	\subfigure[The positive flow]{\includegraphics[width=1.1in]{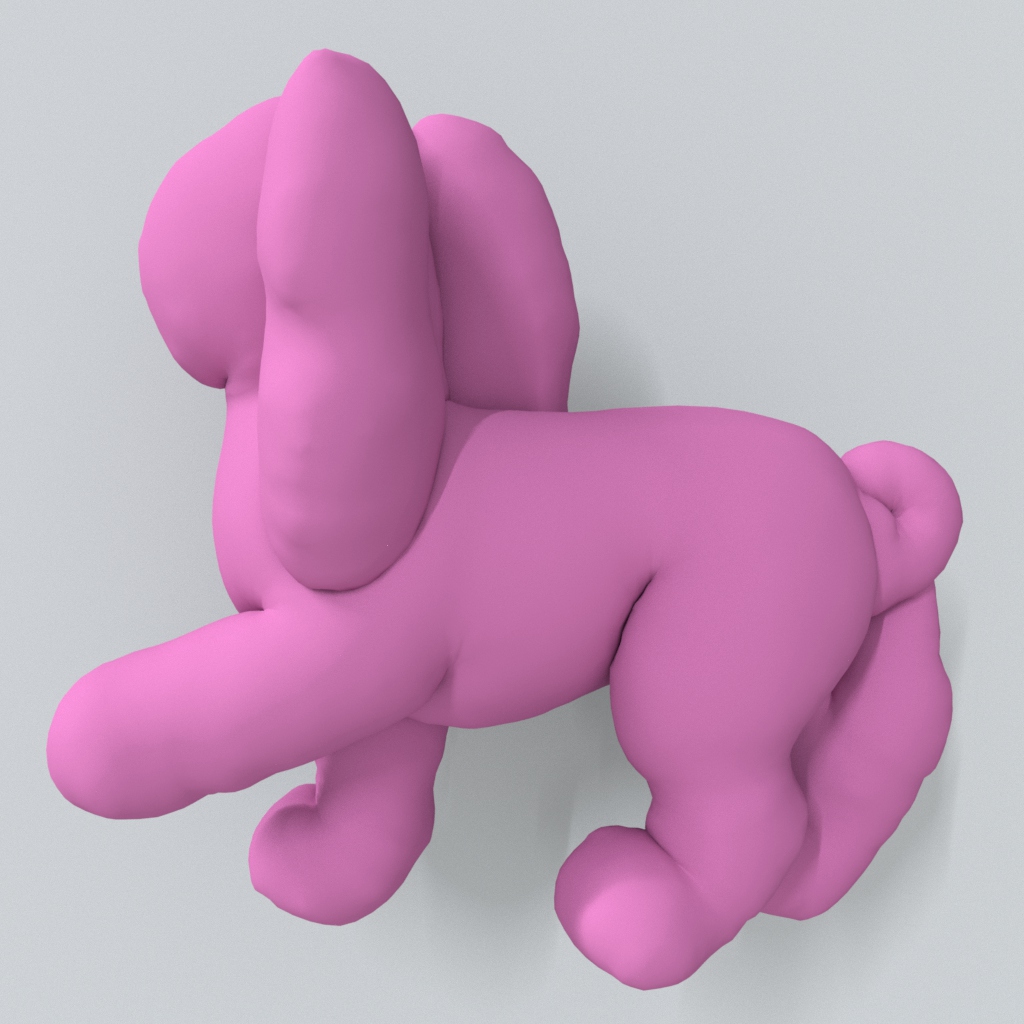}}
	\subfigure[The negative flow]{\includegraphics[width=1.1in]{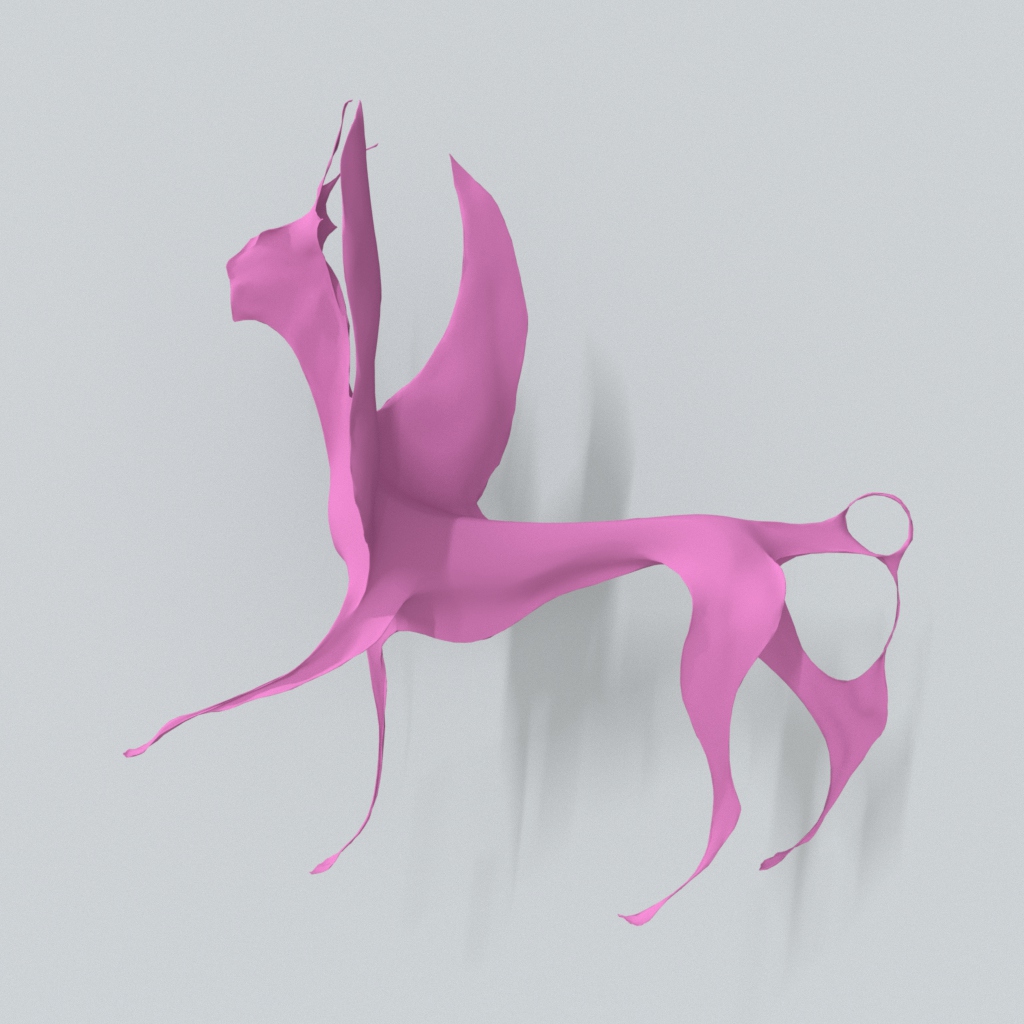}}
	
	\subfigure[The initial state]{\includegraphics[width=1.1in]{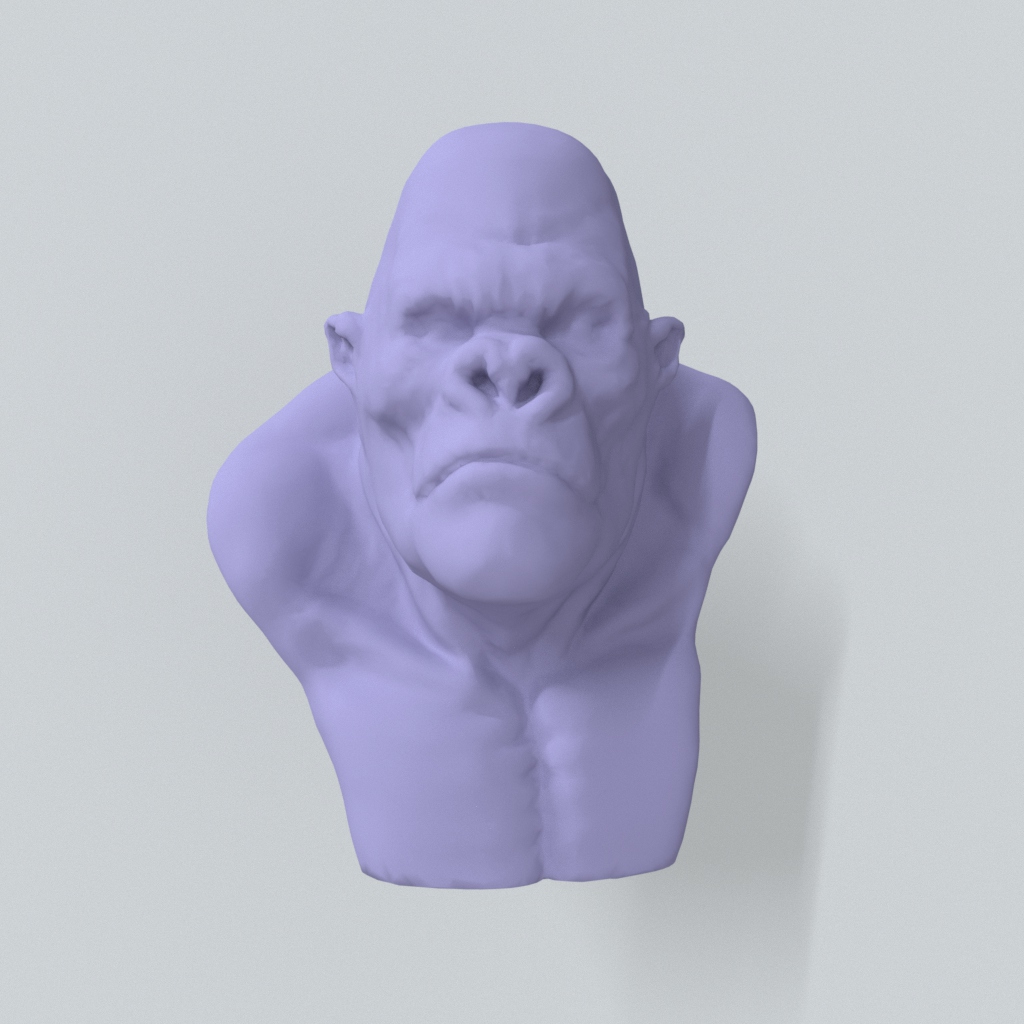}}	
	\subfigure[The positive flow]{\includegraphics[width=1.1in]{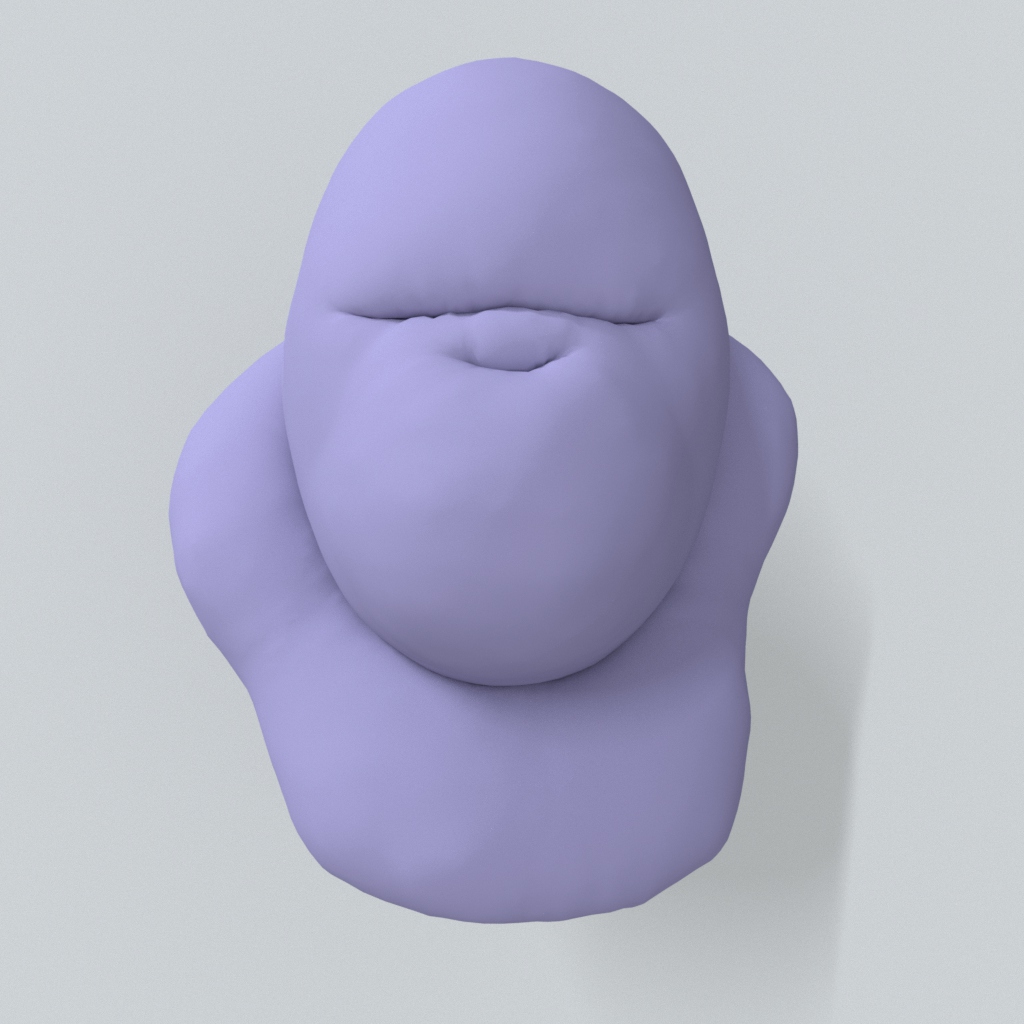}}
	\subfigure[The negative flow]{\includegraphics[width=1.1in]{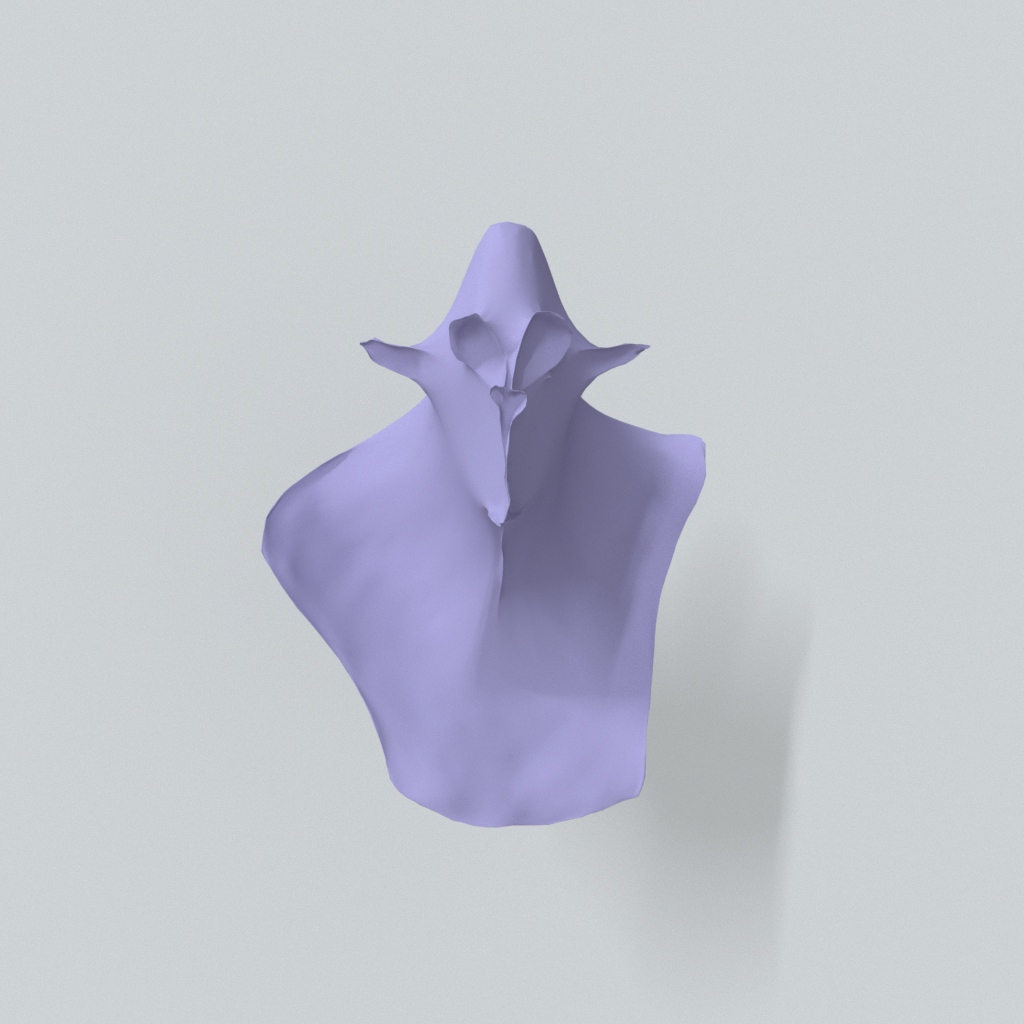}}
	\vspace{-0.12in}
	\caption{\textbf{Globally injective normal flow.} Our method can be also applied to enforce global injectivity in both positive and negative normal flows.}
	\label{fig:normal_flow}
	\vspace{-0.12in}
\end{figure}

\subsection{Geometry Processing}
\label{sec:geometry}
Since our method works as a standalone module to resolve collisions, it can be seamlessly integrated into geometry processing applications for intersection-free results. For instance, \textcolor{black}{ our method can fix seriously self-intersecting \textcolor{black}{animation} frames robustly, such as the blade example shown in Fig.~\ref{fig:blade}.} We can also apply our method to enforce global injectivity in normal flow computations, such as the ones shown in Fig.~\ref{fig:normal_flow}.
        
For the globally injective normal flow application, we first compute ${\mathbf y}^{[k+1]}$ as below:
\begin{equation}
        \left\{
	\begin{array}{l}
		{\mathbf y}_{0}^{[k+1]}={\mathbf x}^{[k]} \pm \beta {\mathbf n}^{[k]}, \\
		{\mathbf y}_{i+1}^{[k+1]}(v)={\mathbf y}_{i}^{[k+1]}(v) + \alpha \Laplace {\mathbf y}_{i}^{[k+1]}(v), \\
		{\mathbf y}^{[k+1]}={\mathbf y}_{3}^{[k+1]},
	\end{array}
        \right.
	\label{eq:normal flow}
\end{equation}
in which $\mathbf{n}^{[k]}$ is the normal vector field of the surface at $\x^{[k]}$, $\beta$ is the flow speed, $\alpha$ is the smoothing intensity and $\Laplace$ is the cotangent-formed \cite{pinkall1993computing,desbrun1999implicit,meyer2003discrete} discretization of the Laplace-Beltrami operator.  We smooth in a simple Jacobi fashion for each vertex $v$ and fix the smoothing iterations be three to get $\y^{[k+1]}$. 
Then we run our two-way method for guaranteeing global injectivity. Finally, we get highly similar results compared with~\cite{fang2021guaranteed}, on all examples from their benchmarks as shown in Fig.~\ref{fig:normal_flow} within less than $1$s.

\begin{figure*}[t]
	\centering	
        \begin{tabular}{c@{ }c@{ }c@{ }c}
          \includegraphics[width=1.72in]{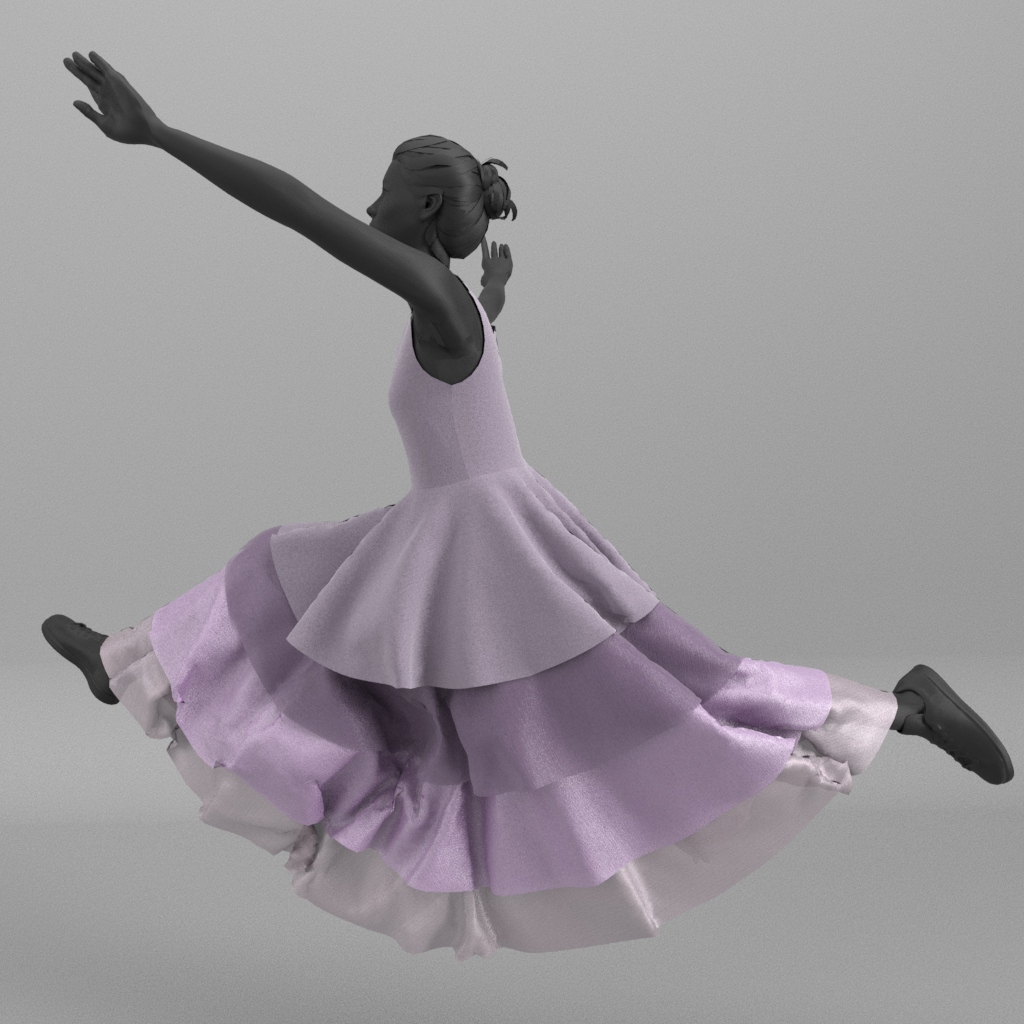}
          &
          \includegraphics[width=1.72in]{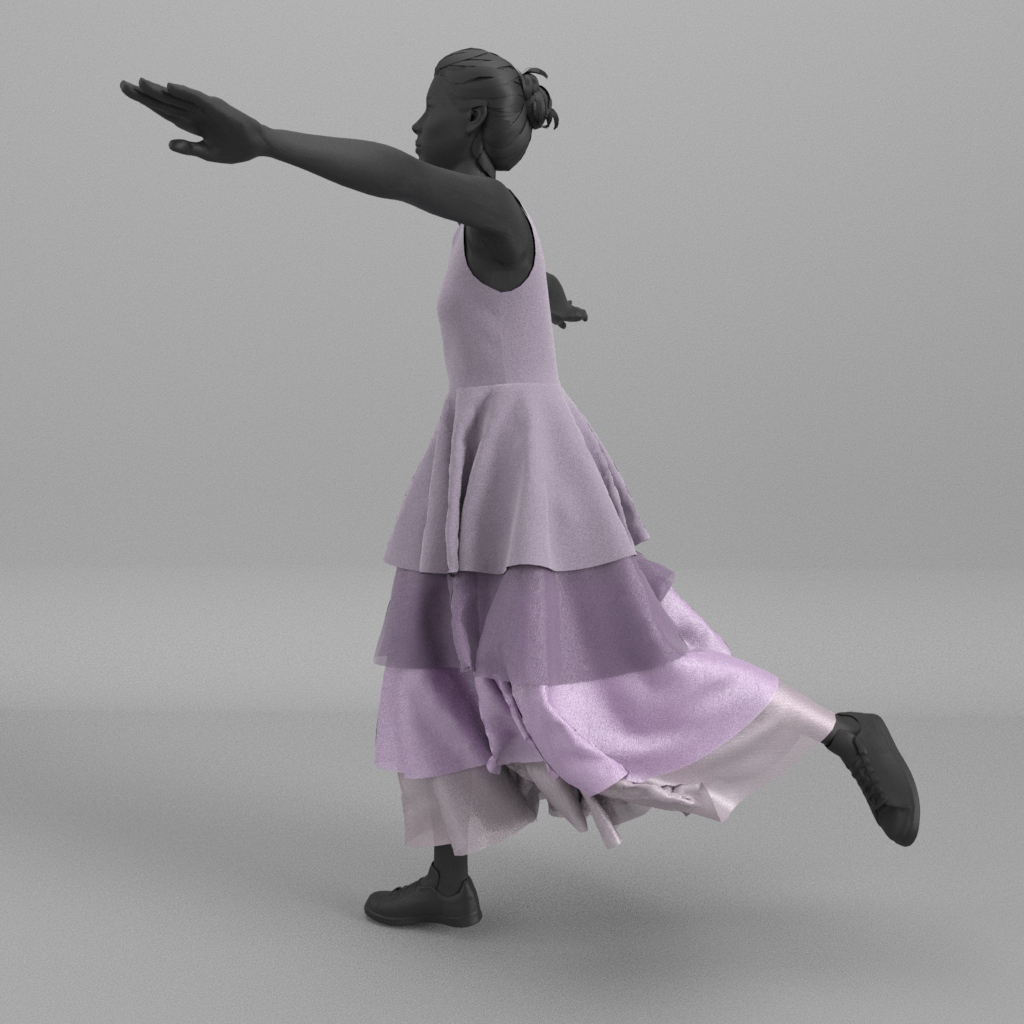}
          &
          \includegraphics[width=1.72in]{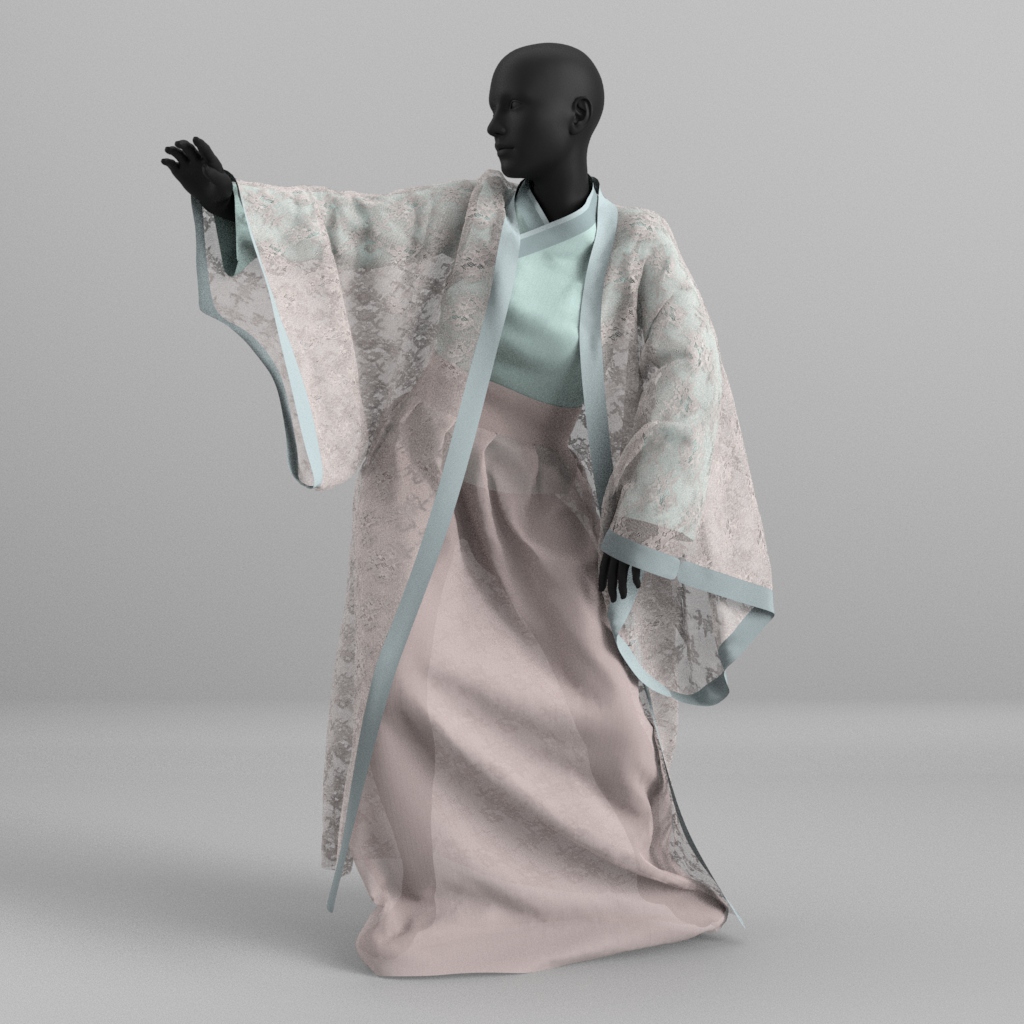}
          &
          \includegraphics[width=1.72in]{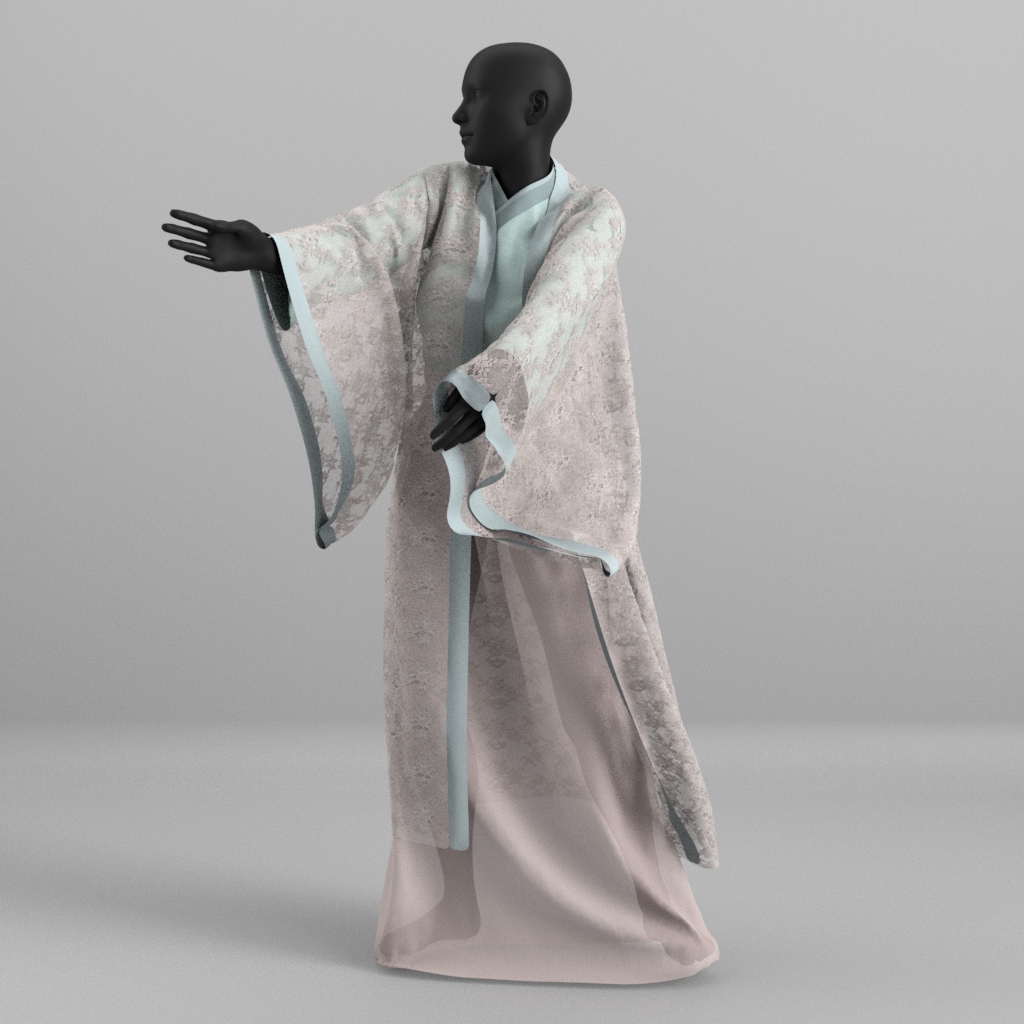}
          \\
          \multicolumn{2}{c}{A multi-layered dress}
          &
          \multicolumn{2}{c}{A multi-layered gown}
        \end{tabular}        
	\vspace*{-0.12in}
	\caption{\textbf{Multi-layered clothing.} Our method is capable of handling complex collisions reliably among multiple layers of clothing, produced by these dancing characters.}
	\label{fig:multi_layer_dress}
\end{figure*}

\section{Limitations and future works}
{
While the experiments demonstrate the efficiency, robustness and good quality of our collision handling algorithm, there is no guarantee of a globally convergent approximate solution to Eq.~\eqref{eq:objective}. In fact, compared to the recently developed incremental potential contact (IPC) method~\cite{li2020codimensional},
all step-and-project methods seem to lack theoretical guarantee of global convergence \textcolor{black}{in exchange for} numerical efficiency. Therefore, we do not compare our method with IPC directly in terms of performance for fairness reasons.
Although finding a convergent solution to Eq.~\eqref{eq:objective} could be over-demanding and unnecessary for a faithful simulation in graphics, analyzing the compromise made by step-and-project methods and further improving its convergence is definitely a valuable future work.
} Besides, our method does not consider friction modeling yet. Now it
simply imitates frictional effects using an additional velocity
filter. If trying to reproduce more realistic frictional contacts, the
method should handle collisions and frictions jointly instead.
Finally, we are interested in implementing our method on distributed
systems consisting of multiple GPUs for real-time performance on
large-scale scenes. \textcolor{black}{To sufficiently exploit the power
  of distributed systems,} we need to carefully study the
communications between tasks, processes and threads, and thus design
effective policy for parallelization and synchronization.


\bibliographystyle{ACM-Reference-Format}
\bibliography{lcp}

\end{document}